\documentclass{amsart}
\pdfoutput=1

\usepackage{amssymb}
\usepackage{amsthm}
\usepackage{amsfonts}
\usepackage{amsmath}
\usepackage{pmboxdraw}
\usepackage{verbatim} 
\usepackage{graphicx}
\usepackage{color}
\usepackage[colorlinks=true, citecolor=blue, filecolor=black, linkcolor=black, urlcolor=black]{hyperref}
\usepackage{cite}
\usepackage[normalem]{ulem}
\usepackage{subcaption}
\usepackage{bbm}
\usepackage{bm}
\usepackage{mathtools}
\usepackage{todonotes}
\usepackage{array}
\usepackage{multirow}
\usepackage{multirow,makecell}  
\allowdisplaybreaks
\usepackage{tikz}

\usepackage[table]{xcolor}
\usepackage{makecell}
\usepackage{booktabs}

\newcommand{\RN}[1]{%
	\textup{\uppercase\expandafter{\romannumeral#1}}%
}

\DeclareMathOperator{\Res}{Res}

\def\PP{\mathcal{P}}

\def\QQ{\mathcal{Q}}

\def\C{\mathbb{C}}
\def\D{\mathbb{D}}

\def\R{\mathbb{R}}

\def\T{\mathbb{T}}

\def\AAA{\mathsf{A}}
\def\QQQ{\mathsf{Q}} 
\def\RRR{\mathsf{R}}

\newcommand{\re}{\operatorname{Re}}
\newcommand{\im}{\operatorname{Im}}

\def\bigO{{\mathcal O}}

	\def\XXint#1#2#3{{\setbox0=\hbox{$#1{#2#3}{\int}$ }
			\vcenter{\hbox{$#2#3$ }}\kern-.59\wd0}}

\theoremstyle{plain}
\newtheorem{thm}{Theorem}[section]

\newtheorem{lem}[thm]{Lemma}

\newtheorem{prop}[thm]{Proposition}
\newtheorem*{assump}{Assumption}

\theoremstyle{remark}
\newtheorem{defn}{Definition}

\newtheorem{rem}{Remark}

\numberwithin{equation}{section}

\begin{document}

\title[Truncated unitary ensemble with a point charge]{Equilibrium measure for the truncated \\ unitary ensemble with a point charge}
\author{Sung-Soo Byun} 
\address{Department of Mathematical Sciences and Research Institute of Mathematics, Seoul National University, Seoul 08826, Republic of Korea;
	School of Mathematics, Korea Institute for Advanced Study, 85 Hoegi-ro, Dongdaemun-gu, Seoul 02455, Republic of Korea}
\email{sungsoobyun@snu.ac.kr}

\author{Christophe Charlier}
\address{Centre for Mathematical Sciences, Lund University, 22100 Lund, Sweden}
\email{christophe.charlier@math.lu.se}

\author{Philippe Moreillon}
\address{Section of Mathematics, University of Geneva, 1205 Geneva, Switzerland}
\email{Philippe.Moreillon@uclouvain.be}
 

\begin{abstract}
We explicitly determine the equilibrium measure associated with the truncated unitary ensemble in the presence of a point charge, in both the weak and strong non-unitarity regimes. Its support undergoes a topological phase transition as the location or strength of the point charge crosses a critical value. In the strong non-unitarity regime, the support changes from a doubly connected to a simply connected planar domain. In the weak non-unitarity regime, the corresponding transition is from the full unit circle to an arc. 
\end{abstract}

\maketitle

\section{Introduction}

Given a potential $V: \C \to \R \cup\{+\infty\}$, we consider $N$ particles $z_1,\ldots,z_N\in\C$ distributed according to the two-dimensional Coulomb gas measure \cite{Fo10,Se24,Le22,BF25a,BF25}
\begin{align}
\frac{1}{Z_{N,V}}  \prod_{1\le j<k \le N} |z_j-z_k|^\beta \prod_{j=1}^N e^{ -\frac{ \beta N}{2} V(z_j) }\,dA(z_j), \label{Coulomb gas 1} 
\end{align} 
where $\beta>0$ is the inverse temperature, $dA(z)=d^2z/\pi$ is the normalised area measure and $Z_{N,V}$ is the normalisation constant. 
When $\beta=2$, this ensemble corresponds to the eigenvalues of random normal matrices. 

The macroscopic behaviour of Coulomb gas ensembles is described by logarithmic potential theory via the notion of \emph{equilibrium measures} \cite{ST97}.
For a compactly supported probability measure $\mu$ on $\mathbb{C}$, we define its weighted logarithmic energy $I_V[\mu]$ by 
\begin{equation} \label{energy}
	I_V[\mu]:= \int_{ \mathbb{C}^2 } \log \frac{1}{ |z-w| }\, d\mu(z)\, d\mu(w) +\int_{ \mathbb{C} } V(z) \,d\mu(z) .
\end{equation} 
Under mild assumptions on \( V \), there exists a unique probability measure \( \mu_V \), called the equilibrium measure associated with \( V \), which minimises \( I_V[\mu] \) over all compactly supported Borel probability measures on $\C$.
Its support $S_V$ is called the droplet. 
It is well known that the empirical measure $\frac{1}{N}\sum_{j=1}^N\delta_{z_j}$ converges weakly almost surely to the equilibrium measure; see e.g. \cite[Theorem 3.3]{Se24}.  
A classical example is $V(z) = |z|^2$, for which the equilibrium measure is uniform on the unit disc (the so-called circular law of the Ginibre ensemble).

\medskip 

In this work, we determine the equilibrium measures associated with the following family of potentials.
For $\rho > 0$, $c \ge 0$ and $a \in [0,\sqrt{1+\rho}]$, we define the potential 
\begin{equation} \label{def of potential W strong}
	W(z) \equiv W_\rho(z) := 
	\begin{cases}
		-\rho \log \Big( 1- \dfrac{|z|^2}{1+\rho} \Big) -2c \log|z-a| &\textup{if } |z| < \sqrt{1+\rho},
		\smallskip 
		\\
		+\infty & \textup{if } |z| \ge \sqrt{1+\rho} .  
	\end{cases}
\end{equation}
Since $W$ is $+\infty$ outside the centred disc of radius $\sqrt{1+\rho}$, the particles are confined to this disc.

The first term in $W$ repels the particles from the circle $|z|=\sqrt{1+\rho}$, while the second repels them from the point $a$. The parameters $\rho$ and $c$ control the strengths of these two repulsions, respectively. These are sometimes called circular root-type and pointwise root-type singularities, respectively \cite{BC24,Noda25}.

The interplay between these two competing repulsive forces determines the
shape of the droplet. 
This suggests a topological phase transition. 
For ease of discussion, fix $\rho>0$ and $a\in(0,1)$. When $c$ is sufficiently small, the hole created around the point charge at $a$ remains separated from the exterior of the droplet. 
Consequently, the particles are expected to occupy a doubly connected domain; see
Figure~1~(A). As $c$ increases, the inner hole eventually merges with the exterior component, resulting in a simply connected droplet; see Figure~1~(B). We refer to these two regimes as the post-critical and pre-critical regimes, respectively. The transition can be induced by varying either the strength $c$ or the location $a$ of the point charge.
For convenience, we shall describe the phase boundary in terms of the
critical location $a_{\rm cri}$, which is given explicitly
in \eqref{def of c cri}.

To handle the case $\rho=0$, we also define
\begin{equation} \label{def of W weak}
	\mathsf{W}(z) := \begin{cases}
		-2c\log|z-a|  &\textup{if } |z|\leq 1,
		\smallskip 
		\\
		+\infty & \textup{if } |z| > 1 .  
	\end{cases}
\end{equation}
This case requires separate consideration, because the potential no longer repels the particles from the circle. Consequently, the particles are expected to accumulate near the unit circle (which acts as a hard wall). Thus, one expects the droplet to be two-dimensional for $\rho>0$ and one-dimensional for $\rho=0$.

\begin{figure}[t]
	\begin{subfigure}{0.32\textwidth}
		\begin{center}	
			\begin{tikzpicture}
				\begin{scope}
					\clip (0,0) circle (1.5);     
					\fill[gray!15] (0,0) circle (1.5); 
					\fill[white] (0.75,0) circle (0.5); 
				\end{scope}
				\draw[thick] (0,0) circle (2);
				\foreach \angle in {0, 30, 60, ..., 330} {
					\draw[->, thick, blue] ({2*cos(\angle)}, {2*sin(\angle)}) -- ({1.5*cos(\angle)}, {1.5*sin(\angle)});
				}
				\node at (0.75,0) {$a$}; \fill (0.75,0) circle (2pt);
				\foreach \angle in {0, 30, 60, ..., 330} {
					\draw[->, thick, red] (0.75,0) -- ({0.75 + 0.5*cos(\angle)}, {0.5*sin(\angle)});
				} 
			\end{tikzpicture}
		\end{center}
		\subcaption{Post-critical regime}
	\end{subfigure}	 \qquad 
	\begin{subfigure}{0.32\textwidth}
		\begin{center}		\begin{tikzpicture}
				\begin{scope}
					\clip (0,0) circle (1.5);     
					\fill[gray!15] (0,0) circle (1.5);  
					\fill[white] (0.75,0) circle (1);  
				\end{scope}
				\draw[thick] (0,0) circle (2);
				\foreach \angle in {0, 30, 60, ..., 330} {
					\draw[->, thick, blue] ({2*cos(\angle)}, {2*sin(\angle)}) -- ({1.5*cos(\angle)}, {1.5*sin(\angle)});
				}
				\node at (0.75,0) {$a$}; \fill (0.75,0) circle (2pt);
				\foreach \angle in {0, 30, 60, ..., 330} {
					\draw[->, thick, red] (0.75,0) -- ({0.75 + 1*cos(\angle)}, {1*sin(\angle)});
				} 
			\end{tikzpicture}
		\end{center}
		\subcaption{Pre-critical regime}
	\end{subfigure}	 
	\caption{The parameters $\rho$ and $a$ are the same in both images. The grey area represents the droplet, which is doubly connected for small $c>0$ (left) and simply connected for sufficiently large $c$ (right).} \label{Fig_competing forces}
\end{figure}
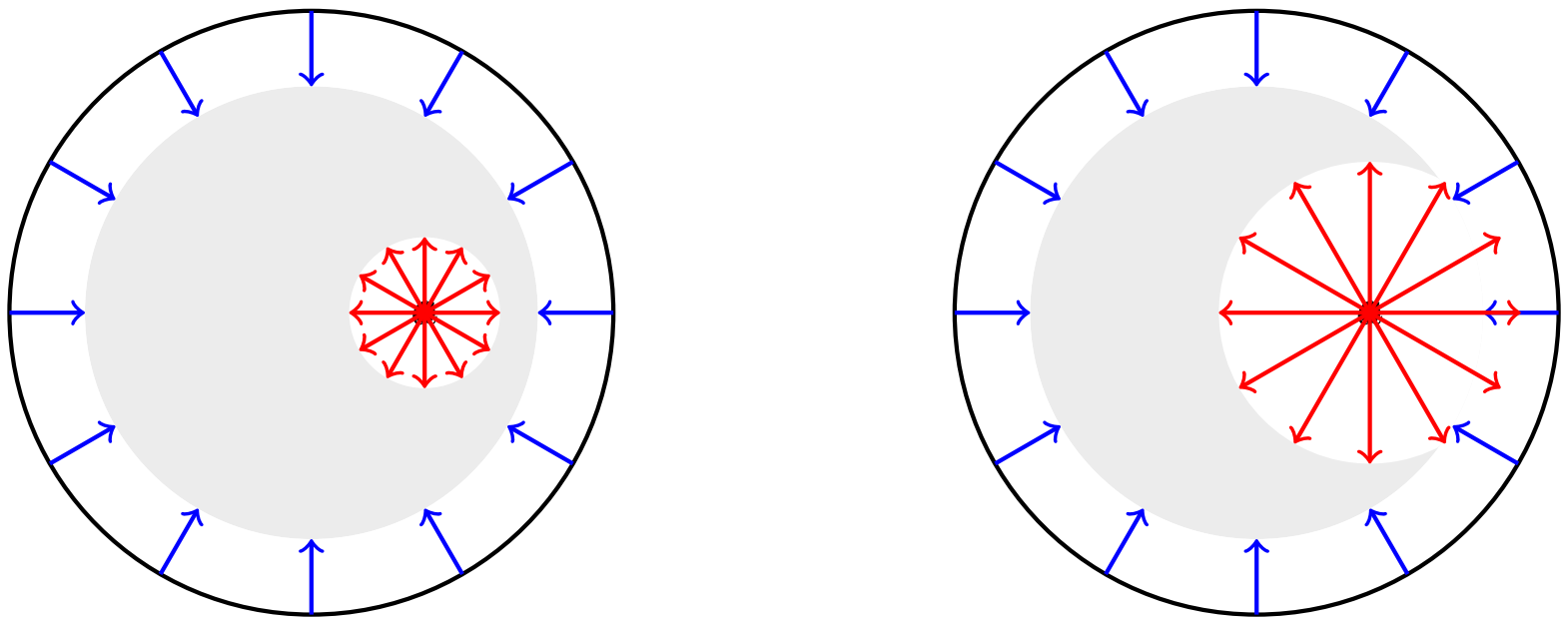

The potentials $W$ and $\mathsf{W}$ also arise naturally in random matrix theory. For $\beta=2$ and $V=W$ with $c = 0$, the Coulomb gas \eqref{Coulomb gas 1} describes the eigenvalues of the truncated unitary ensemble (rescaled by $\sqrt{1+\rho}$). 
This ensemble is obtained by taking the top-left \(N\times N\) submatrix of an \(M\times M\) Haar-distributed unitary matrix.
The parameter $\rho$ in \eqref{def of potential W strong} is related to $M$ via 
$
\rho +1= (M-1)/N.
$ 
Furthermore, the truncated ensemble interpolates between the circular unitary ensemble ($\rho \to 0$) and the Ginibre ensemble ($\rho \to \infty$). We refer to \cite[Sections 2.6 and 11.4]{BF25} for more background on these models.

The potentials $W_\rho$ ($\rho>0$) and $\mathsf{W}$ ($\rho=0$) correspond to the \emph{strong non-unitarity} and \emph{weak non-unitarity} regimes, respectively. The scaling in \eqref{def of potential W strong} is chosen so that the droplet is the unit disc when $c=0$.
When $\beta=2$ and $cN$ is a positive integer, these Coulomb gases also correspond to the $N$ eigenvalues of truncated ensembles of size $(c+1)N$, conditioned to have $cN$ eigenvalues at $a$. 
The case $c>0$ also naturally arises in the study of moments of characteristic polynomials; see e.g. \cite{BSY24} and the references therein.  

In this work, we determine the equilibrium measures associated with
the potentials \(\mathsf W\) and \(W\). Our main results are
summarised in Table~\ref{Table_summary} and will be presented in
detail in the following section.

\begin{table}[b]
	\centering
	\renewcommand{\arraystretch}{1.35}
	\setlength{\tabcolsep}{14pt}
	
	\begin{tabular}{lcc}
		\toprule
		\rowcolor{gray!12}
		&
		Post-critical regime ($a<a_{\rm cri}$)
		&
		Pre-critical regime ($a>a_{\rm cri}$)
		\\
		\midrule
		
		Weak non-unitarity
		& Theorem~\ref{Thm_1D eq msr} (a)
		& Theorem~\ref{Thm_1D eq msr} (b)
		\\
		
		\multicolumn{1}{c}{\textit{Droplet}}
		& \textit{full unit circle}
		& \textit{subarc of the unit circle}
		\\[0.15em]
		
		\midrule
		
		Strong non-unitarity
		& Theorem~\ref{Thm_2D eq msr} (a)
		& Theorem~\ref{Thm_2D eq msr} (b)
		\\
		
		\multicolumn{1}{c}{\textit{Droplet}}
		& \textit{doubly connected}
		& \textit{simply connected}
		\\
		
		\bottomrule
	\end{tabular}
	
	\caption{Summary of results.}
	\label{Table_summary}
\end{table}

\bigskip

\section{Main results}

\subsection{Equilibrium measure at weak non-unitarity}

We first discuss the weak non-unitarity regime, corresponding to $\rho=0$. For fixed $c$, we find that the droplet is the whole unit circle for sufficiently small values of $a$, whereas it is an arc of the unit circle for sufficiently large values of $a$.

\begin{thm}[\textbf{Equilibrium measure at weak non-unitarity}] \label{Thm_1D eq msr}
	 Let $c \ge 0$ and $a \in [0,1]$.  The equilibrium measure $\mu_{ \mathsf{W} }$ associated with the potential \eqref{def of W weak} is given as follows.
	\begin{itemize}
		\item[(a)] \textup{\textbf{(Post-critical regime)}} If $a \leq \frac{1}{1+2c}$, $\mu_{ \mathsf{W} }$ is supported on the entire unit circle and is given by 
\begin{equation} \label{def of eq msr 1D post}
d\mu_{ \mathsf{W} }(\theta) = \frac{1}{2\pi} \Big( 1+c - \frac{c(1-a^2)}{1+a^2-2a\cos\theta} \Big) \cdot \mathbbm{1}_{ [-\pi,\pi] }(\theta) \,d\theta. 
\end{equation}
		\item[(b)] \textup{\textbf{(Pre-critical regime)}} If $\frac{1}{1+2c} < a  \leq 1$, $\mu_{ \mathsf{W} }$ is supported on an arc of the unit circle and is given by
		\begin{equation} \label{def of eq msr 1D pre}
			d\mu_{ \mathsf{W} }(\theta) =  \frac{1}{\pi} \,  \sqrt{a} \, \frac{  |\sin(\theta/2)|  \sqrt{ (a+1)^2(2c+1) -4a(c+1)^2 \cos^2(\theta/2)  }   }{    (a+1)^2 -4a  \cos^2(\theta/2) }  \cdot \mathbbm{1}_{ [-\pi,\pi] \setminus (-\theta_*,\theta_*) }(\theta) \,d\theta, 
		\end{equation}    
		where 
		\begin{equation} \label{def of theta star support}
			\theta_*  := 2 \, \arccos \Big( \frac{ a+1 }{ 2\sqrt{a} }  \frac{ \sqrt{2c+1} }{c+1} \Big) = 2\, \arctan \Big(  \frac{1}{a+1} \sqrt{ \frac{ (2c+1-a)( a(2c+1)-1) }{2c+1}  }  \Big) .   
		\end{equation}  
	\end{itemize}
\end{thm}

Note that the formulas \eqref{def of eq msr 1D post} and \eqref{def of eq msr 1D pre} agree when $a=1/(1+2c)$.

\begin{rem}[Monotonicity of the density]
	The densities in \eqref{def of eq msr 1D post} and \eqref{def of eq msr 1D pre} are even in $\theta$. Moreover, a direct computation shows that for $a>0$ and $c>0$, 
	$$
	\frac{\partial }{\partial \theta} \frac{d\mu_{ \mathsf{W} }(\theta)}{ d\theta } > 0 
	$$
	for $0<\theta<\pi$ in the post-critical regime and $\theta_*<\theta<\pi$ in the pre-critical regime.
Therefore, both densities in \eqref{def of eq msr 1D post} and \eqref{def of eq msr 1D pre} increase strictly as $|\theta|$ grows. This behaviour is expected, since $1$ is the point on the unit circle closest to the repulsive charge at $a>0$.
\end{rem}

\begin{rem}[Harmonic measure interpretation]
In the post-critical regime, the equilibrium measure
\eqref{def of eq msr 1D post} can be written as 
\begin{equation} \label{eq msr 1D post decomp}
  d\mu_{\mathsf W}=(1+c)\,d\sigma-c\,d\omega_a,
\end{equation} 
where $d\sigma(\theta)$ is the uniform probability measure on the unit circle and
\begin{equation}
    d\omega_a(\theta)
    =
    \frac{1}{2\pi}
    \frac{1-a^2}{1+a^2-2a\cos\theta}\,d\theta
\end{equation} 
is the harmonic measure of the unit disc viewed from the point
$a\in\mathbb D$. From \eqref{eq msr 1D post decomp}, the effect of the point charge at $a$ can be interpreted as removing \(c\) times the harmonic measure $\omega_a$ from \((1+c)\) times  the uniform measure $\sigma$. Note also that when \(c=0\), the equilibrium measure reduces to $\sigma$. 
\end{rem}
 
\begin{rem}[Stereographic representation] \label{Rem_stereographic}
Consider the stereographic parametrisation
\[
    e^{i\theta}=\frac{1+ix}{1-ix},
    \qquad
    x=\tan\frac{\theta}{2},
    \qquad
    -\pi<\theta<\pi.
\]
Let \(\nu\) be the push-forward of \(\mu_{\mathsf W}\) under the map
\(x=\tan(\theta/2)\), i.e. 
$d\nu(x)=\frac{d\mu_{\mathsf W}(\theta)}{d\theta}|_{\theta = 2\arctan x}\,\frac{2\,dx}{1+x^2}.$
Then we have the following representations.

\begin{itemize}
    \item In the post-critical regime,
    \begin{equation} \label{weak post density real line}
       d\nu(x)
        =
        \frac{1}{\pi}
        \left(
            \frac{1+c}{1+x^2}
            -
            \frac{c\, b}{x^2+b^2}
        \right)\,dx, \qquad    b:=\frac{1-a}{1+a}. 
    \end{equation} 

    \item In the pre-critical regime,
    \begin{equation}\label{weak pre density real line}
        d\nu(x)
        =
        \frac{2\sqrt{a(2c+1)}}{\pi(1+a)}
        \frac{|x|\sqrt{x^2-x_*^2}}
        {(x^2+b^2)(x^2+1)}
        \mathbbm{1}_{\{|x|\geq x_*\}}\,dx,
    \end{equation}
    where $b$ is given in \eqref{weak post density real line}, and 
    \begin{equation} \label{def of x star real line}
      x_*:=\tan\frac{\theta_*}{2}
        =
        \frac{1}{1+a}
        \sqrt{
            \frac{(2c+1-a)(a(2c+1)-1)}{2c+1}
        }.
    \end{equation} 
\end{itemize}
The push-forward measure $\nu$ will play an important role in the proof of Theorem~\ref{Thm_1D eq msr} (b). We also note that elementary integration of \eqref{weak post density real line} and \eqref{weak pre density real line} gives $\nu(\mathbb R)=1$, which implies $\mu_{\mathsf W}([-\pi,\pi])=1$.
\end{rem}

\subsection{Equilibrium measure at strong non-unitarity}
We now consider the case of strong non-unitarity, in which the equilibrium measure is supported on a two-dimensional domain. This regime is significantly more complicated than the weak non-unitarity regime.
 
We first recall the known results for the Ginibre ensemble, corresponding to the limit $\rho\to\infty$. 
In this case, the droplet was obtained explicitly in \cite{BBLM15}. 
It undergoes a topological phase transition at the critical value
\begin{equation} \label{critical values Ginibre}
a_{ \rm cri }^{ \rm g }:= \sqrt{c+1}-\sqrt{c} .
\end{equation} 
For $c>0$, the droplet is given as follows (see \cite[Section 2]{BBLM15}). 
\begin{itemize}
		\item \textbf{Post-critical regime for $\rho=+\infty$}: for  $a \leq  a_{\rm cri}^{ \rm g }$, the droplet is given by 
		\begin{equation} \label{droplet post Ginibre}
			S=\overline{\mathbb{D}(0,\sqrt{1+c})} \setminus \mathbb{D}(a,\sqrt{c}),
		\end{equation}
		where $\mathbb{D}(z,r)$ is the open disc with centre $z$ and radius $r>0.$
		\smallskip 
		\item \textbf{Pre-critical regime for $\rho=+\infty$}: for $a > a_{\rm cri}^{ \rm g }$, the droplet is a simply connected domain whose conformal radius $R$ at $\infty$ is given by 
		\begin{equation} \label{def of R f Ginibre}
			R=\frac{1+a^2q^2}{2a q}. 
		\end{equation}   
		The boundary of the droplet is the image of the unit circle under the  exterior conformal map 
		\begin{equation} \label{f conformal Ginibre}
			f(z)=R\,z-\frac{\kappa}{z-q}-\frac{\kappa}{q}, \qquad \kappa=(R-aq)(1-q^2). 
		\end{equation}
		Here, $q$ is the unique root of the equation 
		\begin{equation} \label{q equation}
			q^6 - \frac{a^2+4c+2}{2a^2}  q^4+\frac{1}{2a^4} =0  
		\end{equation}
		such that $0<q<1$ and $\kappa>0.$ It also satisfies $f(1/q)=a$.
	\end{itemize}

We now discuss our results, valid for $\rho\in (0,+\infty)$ and $c>0$. Recall that $W$ is given by \eqref{def of potential W strong}.  
The topological transition of the droplet can be described in terms of the location $a$. More precisely, we define 
\begin{equation} \label{def of c cri} 
    a_{ \rm cri } := \frac{ \sqrt{\rho+1}    }{ \sqrt{ ( \rho+c+1 )(c+1) }+\sqrt{ (\rho+c)c } }. 
\end{equation}
Note that as $\rho \to +\infty$, $a_{\rm cri}$ converges to $a_{\rm cri}^{\rm g}$ (defined in \eqref{critical values Ginibre}). By analogy with \cite{BBLM15}, we call the regime $a\leq a_{\rm cri}$ the post-critical regime and the regime
$a > a_{\rm cri}$ the pre-critical regime.

\begin{defn}[Parameters in the description of the droplet] Let $\rho>0$, $c>0$ and $a\in[0,\sqrt{1+\rho})$.
	
	For the post-critical regime, we define 
	\begin{align}
		& \mathcal{A} \equiv \mathcal{A}(a,\rho,c) := \frac{ a \rho(\rho+1) }{ \rho( \rho+1) +  c(\rho+1-a^2) }, \label{def of A center}
		\\
		& r_0 \equiv r_0(a,\rho,c) := \frac{ (1+\rho-a^2)  \sqrt{ c (\rho+1)(\rho+c) } }{ \rho( \rho+1) +  c(\rho+1-a^2)  },  \label{def of r0}
		\\
		& r_1 \equiv r_1( \rho,c ) :=  \sqrt{  \frac{(c+1)(\rho+1)}{ \rho+c+1 } }. \label{def of r1}  
	\end{align} 
    
For the pre-critical regime, the description of the droplet is much more complicated and given in terms of interrelated parameters. Let $\QQQ\in(0,\min\{1,a^{-1}\})$ be a parameter to be determined below. We first define a polynomial $\PP$ of degree at most six by
	\begin{align} \label{def of polynomial PP}
		\PP(y) := \PP_0(y)+ \frac{\PP_1(y)}{ 1+\rho-a^2 }, 
	\end{align}
	where 
	\begin{align}
		\begin{split}
			\PP_0(y)& :=  (1-a^2)  \Big(y-\frac{ 1-\QQQ }{2}  \Big)^2 \Big(y+\frac{ 1+\QQQ }{2}  \Big) ^2 \Big(  y +\frac{1+a}{1-a}\frac{ 1+\QQQ}{2} \Big)   \Big( y +\frac{1-a}{1+a}\frac{1+\QQQ}{2} \Big) 
			\\
			&\quad + 4 a^2 \QQQ y  \Big(y-\frac{ 1-\QQQ }{2}  \Big) \Big(y+\frac{ 1+\QQQ }{2}  \Big)  \Big(y-\frac{ 1+\QQQ }{2}  \Big)^2,
			\\
			\PP_1(y) &: = \QQQ a^2 y^2 \bigg[  (a^2+1)\QQQ(2y-1-\QQQ)^2+2(2y+1+\QQQ)(2y-1-\QQQ)+8\QQQ(1+\QQQ)y \bigg] .  
		\end{split}
	\end{align}  
	When $a=1$, the polynomial $\PP$ is understood by continuous extension. 
	Let 
	\begin{equation} \label{relation X and Y}
		X=(1+\rho-a^2)Y, 
	\end{equation}
	where $Y$ is the unique root of $\PP$ in the interval $ ( \frac{1-\QQQ}{2}, \frac{1+\QQQ}{2} )$. (The existence and uniqueness of $Y$ follow from Lemma~\ref{Lem_poly PP unique root}.)
	With $X$ given by \eqref{relation X and Y}, we define 
	\begin{equation}
	 \label{RRR in terms of X}
	\RRR  = \frac{1}{2a X} \Big( X^2  + \QQQ(1+\rho+a^2)  X  - \frac{ (1-\QQQ^2)(1+\rho-a^2)^2 }{ 4  }  \Big),		
	\end{equation}
	and 
	\begin{align} 
		\begin{split} \label{AAA in terms of X}
		\AAA&= \frac{1}{ 2a  X \QQQ(1-\QQQ)  }    \Big( X^2 - \frac{ (1+\QQQ)^2(1+\rho-a^2)^2 }{ 4 } \Big)       ^{-1} 
		\\
		&\quad \times  \Big(  X^2   + \QQQ(\sqrt{1+\rho} + a )^2  X  - \frac{ (1-\QQQ^2)(1+\rho-a^2)^2 }{ 4   }  \Big) 
		\\
		&\quad \times \Big(  X^2   + \QQQ(\sqrt{1+\rho} - a )^2  X  - \frac{ (1-\QQQ^2)(1+\rho-a^2)^2 }{ 4   }  \Big). 
	\end{split}
	\end{align}
	Finally, $\QQQ$ is determined by the equation
	\begin{equation} \label{c in terms of R Q A}
		c=-(1+\rho)  \frac{   \RRR( \AAA \QQQ  - \AAA \QQQ^3 - \RRR)+ \QQQ^2 }{ \RRR( \AAA \QQQ  - \AAA \QQQ^3 - \RRR) + \QQQ^2 (1 + \rho)  } , 
	\end{equation}
	under the constraints that 
	\begin{align}\label{lol18}
	Y\in\Big(\frac{1-\QQQ}{2},\frac{1+\QQQ}{2}\Big),
\qquad
0<\QQQ<\min\{1,a^{-1}\}.
	\end{align}
	 Proposition~\ref{Prop_admissible pre parameters} guarantees that the system consisting of $\PP(Y)=0$ and \eqref{c in terms of R Q A} has a unique solution $(Y,\QQQ)$ satisfying \eqref{lol18}.
\end{defn}

We are now ready to state our result. Recall that $a_{ \rm cri }$ is given by \eqref{def of c cri}.  

	\begin{thm}[\textbf{Equilibrium measure at strong non-unitarity}] \label{Thm_2D eq msr}  Let $\rho > 0$, $c > 0$ and $a \in [0,\sqrt{1+\rho})$. The equilibrium measure $\mu_W$ associated with the potential \eqref{def of potential W strong} is given by
		\begin{equation} \label{def of mu W 2D main}
			d\mu_W(z)= \frac{ \rho (1+\rho) }{ (1+\rho-|z|^2)^2} \cdot \mathbbm{1}_{ \{ z \in S_W \} } \,dA(z).
		\end{equation}
		Here, the droplet $ S_W$ is given as follows.
		\begin{itemize}
			\item[(a)] \textbf{\textup{(Post-critical regime)}} Suppose that $a \le a_{ \rm cri } $. Then the droplet $S_W$ is given by
			\begin{equation} \label{droplet post trunc}
				S_W = \overline{ \mathbb{D}(0,r_1) } \setminus \mathbb{D }(\mathcal{A}, r_0 ),  
			\end{equation}
			where $r_0, r_1$ and $\mathcal{A}$ are given by \eqref{def of r0}, \eqref{def of r1}, and \eqref{def of A center}, respectively. 
			\smallskip 
			\item[(b)] \textbf{\textup{(Pre-critical regime)}} Suppose that $a > a_{ \rm cri } $. Then the droplet $S_W$ is a simply connected compact set whose boundary is given by the image of the unit circle under the conformal map 
			\begin{equation} \label{def of f pre trunc main}
				f_\rho(z)= \RRR \, z - \frac{ \mathcal{K} }{ z-\QQQ }- \frac{ \mathcal{K} }{ \QQQ }, \qquad  \mathcal{K}= ( \RRR-\AAA\QQQ )(1-\QQQ^2) .
			\end{equation}
			Here, $\RRR$ and $\AAA$ are given by \eqref{RRR in terms of X} and \eqref{AAA in terms of X}, respectively.  
		\end{itemize} 
	\end{thm}

\begin{figure}[b]
	\begin{subfigure}{0.32\textwidth}
		\begin{center}	
			\includegraphics[width=\textwidth]{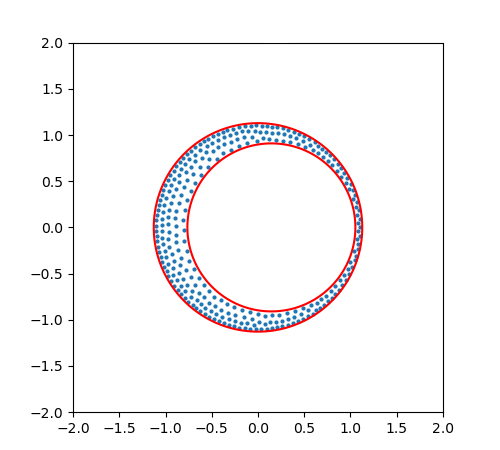}
		\end{center}
		\subcaption{$c=0.75$, $a=0.25$, $\rho=1$}
	\end{subfigure}	
	\begin{subfigure}{0.32\textwidth}
		\begin{center}			\includegraphics[width=\textwidth]{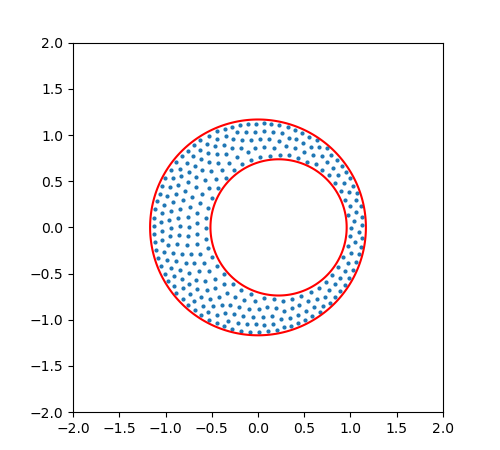}
		\end{center}
		\subcaption{$c=0.5$, $a=0.25$, $\rho=4$}
	\end{subfigure}	
	\begin{subfigure}{0.32\textwidth}
		\begin{center}	
			\includegraphics[width=\textwidth]{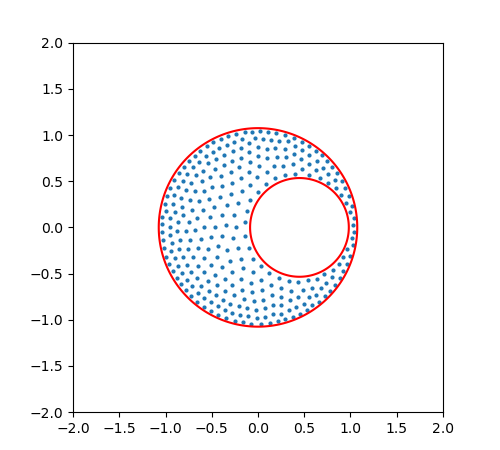}
		\end{center}
		\subcaption{$c=0.25$, $a=0.5$, $\rho=2$}
	\end{subfigure}

	\begin{subfigure}{0.32\textwidth}
		\begin{center}	
			\includegraphics[width=\textwidth]{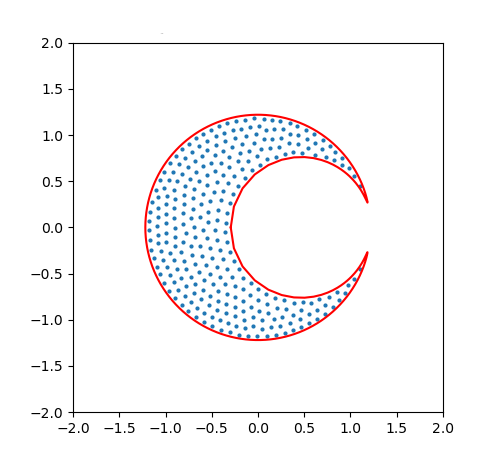}
		\end{center}
		\subcaption{$\QQQ=99/100$, $c \approx 0.58$}
	\end{subfigure}	
	\begin{subfigure}{0.32\textwidth}
		\begin{center}			\includegraphics[width=\textwidth]{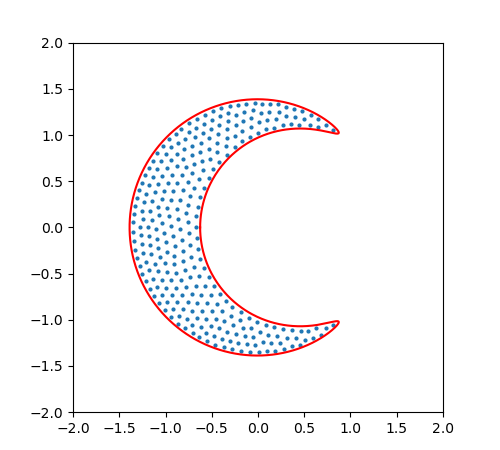}
		\end{center}
		\subcaption{$\QQQ=7/8$, $c \approx 1.20$}
	\end{subfigure}	
	\begin{subfigure}{0.32\textwidth}
		\begin{center}	
			\includegraphics[width=\textwidth]{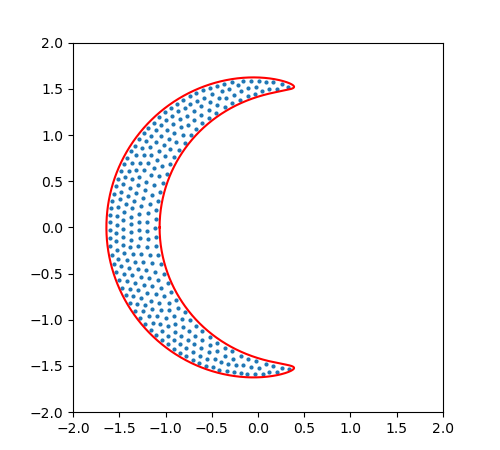}
		\end{center}
		\subcaption{$\QQQ=3/4$, $c \approx 2.47$}
	\end{subfigure}
	\caption{The red curves in (A)---(C) show the boundary of the droplet in the post-critical regime, while the red curves in (D)---(F) show the boundary of the droplet in the pre-critical regime with $\rho=8$ and $a=1/2$. The blue dots are the corresponding Fekete points with $N=400$.} \label{Fig_droplet}
\end{figure}

See Figure~\ref{Fig_droplet} for a numerical illustration of Theorem~\ref{Thm_2D eq msr}, where we compare our analytic results with Fekete point configurations (these configurations maximise the density \eqref{Coulomb gas 1}). 
In practice, it is more convenient to first choose a parameter $\QQQ\in (0,\min\{1,a^{-1}\})$, and then define $X$, $\RRR$, and $\AAA$ as above. Then the associated point charge $c$ is determined by the relation \eqref{c in terms of R Q A}. 
For instance, in Figure~\ref{Fig_droplet} (D)--(F), we first choose $\QQQ$ and then determine $c$ using \eqref{c in terms of R Q A}.

\begin{rem}[Radially symmetric case]
In the radially symmetric case, corresponding to either $a=0$ or $c=0$, 
the droplet is given by the annulus 
\begin{equation}
	S_W\big|_{a=0 \textup{ or } c=0} =  \Big\{  \sqrt{ \tfrac{c(\rho+1)}{ \rho+c } }  \le |z| \le  \sqrt{ \tfrac{ (c+1)(\rho+1) }{ \rho+c+1 }  }\Big\}. 
\end{equation}
This agrees with the droplet obtained from the well-known theory for
radially symmetric potentials; see \cite[Section IV.6]{ST97}. Therefore, throughout
the proof, we restrict ourselves to the non-radially symmetric case
$a>0$ and $c>0$. 
\end{rem}

\subsection{Related work and discussions}

We now discuss some related work and provide further motivation for the problems considered here. 

\begin{rem}[Connection with the companion paper \cite{BCMS25}]
\label{Rem_companion paper}
One of the main motivations for studying the equilibrium problems in
Theorems~\ref{Thm_1D eq msr} and \ref{Thm_2D eq msr} comes from our
companion paper \cite{BCMS25}, where we investigate precise large
deviations for geometric last passage percolation (LPP) and their
connections with random matrix theory.

A key ingredient in \cite{BCMS25} is a duality formula between geometric LPP
and the truncated unitary ensemble. Through this formula, quantities
associated with the distribution of the last passage time are
expressed in terms of moments of characteristic polynomials of
truncated unitary matrices. Equivalently, these moments can be written
as the Coulomb gas partition function \(Z_{N,V}\) in
\eqref{Coulomb gas 1}, with $\beta=2$ and the external potential \(V\) given by
\(W\) in \eqref{def of potential W strong} or by \(\mathsf W\) in
\eqref{def of W weak}, depending on the truncation regime.

The duality formula also relates the truncation regime to the geometry of the LPP lattice.
On the random-matrix side, weak and strong non-unitarity are distinguished by the truncation scale. On the LPP side, the corresponding distinction concerns the asymptotic shape of the underlying lattice: weak non-unitarity corresponds to an
asymptotically square lattice, whereas strong non-unitarity corresponds to a rectangular lattice. 

There is a further, and more significant, correspondence that provides
a central motivation for the present work. For geometric LPP,
Johansson \cite{Joh00} identified the law-of-large-numbers value of
the last passage time (commonly known as the \emph{shape function}; see \cite[Eq.~(1.3)]{BCMS25}). This
typical value serves as the threshold separating the lower- and
upper-tail large-deviation regimes. Through the duality established in
\cite{BCMS25}, these two large-deviation regimes correspond to two
distinct asymptotic regimes for the characteristic polynomial moments
of the truncated unitary ensemble. What remained unclear, however, was how this transition arises in the associated random matrix equilibrium problem. 

The results of the present paper provide this missing
potential theoretic interpretation: the transition between the two
asymptotic regimes corresponds precisely to the topological phase
transition in the support of the equilibrium measure described in
Theorems~\ref{Thm_1D eq msr} and \ref{Thm_2D eq msr}. In particular,
the law-of-large-numbers threshold determined by the LPP shape
function in \cite{Joh00} corresponds, on the random-matrix side, to
the critical value in \eqref{def of c cri}. Thus, the equilibrium
problems studied here reveal the geometric mechanism underlying the
phase transition that appears in the characteristic polynomial
moments in \cite{BCMS25}.
\end{rem}

\begin{rem}[Comparison with the results from \cite{BBLM15}] Another motivation for the present work is to extend the well-known results of \cite{BBLM15} for the Ginibre ensemble to the truncated unitary ensemble. More precisely, Theorem~\ref{Thm_2D eq msr} generalises the equilibrium-measure results of \cite{BBLM15}, which correspond to the limiting case \(\rho=+\infty\).

An important distinction between the two settings lies in the density
of the equilibrium measure. In the Ginibre case \(\rho=+\infty\), the
equilibrium measure has uniform density on its support, whereas for $0<\rho<+\infty$ the density is non-uniform; see
\eqref{def of mu W 2D main}. The uniform density in the Ginibre case considerably simplifies the analysis and allows one to employ a broader range of techniques; see, for instance, \cite{Sa92,LM16,RM25}. For finite $\rho$, the analysis is therefore more involved, although the droplet undergoes a similar topological phase transition. Below, we provide a more detailed comparison between Theorem~\ref{Thm_2D eq msr} and the results of \cite{BBLM15}.

We first discuss the post-critical regime. For both $\rho \in (0,+\infty)$ and $\rho=+\infty$, the droplet is a closed disc with an open disc removed; see \eqref{droplet post Ginibre} and \eqref{droplet post trunc}. However, for $\rho=+\infty$, the centre of the smaller disc corresponds exactly to the point $a$ where the point charge is inserted, while for finite $\rho$ the centre is generally not $a$ but is more complicated and given by $\mathcal{A}$ in \eqref{def of A center} (with $\mathcal A<a$ whenever $a>0$). 
For the $\rho \in (0,+\infty)$ case, the shape of the hole is a disc (this fact is a priori not obvious since the density of the equilibrium measure is not constant). 
        
        By \eqref{def of A center}, \eqref{def of r0} and \eqref{def of r1}, as $\rho \to + \infty$ we have
        \begin{align}\label{Acal convergence}
        \mathcal{A} = a+ O( \frac{1}{\rho} ), \qquad r_0 = \sqrt{c}+ O( \frac{1}{\rho} ), \qquad r_1 = \sqrt{c+1}+ O( \frac{1}{\rho} ),
        \end{align}
        which shows that the droplet \eqref{droplet post trunc} tends to \eqref{droplet post Ginibre} as $\rho \to + \infty$. 

\medskip 

Next, we discuss the pre-critical regime. The conformal maps \eqref{f conformal Ginibre} and
\eqref{def of f pre trunc main} have the same rational structure:
each has a simple pole in the unit disc, located at \(q\) and
\(\QQQ\), respectively, as well as a pole at infinity. Moreover,
by the definitions of \(\kappa\) and \(\mathcal K\), they satisfy $ f(1/q)=a$ and $f_\rho(1/\QQQ)=\AAA.$  
Thus, \(\RRR\) and \(\AAA\) play the roles of \(R\) and \(a\),
respectively, for finite \(\rho\). 
While $R$ is expressed directly in terms of $a$ and $q$ by \eqref{def of R f Ginibre}, the parameters \(\RRR\) and \(\AAA\) are determined through the auxiliary parameter \(X\) by \eqref{RRR in terms of X} and \eqref{AAA in terms of X}.

We now show that the pre-critical result of \cite{BBLM15} is recovered
as \(\rho\to+\infty\). We make the dependence on \(\rho\) explicit by
writing $\QQQ=\QQQ_\rho, Y=Y_\rho, X=X_\rho, \RRR=\RRR_\rho,$ and $\AAA=\AAA_\rho.$
Fix \(a\) and \(c\) in the strict Ginibre pre-critical regime, and let
\(q\) be the unique admissible solution of \eqref{q equation}.

By \eqref{def of polynomial PP}, a direct computation gives 
\begin{align*}
 &\PP_0\Big(\frac{1-\QQQ_\rho}{2}\Big)=0, \qquad  \PP_0'\Big(\frac{1-\QQQ_\rho}{2}\Big) =2a^2\QQQ_\rho^3(1-\QQQ_\rho), 
 \qquad \PP_1\Big(\frac{1-\QQQ_\rho}{2}\Big)
    =
    -a^2\QQQ_\rho^2(1-\QQQ_\rho)^2
    (1-a^2\QQQ_\rho^2).
\end{align*} 
Therefore, expanding $\PP(Y_\rho)=0$ as $\rho\to+\infty$, we obtain
\[
    Y_\rho
    =
    \frac{1-\QQQ_\rho}{2}
    +
    \frac{
        (1-\QQQ_\rho)(1-a^2\QQQ_\rho^2)
    }{
        2\QQQ_\rho(1+\rho-a^2)
    }
    +O(\rho^{-2}), \qquad \mbox{as } \rho \to + \infty.
\]
By \eqref{relation X and Y}, it follows that
\[
    X_\rho
    =
    \frac{1-\QQQ_\rho}{2}(1+\rho-a^2)
    +
    \frac{
        (1-\QQQ_\rho)(1-a^2\QQQ_\rho^2)
    }{
        2\QQQ_\rho
    }
    +O(\rho^{-1}), \qquad \mbox{as } \rho \to + \infty.
\]

Substituting this expression into \eqref{RRR in terms of X} and
\eqref{AAA in terms of X}, we find
\[
    \RRR_\rho
    =
    \frac{1+a^2\QQQ_\rho^2}{2a\QQQ_\rho}
    +O(\rho^{-1}),
    \qquad
    \AAA_\rho=a+O(\rho^{-1}), \qquad   \mathcal K_\rho
    =
    \frac{
        (1-\QQQ_\rho^2)(1-a^2\QQQ_\rho^2)
    }{
        2a\QQQ_\rho
    }
    +O(\rho^{-1}). 
\]  
It remains to determine the limit of \(\QQQ_\rho\). Rewriting
\eqref{c in terms of R Q A}, we have
\[
    \frac{
        \RRR_\rho
        \bigl(
            \RRR_\rho
            -\AAA_\rho\QQQ_\rho(1-\QQQ_\rho^2)
        \bigr)
    }{
        \QQQ_\rho^2
    }
    =
    \frac{(1+c)(1+\rho)}{1+\rho+c}.
\]
Using the above asymptotics, we obtain  
\begin{align*}
\mathsf Q_\rho^6
-\frac{a^2+4c+2}{2a^2}\mathsf Q_\rho^4
+\frac{1}{2a^4}
=O(\rho^{-1}), \qquad \mbox{as } \rho \to + \infty.
\end{align*} 
Hence every accumulation point of \(\QQQ_\rho\) as $\rho\to+\infty$ satisfies \eqref{q equation}. Since the admissible solution
of \eqref{q equation} is unique, it follows that $ \QQQ_\rho\to q$ as $\rho \to +\infty$. 
Moreover, since the admissible root is simple, we have $\QQQ_\rho=q+O(\rho^{-1})$ as $\rho\to+\infty$.
Therefore,
\begin{equation}\label{AAA convergence}
    \RRR_\rho=R+O(\rho^{-1}),
    \qquad
    \AAA_\rho=a+O(\rho^{-1}),
    \qquad
    \mathcal K_\rho=\kappa+O(\rho^{-1}), \qquad \mbox{as } \rho \to + \infty,
\end{equation}
where $R$ and $\kappa$ are given by \eqref{def of R f Ginibre} and \eqref{f conformal Ginibre}.  
It follows that $f_\rho(z)=f(z)+O(\rho^{-1})$ uniformly for \(z\) on the unit circle. Consequently, the conformal map \eqref{def of f pre trunc main}, and hence the corresponding droplet, converges as \(\rho\to+\infty\) to its Ginibre counterpart obtained in \cite{BBLM15}.
\end{rem}

\begin{rem}[Other planar equilibrium problems]
Explicitly solvable planar equilibrium problems beyond the radially
symmetric setting are relatively scarce. Among the known examples are
models for which the equilibrium measure has constant density, either
on the complex plane \cite{By24,BY26,BBLM15,KKL24} or on the Riemann
sphere \cite{BFL25,LD21,CK22,CK19}. Other examples include equilibrium
problems giving rise to multi-component droplets
\cite{ABK21,BM15,MS19,BK12,KL15,KT15,BS20}, as well as those involving hard-wall constraints \cite{Ch23a,CL24,CL24a,Ad18}. 
The present work contributes to this line of research by providing an
explicitly solvable planar equilibrium problem with a non-uniform
density.  

Such equilibrium measures also arise naturally in non-Hermitian random matrix theory. In particular, the present work
continues the study of equilibrium measures arising from moments of
characteristic polynomials. This direction was initiated for the
Ginibre ensemble in \cite{BBLM15} and has subsequently been developed
for the spherical ensemble in \cite{BFL25} and for the elliptic
Ginibre ensemble in \cite{By24,BY26}. The present results provide the
corresponding analysis for the truncated unitary ensemble.  

In addition, we also determine the equilibrium measure in the weak non-unitarity regime of the truncated unitary ensemble; see Theorem~\ref{Thm_1D eq msr}. Its support lies on the unit circle, in contrast to the two-dimensional supports in the spherical and elliptic Ginibre cases discussed above. We obtain this measure by taking the limit $\rho\to0$ of the planar equilibrium measures in Theorem~\ref{Thm_2D eq msr}. In this limit, the droplet shrinks towards the unit circle and its planar density diverges. Integrating this density in the radial direction gives the limiting density on the circle. We explain this calculation in Section~\ref{Section_conformal weak}.
\end{rem}

 \begin{rem}[Free energies and orthogonal polynomials]
As mentioned above, models with logarithmic insertions of the form
appearing in \eqref{def of potential W strong} arise naturally in the
study of moments of characteristic polynomials. In such problems, a
precise description of the equilibrium measure is an essential first
step in the asymptotic analysis. In particular, both the density and the geometry and topology of its support can affect the asymptotics of these moments.

Beyond the equilibrium problem itself, the asymptotics of the
corresponding orthogonal polynomials have been extensively studied;
see, for instance,
\cite{BBLM15,BFKL26,DMMS25,KKL24,LY17,LY23,BEG18,BGM17,BKP23,AFLS25,KLY25,LY19}.
Closely related models have also been investigated in connection with
free energy expansions \cite{BSY24,BYY26}; see also
\cite{WW19,DS22,DMMS25,BDHK25,CL26} for related recent developments.

In this context, understanding how the topology of the droplet is reflected in the asymptotic expansion has been a longstanding problem \cite{CFTW15,JMP94,TF99,ZW06}. For recent progress in this direction, see e.g. \cite{BKS23,AFLS25,ACCL22,ACCL23,ACC24,ACC26,BKSY25,BP26,Ch22,Ch23, Rou25,By26}.

From this perspective, the explicit equilibrium measures obtained in
Theorems~\ref{Thm_1D eq msr} and \ref{Thm_2D eq msr} provide
a natural starting point for a more refined asymptotic analysis of
the corresponding partition functions, characteristic polynomial
moments, and orthogonal polynomials.
\end{rem}

\subsection{Outline of the proof}

It is well known that the equilibrium measure $\mu_V$ and its support $S_{V}$ are characterised by the variational conditions (see \cite[p.27]{ST97})
\begin{align}
	\label{var eq_2D}
	\int_\C \log \frac{1}{|\zeta-z|^2}\,d\mu_V(z)+V(\zeta)= \mathsf{C} , \quad \text{q.e.} \quad \text{if }\zeta \in S_V, 
	\smallskip 
	\\
	\int_\C \log \frac{1}{|\zeta-z|^2}\,d\mu_V(z)+V(\zeta)\ge  \mathsf{C} , \quad \text{q.e.} \quad \text{if }\zeta \notin S_V, \label{var ineq_2D}
\end{align} 
where ``q.e." stands for ``quasi-everywhere" and $\mathsf{C}$ is called the modified Robin's constant, or the Euler-Lagrange constant (\eqref{var eq_2D}--\eqref{var ineq_2D} are also sometimes called the Euler-Lagrange conditions). 
Furthermore, if $V$ is $C^2$-smooth in a neighbourhood of $S_V$, it follows from Frostman's theorem that $\mu_V$ is absolutely continuous with respect to $dA$ and takes the form 
\begin{equation} \label{Frostman thm}
	d\mu_V(z)= \Delta V(z) \cdot \mathbbm{1}_{ \{ z \in S_V \} } \,dA(z), 
\end{equation}
where $\Delta := \partial \bar{\partial}$ is the quarter of the usual Laplacian; see \cite[Theorem II.1.3]{ST97}. Notice that \eqref{def of mu W 2D main} in Theorem~\ref{Thm_2D eq msr} follows immediately from \eqref{Frostman thm} and
\begin{equation} \label{Lap of potential W}
	\Delta W(z) = \frac{ \rho (1+\rho) }{ (1+\rho-|z|^2)^2},
\qquad |z|<\sqrt{1+\rho}, \quad z\ne a. 
\end{equation} 

The main difficulty in proving Theorem~\ref{Thm_2D eq msr} is therefore to determine the droplet. In general, however, it is difficult to ``guess" the droplet. The \emph{conformal mapping method} is an efficient method to address this problem. However, this method still requires one to ``guess" the topology of the droplet (such as the number of connected components, whether it is simply connected or not, etc). We refer to \cite{By24,ABK21,BY26} for further references on this method. Once a candidate droplet has been obtained, one must verify the variational conditions \eqref{var eq_2D}--\eqref{var ineq_2D} for the corresponding measure to establish that it is indeed the equilibrium measure.

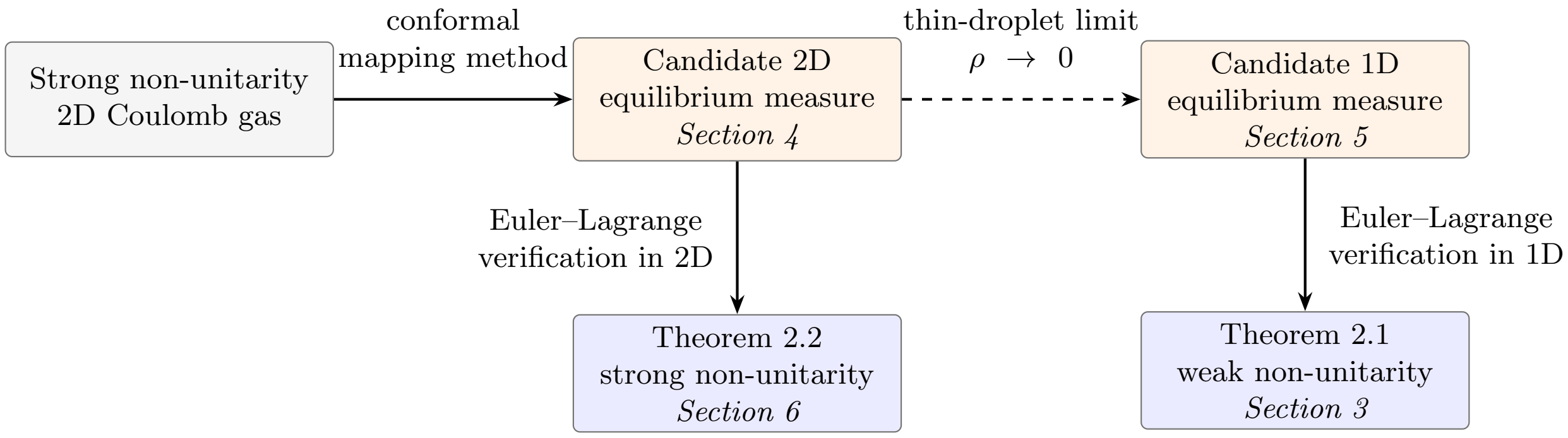
\begin{figure}[t]
\centering
\usetikzlibrary{positioning,arrows.meta}

\begin{tikzpicture}[
  node distance=16mm and 25mm,
  every node/.style={
    align=center,
    rounded corners=2pt,
    draw=black!55,
    minimum height=12mm,
    text width=32mm,
    font=\small
  },
  model/.style={fill=black!4},
  candidate/.style={fill=orange!10},
  theorem/.style={fill=blue!8},
  derive/.style={-{Stealth[length=2mm]}, thick},
  explain/.style={-{Stealth[length=2mm]}, thick, dashed},
  verify/.style={-{Stealth[length=2mm]}, thick}
]

\node[model] (model) {Strong non-unitarity\\ 2D Coulomb gas};

\node[candidate, right=of model] (twoD) {
  Candidate 2D \\
  equilibrium measure 
  \\
  \textit{Section~\ref{Section_conformal strong}}
};

\node[candidate, right=of twoD] (oneD) {
  Candidate 1D equilibrium measure\\
  \textit{Section~\ref{Section_conformal weak}}
};

\node[theorem, below=of twoD] (thm2) {
  Theorem~\ref{Thm_2D eq msr} \\ 
  strong non-unitarity\\
  \textit{Section~\ref{Section_EL strong}}
};

\node[theorem, below=of oneD] (thm1) {
  Theorem~\ref{Thm_1D eq msr} \\ 
  weak non-unitarity\\
  \textit{Section~\ref{Section_EL weak}}
};

\draw[derive]
  (model) -- node[above, draw=none, fill=none, text width=29mm]
  {conformal \\ mapping method} (twoD);

\draw[explain]
  (twoD) -- node[above, draw=none, fill=none, text width=30mm]
  {thin-droplet limit\\$\rho\to0$} (oneD);

\draw[verify]
  (twoD) -- node[left, draw=none, fill=none, text width=27mm]
  {Euler--Lagrange\\verification in 2D} (thm2);

\draw[verify]
  (oneD) -- node[right, draw=none, fill=none, text width=27mm]
  {Euler--Lagrange\\verification in 1D} (thm1);

\end{tikzpicture}

\caption{
Outline of the paper.
}
\label{Fig_roadmap}
\end{figure}

Sections~\ref{Section_conformal strong} and \ref{Section_conformal weak} explain how we obtain the candidate equilibrium measures. The proofs of Theorems~\ref{Thm_1D eq msr} and \ref{Thm_2D eq msr}, given in Sections~\ref{Section_EL weak} and \ref{Section_EL strong}, verify the Euler--Lagrange conditions directly from the explicit formulas and do not rely on these derivations. See Figure~\ref{Fig_roadmap} for an overview.

\subsection*{AI use disclosure}
A first draft of this work was prepared in January 2025. All proofs and text in that draft were written by the authors; it did not include Section~\ref{Section_EL weak} and Proposition~\ref{Prop_admissible pre parameters}. During the final stage of manuscript preparation, we used ChatGPT 5.6 Pro and 6 Astra to help with Section~\ref{Section_EL weak} and Proposition~\ref{Prop_admissible pre parameters}. We also used them throughout the manuscript to check proofs and improve the language.

\subsection*{Acknowledgement}
SSB was supported by National Research Foundation of Korea grants (RS-2025-00516909, RS-2026-25607520, and RS-2026-25518141).
CC is a Research Associate of the Fonds de la Recherche Scientifique - FNRS. CC also acknowledges support from the Swedish Research Council, Grant No. 2021-04626, and from the European Research Council (ERC), Grant Agreement No. 101115687.
PM is a Postdoctoral Researcher of the Fonds de la Recherche Scientifique – FNRS. PM also acknowledges support from the Swedish Foundation for
International Cooperation in Research and Higher Education (PD2023-9315).

\section{Proof of Theorem~\ref{Thm_1D eq msr}}
\label{Section_EL weak}

In this section, we prove Theorem~\ref{Thm_1D eq msr}. Recall from
Remark~\ref{Rem_stereographic} that the measure \(\mu_{\mathsf W}\)
in Theorem~\ref{Thm_1D eq msr} is a probability measure. It therefore remains to verify the Euler--Lagrange conditions \eqref{var eq_2D} and \eqref{var ineq_2D}. Let $\D:=\{z\in\C:|z|<1\}$. 
Since \(\mathsf W=+\infty\) outside $\overline{\D}$, it suffices to consider \(z\in\overline{\D}\).

\subsection{Proof of Theorem~\ref{Thm_1D eq msr} (a):
the post-critical regime}

Recall that $\mathsf W$ is defined in \eqref{def of W weak}. In this subsection, we assume that $  0\leq a\leq\frac{1}{1+2c}.$  
It suffices to show that  
\begin{equation} \label{EL conditions for 1D case post}
  \int_{\T}\log\frac{1}{|z-\zeta|^2}\,
    d\mu_{\mathsf W}(\zeta)+\mathsf W(z)
    \begin{cases}
        =0, &\textup{if } z\in\T,
    \smallskip 
    \\
        \geq0, &\textup{if } z\in\D,
    \end{cases}
\end{equation}  
where $\T= \{ z\in \C: |z|=1\}$ and $d\mu_{\mathsf W}$ is given by \eqref{def of eq msr 1D post}.

For \(z\in\D\), we have the absolutely convergent Fourier expansion 
\begin{equation} \label{expansion log Fourier}
 \log\frac{1}{|z-e^{i\theta}|^2} =   -\log|1-ze^{-i\theta}|^2 =  2\re\sum_{n=1}^{\infty}
    \frac{z^ne^{-in\theta}}{n}. 
\end{equation}  
Moreover, for \(0\leq a<1\), the Poisson kernel admits the absolutely
convergent Fourier expansion 
\begin{equation}
   \frac{1-a^2}{1+a^2-2a\cos\theta}
    =
    1+2\sum_{n=1}^{\infty}a^n\cos(n\theta).
\end{equation} 
Combining these two expansions and integrating term by term, we obtain 
\begin{equation} \label{integral of Poisson kernel}
\frac{1}{2\pi}
    \int_{-\pi}^{\pi}
    \log\frac{1}{|z-e^{i\theta}|^2}
    \frac{1-a^2}{1+a^2-2a\cos\theta}\,d\theta =   2\re\sum_{n=1}^{\infty}\frac{(az)^n}{n}
    =
    \log\frac{1}{|1-az|^2}.
\end{equation}  
On the other hand, \eqref{expansion log Fourier} immediately gives
 \[
    \frac{1}{2\pi}
    \int_{-\pi}^{\pi}
    \log\frac{1}{|z-e^{i\theta}|^2}\,d\theta=0. 
\]
Therefore, by \eqref{def of W weak}, \eqref{def of eq msr 1D post} and
\eqref{integral of Poisson kernel}, the weighted logarithmic potential
takes the explicit form
\begin{equation} \label{eval of log potential 1D post}
\int_{\T}\log\frac{1}{|z-\zeta|^2}\,
    d\mu_{\mathsf W}(\zeta) +\mathsf W(z) 
    =  c\log\frac{|1-az|^2}{|z-a|^2}, \qquad z \in \D.
\end{equation} 

Since $\mu_{\mathsf W}$ has a continuous density on $\T$, \eqref{eval of log potential 1D post} also holds for $z\in \T$.
For \(z\in\T\), we have $|1-az|=|z-a|$, and therefore the equality in \eqref{EL conditions for 1D case post} holds.
On the other hand, for $z \in \D$, we have 
\[
    |1-az|^2-|z-a|^2
    =(1-a^2)(1-|z|^2)\geq0.
\]
Together with \eqref{eval of log potential 1D post}, this yields the inequality in \eqref{EL conditions for 1D case post}. This completes the proof in the post-critical regime. \qed 

\subsection{Proof of Theorem~\ref{Thm_1D eq msr} (b): the pre-critical regime}

We now turn to the pre-critical regime and assume that $\frac{1}{1+2c}<a\leq1.$   

We first verify the Euler--Lagrange conditions on the unit circle, namely, that there exists a constant \(\mathsf C_{\mathsf W}\) such that
\begin{equation} \label{EL conditions for 1D case pre}
\int_{\T}\log\frac{1}{|z-\zeta|^2}\, d\mu_{\mathsf W}(\zeta)+\mathsf W(z)    \begin{cases}
        =\mathsf C_{\mathsf W},
        &\textup{if }z=e^{i\theta},\
        \theta\in[-\pi,\pi]\setminus(-\theta_*,\theta_*),
        \smallskip 
        \\
        \ge \mathsf C_{\mathsf W},
        &\textup{if }z=e^{i\theta},\
        |\theta|<\theta_*,
    \end{cases}
\end{equation} 
where $\mu_{\mathsf W}$ is given by \eqref{def of eq msr 1D pre}.
To verify these conditions, we use the stereographic projection introduced in Remark~\ref{Rem_stereographic} to rewrite the problem on the real line. 
Recall that, in the pre-critical regime, the push-forward measure \(\nu\) is given by
\eqref{weak pre density real line}. Define
\begin{equation} \label{def of H in R}
  H(x)  :=  2\int_{\R}\log\frac{1}{|x-y|}\,d\nu(y) +(c+1)\log(1+x^2)-c\log(x^2+b^2).
\end{equation}
Using the identities
$$
  \left|
        \frac{1+ix}{1-ix}
        -
        \frac{1+iy}{1-iy}
    \right|^2
    =
    \frac{4(x-y)^2}{(1+x^2)(1+y^2)}, \qquad   \left|
        \frac{1+ix}{1-ix}-a
    \right|^2
    =
    (1+a)^2\frac{x^2+b^2}{1+x^2}, 
$$
we find that the weighted logarithmic potential in
\eqref{EL conditions for 1D case pre}, evaluated at $   z=(1+ix)/(1-ix)$,   agrees with \(H(x)\) up to an additive constant independent of \(x\).
Moreover, under the stereographic projection, the supporting arc corresponds precisely to \(\{|x|\geq x_*\}\), where $x_*$ is given by \eqref{def of x star real line}. Consequently, \eqref{EL conditions for 1D case pre} is equivalent to showing that, for some constant \(\mathsf{C}\),
\begin{equation} \label{EL condition pre in terms of H in R}
H(x)  \begin{cases}
=\mathsf{C}, &\textup{if } |x| \ge x_*, 
\smallskip 
\\
\ge \mathsf{C}, &\textup{if } |x| < x_*.
\end{cases}
\end{equation} 

We compute the Cauchy transform of \(\nu\), defined by 
\begin{equation}
  C_\nu(z):=\int_{\R}\frac{d\nu(y)}{z-y},
    \qquad
    z\in
    \C\setminus
    \bigl((-\infty,-x_*]\cup[x_*,\infty)\bigr).
\end{equation} 
Using \eqref{weak pre density real line}, the symmetry of \(\nu\),
and the change of variables \(u=\sqrt{x^2-x_*^2}\), we obtain 
\begin{align*}
   C_\nu(z) &=  \frac{2\sqrt{a(2c+1)}}{\pi(1+a)}
    \int_{x_*}^{\infty}
    \frac{x\sqrt{x^2-x_*^2}}
    {(x^2+b^2)(x^2+1)}
    \left(
        \frac{1}{z-x}+\frac{1}{z+x}
    \right)dx
    \\
    &=
    \frac{4\sqrt{a(2c+1)}\,z}{\pi(1+a)}
    \int_{x_*}^{\infty}
    \frac{x\sqrt{x^2-x_*^2}}
    {(x^2+b^2)(x^2+1)(z^2-x^2)}\,dx
    \\
    &=  
    -\frac{4\sqrt{a(2c+1)}\,z}{\pi(1+a)}
    \int_0^\infty
    \frac{u^2\,du}
    {(u^2+x_*^2+b^2)
     (u^2+x_*^2+1)
     (u^2+x_*^2-z^2)}.
\end{align*}
To evaluate the last integral, we use the elementary identity
\[
    \int_0^\infty
    \frac{u^2\,du}
    {(u^2+A^2)(u^2+B^2)(u^2+D^2)}
    =
    \frac{\pi}
    {2(A+B)(B+D)(D+A)}, \qquad A,B>0, \; \re D>0. 
\]
We apply this identity for $D=\sqrt{x_*^2-z^2}$, where the branch is chosen so that it is analytic on $ \C\setminus \bigl((-\infty,-x_*]\cup[x_*,\infty)\bigr)$ and positive for \(z\in(-x_*,x_*)\). Using 
\[
    \sqrt{x_*^2+b^2}+\sqrt{x_*^2+1}
    =
    \frac{2\sqrt{a(2c+1)}}{1+a},
\]
we obtain
\begin{align}
\begin{split} \label{Cauchy transform weak pre}
 C_\nu(z) 
    & =
    -\frac{z}{
        \bigl(
            \sqrt{x_*^2-z^2}+\sqrt{x_*^2+b^2}
        \bigr)
        \bigl(
            \sqrt{x_*^2-z^2}+\sqrt{x_*^2+1}
        \bigr)
    }
    \\
    &=
    \frac{(c+1)z}{z^2+1}
    -\frac{cz}{z^2+b^2} 
    +
    \frac{2\sqrt{a(2c+1)}}{1+a}
    \frac{z\sqrt{x_*^2-z^2}}
    {(z^2+1)(z^2+b^2)}. 
\end{split}
\end{align}
(The apparent singularities at $z=\pm i$ and $z=\pm ib$ are removable.)
  
Let \(C_{\nu,+}(x)\) and \(C_{\nu,-}(x)\) denote the boundary values
of \(C_\nu\) from the upper and lower half-planes, respectively. For
\(|x|>x_*\), our choice of branch gives 
\[
    \sqrt{x_*^2-(x\pm i0)^2}
    =
    \mp i\,\operatorname{sgn}(x)
    \sqrt{x^2-x_*^2}.
\]
Thus, the two boundary values of the square root have opposite signs,
and it follows from \eqref{Cauchy transform weak pre} that 
\[
    C_{\nu,+}(x)+C_{\nu,-}(x)
    =
    \frac{2(c+1)x}{x^2+1}
    -\frac{2cx}{x^2+b^2},
    \qquad |x|>x_*.
\]
On the other hand, the Sokhotski–Plemelj formula gives 
\[
    \frac{C_{\nu,+}(x)+C_{\nu,-}(x)}{2}
    =
    \mathrm{p.v.}
    \int_{\R}\frac{d\nu(y)}{x-y}.
\]  
Differentiating \(H\) in \eqref{def of H in R} and using the above identities, we obtain, for \(|x|>x_*\),
\[
\begin{split}
    H'(x)
    &=
    -2\,\mathrm{p.v.}
    \int_{\R}\frac{d\nu(y)}{x-y}
    +
    \frac{2(c+1)x}{x^2+1}
    -
    \frac{2cx}{x^2+b^2}
    \\
    &=
    -C_{\nu,+}(x)-C_{\nu,-}(x)
    +
    \frac{2(c+1)x}{x^2+1}
    -
    \frac{2cx}{x^2+b^2}
    =0.
\end{split}
\]
Hence \(H\) is constant on each connected component of the support $ (-\infty,-x_*]\cup[x_*,\infty).$ 
Since both \(\nu\) and the external field in \eqref{def of H in R} are even, \(H\) is even, and therefore the constants on the two components coincide. This proves the equality part of \eqref{EL condition pre in terms of H in R}.

Next, we turn to the inequality part of
\eqref{EL condition pre in terms of H in R}. For \(|x|<x_*\), the
point \(x\) lies outside the support of \(\nu\), so no principal value
is needed. Moreover, \(\sqrt{x_*^2-x^2}\) is real and positive in this
interval. Thus, by \eqref{Cauchy transform weak pre}, 
\[
\begin{split}
    H'(x)
    &=
    -2C_\nu(x)
    +
    \frac{2(c+1)x}{x^2+1}
    -
    \frac{2cx}{x^2+b^2}
  =
    -\frac{4\sqrt{a(2c+1)}}{1+a}
    \frac{x\sqrt{x_*^2-x^2}}
    {(x^2+1)(x^2+b^2)}.
\end{split}
\]
Note that all factors on the right-hand side, apart from \(x\) and the displayed minus sign, are positive. Therefore,
\[
    H'(x)>0 \quad\text{for }-x_*<x<0,
    \qquad
    H'(x)<0 \quad\text{for }0<x<x_*.
\]
Since \(H(\pm x_*)=\mathsf C\), this proves the inequality part of \eqref{EL condition pre in terms of H in R}. Therefore we have shown \eqref{EL conditions for 1D case pre}. 

It remains to extend the variational inequality from the unit circle
to the interior of the unit disc. Consider the weighted logarithmic
potential 
\[
    z\longmapsto
    \int_{\T}\log\frac{1}{|z-\zeta|^2}\,
    d\mu_{\mathsf W}(\zeta)-2c\log|z-a|.
\]
For each fixed \(\zeta\in\T\), the function $ z\mapsto\log\frac{1}{|z-\zeta|^2}$ is harmonic in \(\D\), since its singularity lies on the boundary.
Consequently, its integral against \(\mu_{\mathsf W}\) is harmonic in
\(\D\). The second term is harmonic away from \(z=a\) and tends to
\(+\infty\) as \(z\to a\).

Suppose first that \(a<1\). For sufficiently small \(\varepsilon>0\),
consider the domain $ \D\setminus\overline{\D(a,\varepsilon)}.$ 
On the unit-circle part of its boundary, the weighted potential is at
least \(\mathsf C_{\mathsf W}\) by \eqref{EL conditions for 1D case pre}. 
On \(\partial\D(a,\varepsilon)\), the same inequality holds for
sufficiently small \(\varepsilon\), since $-2c\log|z-a|\to+\infty$ as $z \to a$.  
The minimum principle for harmonic functions therefore yields
\[
    \int_{\T}\log\frac{1}{|z-\zeta|^2}\,
    d\mu_{\mathsf W}(\zeta)+\mathsf W(z)
    \geq\mathsf C_{\mathsf W}
\]
throughout
\(\D\setminus\overline{\D(a,\varepsilon)}\). Letting
\(\varepsilon\to0\) proves the desired inequality throughout
\(\D\).

If \(a=1\), the singularity lies on the boundary of the unit disc.
In the strict pre-critical regime, the support of 
\(\mu_{\mathsf W}\) is separated from \(1\), while $-2c\log|z-1|\to+\infty $ as $z \to 1$.  
We may therefore remove a sufficiently small neighbourhood of the
boundary point \(1\) and apply the same minimum-principle argument.
Letting the size of this neighbourhood tend to zero yields the
variational inequality throughout \(\D\).

Together with \eqref{EL conditions for 1D case pre}, this verifies
the Euler--Lagrange conditions in the pre-critical regime and
completes the proof of Theorem~\ref{Thm_1D eq msr}~(b). \qed

\section{Conformal mapping method: strong non-unitarity} \label{Section_conformal strong}

In this section, we use the conformal mapping method to derive a candidate for the droplet $S_W$. We verify that this candidate is indeed the droplet in Section~\ref{Section_EL strong}.

\begin{rem}[Single pole assumption] \label{Rem_single pole}
We now explain how the parameters $\mathcal{A}$ and $\AAA$ in \eqref{def of A center} and \eqref{AAA in terms of X} arise in the conformal mapping method. 
A key ingredient is the Cauchy transform
	\begin{equation} \label{def of Cauchy transform}
		C_W(z):= \int \frac{ d\mu_W(w) }{z-w}  
	\end{equation}
	of the equilibrium measure. 
	Define
	\begin{equation} \label{def of BW Cauchy}
		B_W(z):=	C_W( z )+ \frac{c}{ z -a} + \frac{\rho}{ z } , 
	\end{equation}
    which appears in the denominator on the right-hand side of \eqref{relation of Schwarz Cauchy}. 
	In addition to specifying the topology of the droplet, we assume that
    \smallskip 
	\begin{center}
		\emph{There exists a unique $z^* \in S_W^c$ satisfying $B_W(z^*)=0$, and this zero is simple}.
	\end{center}
    \smallskip 
    (The right-hand side of \eqref{relation of Schwarz Cauchy} therefore has a single simple pole in $S_W^c$, located at $z^*$.)
    
	This unique point is precisely
	\begin{equation}
		z^* = \begin{cases}
			\mathcal{A} &\textup{in the post-critical regime},
			\smallskip
			\\
			\AAA &\textup{in the pre-critical regime}. 
		\end{cases}
	\end{equation} 
	Note that as \(\rho \to +\infty\), the last term in \eqref{def of BW Cauchy} diverges, while the first term remains bounded. Consequently, for $B_W(z^*)=0$ to hold, $z^*$ must tend to \(a\) as $\rho \to + \infty$ (we already checked that this is indeed the case; see \eqref{Acal convergence} and \eqref{AAA convergence}). 
\end{rem}

Recall that $W$ is given by \eqref{def of potential W strong}.
By \cite[Theorem I.1.3 (c)]{ST97}, $S_{W}$ is compact. 
For sufficiently large $|\zeta|$, the definition \eqref{def of Cauchy transform} gives the convergent Laurent expansion 
\begin{equation} \label{Cauchy infinity asymp}
	C_W(\zeta) = \frac{1}{\zeta}  \sum_{k=0}^\infty \frac{m_k}{ \zeta^k } ,\qquad m_k := \int z^k \,d\mu_W(z). 
\end{equation}
We make the ansatz that \eqref{var eq_2D} holds with ``q.e." replaced by ``for all". Differentiating \eqref{var eq_2D} with respect to $\zeta$ then yields 
\begin{equation} 
	C_W(\zeta)=   \frac{ \rho \,\overline{\zeta} }{1+\rho-|\zeta|^2} - \frac{c}{\zeta-a}, \qquad \textup{for all }\zeta \in S_W. 
\end{equation} 
Solving for $\overline{\zeta}$, we obtain 
\begin{equation} \label{relation of Schwarz Cauchy}
	\begin{split}
		\overline{\zeta}  
		&=  \frac{\rho+1}{\zeta}   \Big( C_W(\zeta)+ \frac{c}{\zeta-a} \Big)  \Big( C_W(\zeta)+  \frac{c}{\zeta-a} + \frac{\rho}{\zeta}   \Big)^{-1} , \qquad  \textup{for all } \zeta \in S_W.  
	\end{split}
\end{equation}

\subsection{Ginibre ensemble with a point charge revisited}\label{subsec: Ginibre revisited}

It is instructive to first consider the simpler case $\rho=+\infty$.
As mentioned above, the uniform density of the equilibrium measure
substantially simplifies the analysis in this setting compared with
the general case $\rho\in(0,\infty)$.

In this case, the relation \eqref{relation of Schwarz Cauchy} reduces to
\begin{equation}  \label{relation of Schwarz Cauchy Ginibre}
	\begin{split}
		\overline{\zeta}  =  C_W(\zeta) + \frac{c}{\zeta-a} , \qquad  \textup{for all } \zeta \in S_W.  
	\end{split}
\end{equation}

\subsubsection{Post-critical case}\label{subsec:Gin post} For the post-critical case, we make the ansatz that the droplet $S_W$ is of the form  
\begin{equation}
	S_W= \overline{ S_1 \setminus S_0 } , \qquad S_0 \subset S_1, \qquad a \in S_0, \qquad 0 \in S_1,  
\end{equation}
where $S_0$ and $S_1$ are bounded simply connected domains in the complex plane. 
We already know from \eqref{droplet post Ginibre} that 
\begin{equation} \label{S0 S1 post Ginibre}
	S_0 = \mathbb{D}(a,\sqrt{c}), \qquad S_1= \mathbb{D}(0,\sqrt{1+c}). 
\end{equation} 
Let us explain how this conclusion can be drawn using the conformal mapping method. 

Let $f$ be the unique conformal map from $\C\setminus\overline{\mathbb D}$ onto $\C\setminus\overline{S_1}$ satisfying
\begin{equation}
	f(z)= R_1 z +R_2 +O( \frac{1}{z} ), \qquad z \to \infty, 
\end{equation} 
for some $R_1>0$ and $R_2 \in \R$ (the fact that $R_{2}$ is real follows from the symmetry $W(z) = W(\overline{z})$).
The analytic continuation $f:\C \to \C$ is then given for $z\in \overline{\mathbb{D}} \setminus \{0 \} $ by
\begin{equation} \label{f analy conti Ginibre}
	f(z) = \overline{ C_W( f(1/\overline{z}) ) + \frac{ c }{ f(1/\overline{z})-a } }, \qquad z \in \overline{\mathbb{D}} \setminus \{0\},
\end{equation}
where we have used \eqref{relation of Schwarz Cauchy Ginibre} and the fact that $f(z) = f(1/\overline{z})$ for $|z|=1$. 
Since $f(1/\overline{z}) \not = a$ for $z \in \overline{\D}\setminus\{0\}$ by assumption, \eqref{f analy conti Ginibre} implies that $f$ does not have a singularity in $\D\setminus\{0\}$. By \eqref{Cauchy infinity asymp} and \eqref{f analy conti Ginibre}, we have 
\begin{align} \label{f asymp 0 Ginibre}
	f(z) = 
	\frac{1+c}{R_1}\, z +O(z^2) , \qquad z \to 0 . 
\end{align}
Thus, the singularity at $0$ is removable, and $f$ is entire. Together with its expansion at infinity, this implies that $f(z)=R_1z+R_2$. Since $f(0)=R_2$, we conclude from \eqref{f asymp 0 Ginibre} that $R_1=\sqrt{1+c}$ and $R_2=0$, and we find $S_1$ in \eqref{S0 S1 post Ginibre}. 

For $S_0$, let $g$ be the unique conformal map from $\mathbb{D}$ onto $S_0$ such that 
\begin{equation}
	g(z) = a+ \mathcal{R}_1 z +O(z^2), \qquad z \to 0, 
\end{equation}
where $\mathcal{R}_1>0$. The analytic continuation $g:\C\to\C$ is given for $z\in \D^{c}$ by
\begin{equation}\label{g analytic continuation}
	g(z) =  \overline{ C_W( g(1/\overline{z}) ) + \frac{ c }{ g(1/\overline{z})-a } }, \qquad z \in \mathbb{D}^c,
\end{equation}
where we have again used \eqref{relation of Schwarz Cauchy Ginibre} and the fact that $g(z) = g(1/\overline{z})$ for $|z|=1$.
Since $g(0)=a$ and $g(1/\overline{z}) \neq a$ for $z\in \D^{c}$, \eqref{g analytic continuation} implies that $g$ has a simple pole at infinity and no other poles. Thus $g$ is a polynomial of the form $g(z) = a+ \mathcal{R}_1 z.$ 
Note from \eqref{g analytic continuation} that 
\begin{align*}
	g(z) = C_W(a) + \frac{c z}{ \mathcal{R}_1 }, \qquad z \to \infty.  
\end{align*}
Thus we have $\mathcal{R}_1 = \sqrt{c}$, which gives $S_0$ in \eqref{S0 S1 post Ginibre}.

\subsubsection{Pre-critical case}\label{subsec: Ginibre pre}
 For the pre-critical case, we assume that the droplet $S_W$ is of the form  
	\begin{equation}
		S_W= S_1, \qquad a \not \in S_1,
	\end{equation}
	where $S_1$ is a simply connected set. We define the conformal map $f:\C\setminus\overline{\mathbb D} \to S_1^c$ and its analytic continuation as in \eqref{f analy conti Ginibre}. Since $a \in S_1^c$, there exists a unique $q \in \mathbb{D}$ such that $f(1/q)=a$. Therefore, the meromorphic function $f$ is of the form
	$$
	f(z) = R_1 z + R_2 + \frac{R_3}{z-q}, \qquad f(1/q)=a,
	$$
	for some $R_{1},R_{2},R_{3},q\in \R$ (the fact that these parameters are real follows from the symmetry $W(z) = W(\overline{z})$).
	It is more convenient to write
	\begin{align}\label{def of Gin pre criti}
	f(z) = \Big( z- \frac{1}{q} \Big) \Big( R + \frac{ A }{ z-q  } \Big)+a, 
\end{align}	
for some $A\in \R$ and $R>0$. It follows from \eqref{def of Gin pre criti} that 
	\begin{align}\label{lol1}
	f(z) = a+ \frac{A-qR}{ q^2 } + \Big( \frac{A(1-q^2)}{q^3} +R \Big) z  +O(z^2), \qquad z \to 0. 
	\end{align}
On the other hand, using \eqref{f analy conti Ginibre} and \eqref{Cauchy infinity asymp}, we find that $f$ also satisfies \eqref{f asymp 0 Ginibre} (with $R_1=R$). Comparing the constant and linear terms in \eqref{f asymp 0 Ginibre} and \eqref{lol1} yields
	\begin{equation} \label{relation A q Ginibre}
		A= q (R-aq), \qquad  \frac{ (R-a q)(1-q^2)  }{ q^2 } + R= \frac{1+c}{R}. 
	\end{equation}
    Note that 
    $$
       \underset{ w=0 }{ \Res } \, \Big[  f(1/w) \,f'(w) \Big]  =  \frac{ R (R-aq +aq^3) }{ q^2 }, \qquad   \underset{ w=q }{ \Res } \, \Big[  f(1/w) \,f'(w) \Big]  = - \frac{  (R-aq)(R-aq^3) }{ q^2 }.
    $$
	Using these identities together with Green's formula and the fact that $\overline{f(w)} = f(1/w)$ for $|w|=1$, we have 
	\begin{align*}
		1& = \int_{S_W} \,dA(z) =\frac{1}{2\pi i} \int_{  \partial S_W } \overline{z} \,dz = \frac{1}{2\pi i } \int_{ |w|=1 } f(1/w) \,f'(w) \,dw 
		\\
		& = \underset{ w=0 }{ \Res } \, \Big[  f(1/w) \,f'(w) \Big]  + \underset{ w=q }{ \Res } \,  \Big[  f(1/w) \,f'(w) \Big] = a q (2R-aq). 
	\end{align*}
	Thus $	R= (1+a^2q^2)/( 2a q ). $ Substituting this into the second equation in \eqref{relation A q Ginibre} gives \eqref{q equation}. 
	Using $A=q(R-aq)$ in \eqref{def of Gin pre criti}, we conclude that $f$ is given by \eqref{f conformal Ginibre}.

\subsection{Conformal mapping method: post-critical regime}

We now discuss the truncated unitary ensemble with a point charge in the post-critical regime.  
As in Subsection \ref{subsec:Gin post}, we make the ansatz that the droplet $S_W$ is of the form 
\begin{equation}
	S_W= \overline{S_1 \setminus S_0}, \qquad S_0 \subset S_1, \qquad a \in S_0, \qquad 0 \in S_1,
\end{equation}
where $S_0$ and $S_1$ are bounded simply connected domains.

We first investigate $S_1$. 
Let $f$ be the unique conformal map from $\C\setminus\overline{\mathbb D}$ onto $\C\setminus\overline{S_1}$ satisfying
\begin{equation}\label{lol19}
	f(z)= R_1 z +R_2 +O( \frac{1}{z} ), \qquad z \to \infty, 
\end{equation}
where $R_1>0$ and $R_{2} \in \mathbb{R}$.
According to \eqref{relation of Schwarz Cauchy}, the meromorphic continuation of $f$ into $ \overline{\mathbb{D}} \setminus \{0 \}$ is given by 
\begin{align}
\begin{split}
 \label{def of analy conti f1}
f(z)  =    \overline{   \frac{\rho+1}{ f(1/\overline{z}) }   \Big(  1 - \frac{1}{ B_W(f(1/\overline{z})) } \frac{\rho}{f(1/\overline{z})}   \Big)    } , 
\end{split}
\end{align}
where  $B_W$ is given by \eqref{def of BW Cauchy}.  

Following Remark~\ref{Rem_single pole}, we make the ansatz that $z \mapsto B_W(z) $ has a unique zero for $z \in S_W^c$ and that this zero is simple. 
 
Suppose that $B_W(z) \not =0$ if $ z \not \in S_1.$ By \eqref{def of analy conti f1}, $f$ has no singularities in $\mathbb D\setminus\{0\}$. Using \eqref{Cauchy infinity asymp} in \eqref{def of analy conti f1}, we obtain
\begin{equation}\label{lol20}
    f(z)=\frac{(1+c)(1+\rho)}{R_1(\rho+c+1)}\,z+O(z^2),
    \qquad z\to0.
\end{equation}
Thus the singularity at $0$ is removable, and $f$ is entire.
Together with \eqref{lol19}, this implies that $f(z)=R_1z+R_2$, $R_{2}=0$ and $R_{1}=r_{1}$,  where $r_1$ is given by \eqref{def of r1}. This already shows that $S_{1} = \D(0,r_{1})$.

For later purposes, we also note that 
\begin{equation} \label{mass_outer droplet_post}
	\int_{ |z|< r_1 }  \Delta W(z) \,dA(z) = c+1, 
\end{equation}
which follows from \eqref{Lap of potential W} and a computation using polar coordinates. 

\medskip 

Next, we investigate $S_0$. 
Let $g$ be the unique conformal map from $\mathbb{D}$ to $S_0$ satisfying 
\begin{equation}
	g(z) = \mathcal{A}+ \mathcal{R} z  +O(z^2), \qquad z \to 0, 
\end{equation}
where $\mathcal{A}>0$ is such that $B_W(\mathcal{A})=0$ and $\mathcal{R}>0$. By \eqref{relation of Schwarz Cauchy}, the meromorphic continuation
of $g$ to $\mathbb D^c$ is given by
\begin{align}\label{g outside D truncated post}
\begin{split}
g(z) & =   \overline{   \frac{\rho+1}{ g(1/\overline{z}) }   \Big( C_W( g(1/\overline{z}) )+ \frac{c}{g(1/\overline{z})-a} \Big)  \Big( C_W(g(1/\overline{z}))+  \frac{c}{g(1/\overline{z})-a} + \frac{\rho}{g(1/\overline{z})}   \Big)^{-1}   } \\
&=  \overline{   \frac{\rho+1}{ g(1/\overline{z}) }   \Big(1 - \frac{1}{ B_W( g(1/\overline{z}) ) } \frac{\rho}{g(1/\overline{z})}  \Big)    } .
\end{split}
\end{align}
For $\rho<+\infty$, we have $\mathcal A\neq a$, since $B_W$ has a pole at $a$ and a zero at $\mathcal A$. Then it follows from \eqref{g outside D truncated post} that $g$ has a simple pole at infinity and no other poles. (Indeed, if $0 \notin S_{0}$, then this directly follows from \eqref{g outside D truncated post}, while if $0\in S_{0}$, then $g(1/\overline{z_{1}})=0$ for some $z_{1}\in \D^{c}$, but the first line in \eqref{g outside D truncated post} implies that $z_{1}$ is not a pole of $g$.) Therefore $g$ is of the form $g(z) = \mathcal{A}+\mathcal{R}z.$ 
This shows that $S_{0}=\D(\mathcal{A},\mathcal{R})$. 

The quantity $\frac{\mathcal{R}}{a-\mathcal{A}}$ is the preimage of $a$ under $z\mapsto g(1/\overline{z})$. Indeed, since $g(z)=\mathcal{A}+\mathcal{R}z$, we have $g(\frac{a-\mathcal{A}}{\mathcal{R}})=a.$ 
Since $a\in S_0=g(\mathbb D)$, we have $\frac{\mathcal R}{a-\mathcal A}\in\mathbb D^c$. Then \eqref{g outside D truncated post} implies that
\begin{equation}
	g(z)= \frac{\rho+1}{ a } +o(1) ,\qquad \mbox{as } z \to \frac{\mathcal{R} }{ a-\mathcal{A} },
\end{equation} 
and therefore 
\begin{equation} \label{R A rel post}
	\mathcal{R}^2 = (1-\mathcal{A}/a) (1+\rho-a \mathcal{A}). 
\end{equation}

Using \eqref{mass_outer droplet_post} together with $\int_{S_{W}}\Delta W(z) dA(z)=1$ and $S_{W} = \overline{S_1\setminus S_0}$, we obtain
\begin{equation} \label{cond_mass of inner hole}
	\int_{   |z-\mathcal{A}| < \mathcal{R}   }  \frac{ \rho (1+\rho) }{ (1+\rho-|z|^2)^2} \,dA(z) = c. 
\end{equation}   
On the other hand, by applying Green’s formula and the change of variables $z= \mathcal{A}+\mathcal{R}w$, we have 
\begin{align*}
	&\quad \int_{   |z-\mathcal{A}| < \mathcal{R}   }  \frac{ \rho (\rho+1) }{ (1+\rho-|z|^2)^2} \,dA(z)  = \frac{1}{2\pi i} \int_{   |z-\mathcal{A}| = \mathcal{R}   }  \frac{ \rho \, \overline{z} }{ 1+\rho-|z|^2 } \,dz 
	\\
	& =  \frac{1}{2\pi i} \int_{   |w|=1   }   \frac{  \rho\,\mathcal{R}   ( \mathcal{A}+\mathcal{R}/w )    }{ 1+\rho-  ( \mathcal{A}+\mathcal{R} w )  ( \mathcal{A}+\mathcal{R}/w )    } \,dw  =  - \frac{\rho}{ \mathcal{A} } \frac{1}{2\pi i} \int_{   |w|=1   }   \frac{    \mathcal{A}w+\mathcal{R}     }{ (w-w_+)(w-w_-) } \,dw ,
\end{align*}  
where $w_+$ and $w_-$ are defined by
$$
\mathcal{A} \mathcal{R} w^2+( \mathcal{A}^2+\mathcal{R}^2-\rho-1  ) w + \mathcal{A } \mathcal{R } =  \mathcal{A} \mathcal{R} ( w- w_+ ) (w-w_-). 
$$

Note that 
$$
w_+ + w_- = -\frac{  \mathcal{A}^2+\mathcal{R}^2-\rho-1 }{ \mathcal{A} \mathcal{R} }, \qquad w_+ w_- =1 
$$
and 
$$ 
w_ \pm = \frac{ -( \mathcal{A}^2+\mathcal{R}^2-\rho-1  )  \pm \sqrt{ ( \mathcal{A}^2+\mathcal{R}^2-\rho-1  ) ^2- 4 \mathcal{A}^2 \mathcal{R}^2  } }{ 2 \mathcal{A} \mathcal{R} }.
$$ 
Since $S_0\subset S_1$, we have $\mathcal A+\mathcal R\le r_1<\sqrt{1+\rho}$.
Consequently, $1+\rho-\mathcal A^2-\mathcal R^2> 2\mathcal A\mathcal R$, and hence $0<w_-<1<w_+$. 
Write 
$$
F( w) :=  - \frac{\rho}{ \mathcal{A} }   \frac{    \mathcal{A}w+\mathcal{R}     }{ (w-w_+)(w-w_-) }. 
$$
Since $w_-$ is the only pole of $F$ in the unit disc,  
\begin{align}
	\begin{split}
		&\quad \int_{   |z-\mathcal{A}| < \mathcal{R}   }  \frac{ \rho (\rho+1) }{ (1+\rho-|z|^2)^2} \,dA(z)  =  \underset{ w=w_- }{\textrm{Res}} \, F(w)   =  \frac{\rho}{ \mathcal{A} }   \frac{    \mathcal{A}w_- +\mathcal{R}     }{ w_+-w_- } 
		\\
		&=  \frac{\rho}{2} \,    \frac{ \rho+1  -\mathcal{A}^2+\mathcal{R}^2 - \sqrt{ ( \mathcal{A}^2+\mathcal{R}^2-\rho-1  ) ^2- 4 \mathcal{A}^2 \mathcal{R}^2  } }{   \sqrt{ ( \mathcal{A}^2+\mathcal{R}^2-\rho-1  ) ^2- 4 \mathcal{A}^2 \mathcal{R}^2  }  }. 
	\end{split}
\end{align}
Therefore the condition \eqref{cond_mass of inner hole} gives 
$$
\rho+1  -\mathcal{A}^2+\mathcal{R}^2  = \Big(  \frac{ 2c }{ \rho } +1 \Big)  \sqrt{ ( \mathcal{A}^2+\mathcal{R}^2-\rho-1  ) ^2- 4 \mathcal{A}^2 \mathcal{R}^2  }.
$$
Combining this with \eqref{R A rel post}, we obtain (recall that $1+\rho-a^{2}>0$) 
\begin{equation}
	\mathcal{A}= \frac{ a \rho(\rho+1) }{ \rho( \rho+1) +  c(\rho+1-a^2) }, \qquad  \mathcal{R}= \frac{ 1+\rho-a^2 }{ \rho( \rho+1) +  c(\rho+1-a^2)  } \sqrt{ c (\rho+1)(\rho+c) }.
\end{equation}
Therefore $g(z)= \mathcal{A}+r_0z$, where $r_{0}$ is given by \eqref{def of r0}. This shows that $S_{0}=\D(\mathcal{A},r_{0})$.

\subsection{Conformal mapping method: pre-critical regime}

For the pre-critical case, we assume that 
\begin{equation}
	S_W= S_1, \qquad a \not \in S_1,
\end{equation}
where $S_1$ is a simply connected compact set.
Let  $f$ be the unique conformal map from $\C\setminus \overline{\mathbb{D}}$ onto $S_1^c$ such that 
\begin{equation}\label{f truncated pre asymp inf}
	f(z)= R_1 z +R_2 +O( \frac{1}{z} ), \qquad z \to \infty,
\end{equation}
where $R_1>0$ and $R_2\in\mathbb R$.
We then define the meromorphic continuation of $f$ to
$\overline{\mathbb D}\setminus\{0\}$ using
\eqref{def of analy conti f1}.

We need to specify the poles of $f$ inside $\mathbb{D}$. 
\begin{itemize}
	\item Suppose that $f(1/\overline{z}_*)=0$ for some $z_* \in \mathbb{D}$. This implies $0 \not \in S_1.$ The point $z_*$ is a removable singularity of $f$. Indeed, from \eqref{def of analy conti f1} we have   
	$$
	f(z) = \frac{ \rho+1 }{ \rho } \Big( C_W(0)-\frac{c}{a} \Big)+o(1), \qquad \mbox{as } z \to z_*. 
	$$
	
	\item Suppose that $f(1/\overline{z}_*)=a$ for some $z_* \in \mathbb{D}$. Again, $z_*$ is a removable singularity of $f$ since, by \eqref{def of analy conti f1},
	\begin{align}\label{lol3}
	f(z) = \frac{\rho+1}{a}+o(1) \quad \mbox{as } z\to z_{*}. 
	\end{align}
	\item Suppose that 
	$
	B_W(f(1/\overline{z}_* )) =0 
	$
	for some $z_* \in \mathbb{D}$. Then this point is a pole of $f$. Our main assumption is that there is only one such point and that $f(1/\overline{z}_* )$ is a simple zero of $B_W$; see Remark~\ref{Rem_single pole}. 
\end{itemize}

We now make more precise the assumption of Remark~\ref{Rem_single pole}.

\begin{assump}[Single pole assumption for the pre-critical regime]
Suppose that $B_W$ has a unique zero in $S_W^c$, which we denote by $\AAA$, and that this zero is simple. Thus, 
\begin{equation}\label{lol2}
	C_W(\AAA )= - \frac{\rho}{\AAA } - \frac{c}{\AAA -a}.  
\end{equation}  
We also write $\QQQ$ for the unique point in $\mathbb{D} \setminus\{0\}$ such that $f(1/\QQQ)=\AAA.$ Since $W(\overline z)=W(z)$, we have $B_W(\overline z)=\overline{B_W(z)}$ and $f(\overline z)=\overline{f(z)}$, which imply that $\AAA$ and $\QQQ$ are real. For $ \rho \in (0, \infty)$, we have $\AAA\ne a$, since $B_W$ has a pole at $a$ (recall \eqref{def of BW Cauchy}).
It follows from \eqref{def of analy conti f1} that $\QQQ$ is the only pole of $f(z)$ inside $\mathbb{D}$, and this pole is simple.
\end{assump}

\begin{rem}
Recall from Subsection \ref{subsec: Ginibre pre} that for the pre-critical regime of Ginibre (corresponding to $\rho=+\infty$), $q$ is the only pole of $f$ inside $\D$, and satisfies $f(1/q)=a$. As shown in \eqref{AAA convergence} (and as expected),
\begin{equation} \label{AAA QQQ asymp rho}
	\lim_{ \rho \to +\infty } \AAA= a, \qquad \lim_{ \rho \to +\infty } \QQQ= q. 
\end{equation}
Furthermore, since $C_W(a) < +\infty$, we also expect from \eqref{lol2} that
\begin{equation} \label{AAA asymp rho}
\AAA  = a -\frac{c a}{\rho}+O( \frac{1}{\rho^2} ) \qquad \mbox{as } \rho \to + \infty. 
\end{equation}
\end{rem}

Using \eqref{def of analy conti f1}, \eqref{f truncated pre asymp inf} and \eqref{Cauchy infinity asymp}, we obtain $f(0):=\lim_{z\to 0}f(z)=0$. Therefore, since $\QQQ$ is the only finite pole of $f$ and is simple, \eqref{f truncated pre asymp inf} implies that $f$ is of the form  
\begin{equation} \label{def of f pre trunc}
	f(z)= \RRR \, z - \frac{ \mathcal{K} }{ z-\QQQ }- \frac{ \mathcal{K} }{ \QQQ } =\RRR  \frac{    z (z -\QQQ-\frac{\mathcal{K}}{  \QQQ \RRR  } )  }  { z- \QQQ  },
\end{equation} 
where $\RRR=R_1>0$ and $\mathcal K\in\mathbb R$.
From $f(1/\QQQ)=\AAA$, we obtain 
\begin{equation} \label{Kappa RAQ}
\RRR = \AAA \QQQ+ \frac{ \mathcal{K}  }{ 1-\QQQ^2 }, \qquad \mbox{or equivalently} \qquad \mathcal{K}= ( \RRR-\AAA\QQQ )(1-\QQQ^2) .
\end{equation}
Substituting the second identity in \eqref{Kappa RAQ} into \eqref{def of f pre trunc} yields
\begin{equation}
f(z)-\AAA= \RRR  \frac{ ( z-\frac{1}{\QQQ} ) ( z- \frac{\AAA\QQQ^2}{\RRR}    )  }{  z-\QQQ  }.
\end{equation}
 
By \eqref{def of f pre trunc} and \eqref{Kappa RAQ}, it remains
to determine the three parameters $\RRR$, $\AAA$ and $\QQQ$.
We now derive three algebraic relations for these parameters.
The first two follow from the continuation formula
\eqref{def of analy conti f1}, while the third follows from
the condition that the associated measure has total mass one.

\begin{lem} \label{Lem_c and parameters}
We have 
\begin{equation} \label{f'(0) relation}
 \frac{ \RRR- \AAA \QQQ(1-\QQQ^2) }{ \QQQ^2 } =   \frac{ (1+c)(1+\rho)  }{ \RRR (\rho+c+1) }, 
\end{equation}
which is equivalent to \eqref{c in terms of R Q A}.  
\end{lem}

\begin{proof}
Using \eqref{def of f pre trunc} and \eqref{Cauchy infinity asymp}, we expand the right-hand side of \eqref{def of analy conti f1} to obtain
\begin{equation}
	f(z)  =  \frac{ (1+c)(1+\rho)  }{ \RRR (\rho+c+1) }\, z +  \frac{ (1+\rho)  }{ \QQQ\RRR^2 (\rho+c+1)^2 } \Big( \mathcal{K} (1+c)(\rho+c+1) + \QQQ\rho (ac+ \overline{m_1}) \Big) z^2+ O(z^3) , 
\end{equation}
as $z \to 0$. On the other hand, it follows from \eqref{def of f pre trunc} that
\begin{equation}
f(z) = \frac{ \RRR\QQQ^2+ \mathcal{K} }{ \QQQ^2 } z+ \frac{ \mathcal{K} }{ \QQQ^3 } z^2 +O(z^3), 
\end{equation}   
as $z \to 0$. By comparing the linear coefficients of these two expansions, and substituting $\mathcal{K}$ using the second equation in \eqref{Kappa RAQ}, we obtain \eqref{f'(0) relation}. 
This immediately gives \eqref{c in terms of R Q A}. 
\end{proof}

Lemma~\ref{Lem_c and parameters} gives the first relation by
comparing the two expansions of $f$ at the origin.
We now derive the second relation from \eqref{def of analy conti f1} and the preimages of $a$ under $f$.

\begin{lem}\label{lemma:lemma2}
We have 
\begin{equation} \label{quadratic eq for AAA}
	\QQQ^2(1-\QQQ^2)( a \QQQ-\RRR )( (1+\rho) \QQQ - a \RRR )\AAA^2 + \QQQ ( \QQQ^2(1+\rho)-\RRR^2 )( 2 a \RRR-\QQQ (1+\rho+a^2) ) \AAA +a (  \QQQ^2(1+\rho)-\RRR^2 )^2=0.
\end{equation} 
In particular, selecting the root compatible with the Ginibre asymptotics
\eqref{AAA asymp rho}, we obtain
\begin{align} \label{AAA in terms of RRRR}
	\begin{split}
		\AAA &= \frac{1}{  2  	\QQQ (1-\QQQ^2) }  \frac{   (1+\rho)\QQQ^2-\RRR^2   }{ (  \RRR- a \QQQ )( (1+\rho) \QQQ - a \RRR ) }
		\\
		&\quad \times  \bigg[  2 a \RRR-\QQQ (1+\rho+a^2)  +  \QQQ  \Big(  ( 2 a \RRR-\QQQ(1+\rho+a^2) )^2 +(1-\QQQ^2)(1+\rho-a^2)^2  \Big)^{1/2}   \bigg]. 
	\end{split}
\end{align} 
\end{lem}

\begin{proof}
	Let $z_*$ be the unique point in $\mathbb{D}^c$ such that
	\begin{equation} \label{evaluations of zstar}
		f(z_*)=a, \qquad f(1/z_*) = \frac{\rho+1}{a}. 
	\end{equation} 
(Since $\RRR$, $\mathcal K$ and $\QQQ$ are real, \eqref{def of f pre trunc} gives
$f(\overline{z_*})=\overline{f(z_*)}=a$. The injectivity of $f$ therefore implies that $z_* \in \R$.)
The second identity in \eqref{evaluations of zstar} follows from \eqref{def of analy conti f1} (recall also \eqref{lol3}). 
Substituting \eqref{def of f pre trunc} into the two identities in \eqref{evaluations of zstar} and clearing denominators, we find that $z_*$ is a common root of the equations
	\begin{align}
	\begin{split} \label{zstar quadratic eqns}
		&\QQQ \RRR z^2-(  \QQQ^2 \RRR+ a \QQQ+\mathcal{K} ) z +a \QQQ^2 =0,
		\\
		&\QQQ^2(1+\rho) z^2-( a \QQQ^2 \RRR + a \mathcal{K}+\QQQ(1+\rho) ) z +a \QQQ \RRR=0 . 
	\end{split}
	\end{align}
    In particular, 
	\begin{align}\label{lol21}
	\frac{\QQQ \RRR z_{*}^2-(  \QQQ^2 \RRR+ a \QQQ+\mathcal{K} ) z_{*} +a \QQQ^2}{\QQQ \RRR} = \frac{\QQQ^2(1+\rho) z_{*}^2-( a \QQQ^2 \RRR + a \mathcal{K}+\QQQ(1+\rho) ) z_{*} +a \QQQ \RRR}{\QQQ^2(1+\rho)}.
	\end{align}
	
We first show that $\RRR^2\ne(1+\rho)\QQQ^2$.
Suppose otherwise. Subtracting the two normalised equations
in \eqref{zstar quadratic eqns} and using
\eqref{Kappa RAQ}, we obtain
\[
    \AAA\QQQ(1-\QQQ^2)
    \bigl(a\RRR-(1+\rho)\QQQ\bigr)z_*=0.
\]
The assumed equality and $a^2<1+\rho$ imply that
$a\RRR\ne(1+\rho)\QQQ$. Since $0<|\QQQ|<1$ and
$z_*\ne0$, it follows that $\AAA=0$.
Equation \eqref{f'(0) relation} then gives
\[
    1=\frac{1+c}{1+\rho+c},
\]
contradicting $\rho>0$.
	
Solving the linear equation \eqref{lol21} yields  
	\begin{equation} \label{def of z star}
		z_*= \frac{  a \QQQ  (   (1+\rho)  \QQQ^2 - \RRR^2  ) }{  \QQQ (1+\rho)     (  (\QQQ^2-1) \RRR+ a \QQQ+\mathcal{K} )     -a\RRR(   \QQQ^2 \RRR +  \mathcal{K}  ) }.
	\end{equation}
  Substituting \eqref{def of z star} into any of the equations in \eqref{zstar quadratic eqns}, and substituting also $\mathcal{K}$ using the second equation in \eqref{Kappa RAQ}, we obtain after simplifications (using that $\RRR^2 \neq \QQQ^2 (1 + \rho)$) the desired equation \eqref{quadratic eq for AAA}. Solving this quadratic and selecting the root compatible with \eqref{AAA asymp rho} gives \eqref{AAA in terms of RRRR}.
\end{proof}

We now parametrize $\RRR$ in terms of a parameter $X>0$ as in \eqref{RRR in terms of X}. Then, using \eqref{AAA in terms of RRRR} and noting that 
\begin{align*}
4X^{2}+(1-\mathsf{Q}^{2})(1+\rho)^{2} + a^{2} (1-\mathsf{Q}^{2})\big( a^{2}-2(1+\rho) \big) =4X^2+(1-\mathsf{Q}^2)(1+\rho-a^2)^2 >0,
\end{align*}
we obtain that $\AAA$ is given by \eqref{AAA in terms of X}. 
(The parametrisation \eqref{RRR in terms of X} is chosen to eliminate the square root in \eqref{AAA in terms of RRRR}.) 
Furthermore, by \eqref{def of z star}, we have 
\begin{equation} \label{zstar in terms of X}
	z_* = -  \frac{ X+(1+\QQQ) (1+\rho-a^2)/2  }{ X-(1+\QQQ) (1+\rho-a^2)/2 } .	
\end{equation} 
Lemmas \ref{Lem_c and parameters} and \ref{lemma:lemma2} imply that $\QQQ, \AAA, \RRR$ satisfy \eqref{RRR in terms of X}--\eqref{c in terms of R Q A} for some $X>0$. It remains to derive an equation for $X$ from the condition that the associated measure has total mass one. We first observe the following.

\begin{lem} \label{Lem_F pole residue}
Let 
\begin{equation} \label{def of F capital}
	F(w) :=  \frac{ \rho\, f(1/w) \,f'(w)   }{  1+\rho-f(w)f(1/w)  },  
\end{equation}
where $f$ is given by \eqref{def of f pre trunc}. Then inside $\mathbb{D}$, $F$ has simple poles at $0, \QQQ$ and $1/z_*$, where $z_*$ is given by \eqref{def of z star}.
Furthermore, we have 
\begin{align}
		\underset{ w=\QQQ  }{ \Res } \, F(w) &=  \rho, \label{residue of F at QQQ}
		\\ 
	\underset{ w=0 }{ \Res } \,F(w) & 
	 =  \frac{ \RRR ( \RRR-\AAA \QQQ + \AAA \QQQ^3) \rho }{ \RRR( \AAA \QQQ  - \AAA \QQQ^3 - \RRR) + \QQQ^2 (1 + \rho)  } =1+c,   \label{residue of F at 0}
	 \\
	 \underset{ w=1/z_* }{ \Res } \,  F(w) & = \frac{ a\rho \, \QQQ   }{  \RRR( \AAA \QQQ  - \AAA \QQQ^3 - \RRR)  + \QQQ^2 (1 + \rho)    }  \frac{    z_*-\QQQ }{ z_*^2-1 }   \frac{  (\RRR+\AAA\QQQ^3-\AAA\QQQ) z_*^2 -2 \QQQ \RRR z_*+\RRR }{ z_*(1-\QQQ z_*)  } .  \label{residue of F at zstar}
\end{align}
\end{lem}

\begin{proof} 
By \eqref{evaluations of zstar}, the function $1+\rho-f(w) f(1/w)$ has zeros at $z_{*},z_{*}^{-1}$. Using also \eqref{def of f pre trunc}, we obtain 
\begin{equation} \label{factorisation denominator F}
	1+\rho-f(w) f(1/w) =  \frac{ \QQQ^2(1+\rho) -\RRR( \mathcal{K}+\QQQ^2 \RRR )  }{ \QQQ^2 }    \frac{   (w-z_*)(w-1/z_*)  }{ (w-\QQQ) (w-1/\QQQ) } .
\end{equation}
Then the lemma follows by direct computations using \eqref{def of f pre trunc}, \eqref{Kappa RAQ} and \eqref{factorisation denominator F}. The second equality in \eqref{residue of F at 0} follows from \eqref{f'(0) relation}.
\end{proof}

\begin{rem}
As $\rho \to \infty$, we expect $F(w)$ to tend to $f_{\infty}(1/w)f_{\infty}'(w)$, where $f_{\infty}$ is given by the right-hand side of \eqref{f conformal Ginibre}.
Inside $\mathbb{D}$, the function $f_{\infty}(1/w)f_{\infty}'(w)$ has a simple pole at $0$ and a double pole at $q$. Hence we expect
\begin{equation}\label{lol4}
\lim_{ \rho \to \infty } \QQQ = q, \qquad  \lim_{ \rho \to \infty } z_* = \frac{1}{q}. 
\end{equation}
The first limit in \eqref{lol4} is consistent with \eqref{AAA QQQ asymp rho}.
In other words, the simple poles $\QQQ$ and $1/z_*$ of $F$ are expected to merge when $\rho \to +\infty$, to become a double pole at $q$.   
\end{rem}

\begin{lem}\label{Lem_Mass one} Let $f$ be given by \eqref{def of f pre trunc}, and let $S_W$ be the closed region bounded by $f(\T)$. Then the measure defined by \eqref{def of mu W 2D main} has total mass one if and only if 
\begin{equation}\label{mass one Res eqn}
  \underset{w=1/z_*}{\Res}\,F(w)=-c-\rho.
\end{equation}
\end{lem}

\begin{proof}
Since $f(\T)$ is contained in $\D(0,\sqrt{1+\rho})$, Green's formula
and the change of variables $z=f(w)$, together with the identity $\overline{f(w)}=f(1/w)$ for $|w|=1$, give
$$
    \int_{S_W}\frac{\rho(1+\rho)}{(1+\rho-|z|^2)^2}\,dA(z) =\frac{1}{2\pi i}\int_{\partial S_W}
\frac{\rho\,\overline z}{1+\rho-|z|^2}\,dz
=\frac{1}{2\pi i}\int_{|w|=1}F(w)\,dw.
$$ 
By Lemma~\ref{Lem_F pole residue}, the poles of $F$ in $\D$ are
$0$, $\QQQ$, and $1/z_*$, $\operatorname{Res}_{w=\QQQ}F(w)=\rho$ and $\underset{ w=0 }{ \Res } \,F(w) =1+c$.  Hence
\[
    \int_{S_W}\frac{\rho(1+\rho)}{(1+\rho-|z|^2)^2}\,dA(z)
    =\rho+1+c
    +\underset{w=1/z_*}{\Res}\,F(w).
\]
The assertion follows.   
\end{proof}

\begin{lem} \label{Lem_alg eq for X}
The relation \eqref{mass one Res eqn} can be written as 
\begin{equation}
\mathcal{P}\Big( \frac{ X }{ 1+\rho-a^2 } \Big) =0, 
\end{equation}
where $\PP$ is the polynomial given in \eqref{def of polynomial PP}.  
\end{lem}
The proof of Lemma \ref{Lem_alg eq for X} is given in Appendix~\ref{Appendix_lemmas} (see \ref{proof lemma appendix}). 

Combining Lemmas~\ref{Lem_c and parameters}, \ref{lemma:lemma2},
\ref{Lem_Mass one}, and \ref{Lem_alg eq for X}, we obtain the equations $\PP(Y)=0$ and \eqref{c in terms of R Q A} for $Y$ and $\QQQ$, with $X=(1+\rho-a^2)Y$ and $\RRR,\AAA$ given by \eqref{RRR in terms of X} and
\eqref{AAA in terms of X}.
Proposition~\ref{Prop_admissible pre parameters} shows that this
system has a unique solution $(Y,\QQQ)$ satisfying \eqref{lol18}. We will use Lemma~\ref{Lem_F pole residue} and the mass relation \eqref{mass one Res eqn} again in Section~\ref{Section_EL strong} to verify the Euler--Lagrange conditions in the pre-critical regime.

\section{Candidate equilibrium measure in the weak non-unitarity regime} \label{Section_conformal weak}

We derive the equilibrium measure in Theorem~\ref{Thm_1D eq msr}
by taking the limit $\rho\to0$ of the planar equilibrium
measures in Theorem~\ref{Thm_2D eq msr}. 
In this limit, the planar droplet collapses onto the unit circle: its radial thickness is of order $\rho$, whereas the planar density is of order $\rho^{-1}$. Their combined effect produces a non-trivial limiting density on the unit circle.
Following the idea of cross-section convergence in~\cite[Theorem 2.2]{AB23}, we recover this limiting density by integrating the planar equilibrium density in the radial direction.

\begin{figure}[t]
	\begin{subfigure}{0.32\textwidth}
		\begin{center}	
			\includegraphics[width=\textwidth]{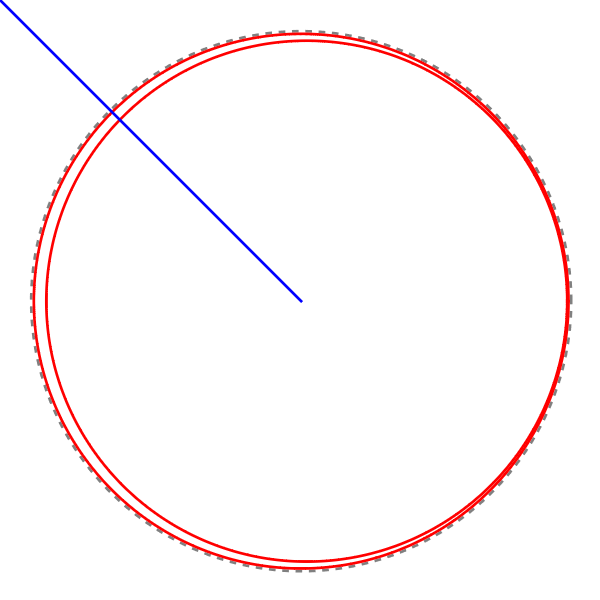}
		\end{center}
		\subcaption{$a=1/3$}
	\end{subfigure}	 \qquad 
	\begin{subfigure}{0.32\textwidth}
		\begin{center}			\includegraphics[width=\textwidth]{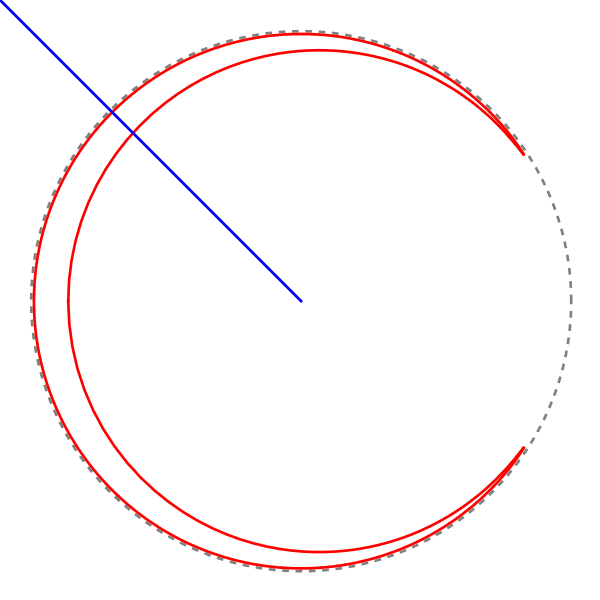}
		\end{center}
		\subcaption{$a=2/3$}
	\end{subfigure}	 
	\caption{Illustration of the droplet for \( c = 1/2 \) and \( \rho = 0.03 \). The red curves represent the boundary \( \partial S_W \), and the dashed circle represents \( \partial\mathbb{D}(0,\sqrt{1+\rho}) \). The blue ray $\{re^{i\theta}:r>0\}$ intersects $\partial S_W$ at two points whose moduli are approximately $1+\rho x_1(\theta)$ and $1+\rho x_2(\theta)$.} \label{Fig_droplet limiting}
\end{figure}

Let $\theta\in[-\pi,\pi]$. We impose no further restriction
in the post-critical regime, and require $|\theta|>\theta_*$
in the pre-critical regime. Define $r_{1,\rho}(\theta)$ and
$r_{2,\rho}(\theta)$ by
\[
    S_W\cap\{re^{i\theta}:r>0\}
    =
    \left\{
        re^{i\theta}:
        r_{2,\rho}(\theta)\leq r\leq r_{1,\rho}(\theta)
    \right\},
\]
where $S_{W}$ is as in Theorem \ref{Thm_2D eq msr}; see Figure~\ref{Fig_droplet limiting}. 
We assume that, for each fixed $\theta$ as above, the radii admit
the expansions
\begin{equation} \label{def of x1 x2}
 r_{j,\rho}(\theta)
    =
    1+\rho x_j(\theta)+O(\rho^2),
    \qquad j=1,2,\qquad \rho\to 0,
\end{equation} 
where the coefficients $x_1(\theta)$ and $x_2(\theta)$ will be determined below.
Since $S_W \subset \mathbb{D}(0,\sqrt{1+\rho})$ and $\sqrt{1+\rho}=1+\rho/2+O(\rho^2)$, we have 
$x_2 \leq x_1 \leq \frac12.$

Note that by \eqref{Lap of potential W}, 
\begin{equation}\label{density weak scaling}
|z|\Delta W(z) \Big|_{ |z|= 1+ \rho x    } = \frac{1}{(1-2x)^2} \frac{1}{\rho} +O(1) \qquad \mbox{as } \rho \to 0, 
\end{equation}  
uniformly for $x$ in compact subsets of $(-\infty,\frac12)$.
Thus, we expect the limiting equilibrium measure to be given by
\begin{equation} \label{def of Upsilon}
d \mu_{\mathsf W}(\theta) = \frac{ \Upsilon(\theta) }{\pi} \, d\theta, \qquad  \Upsilon(\theta):=	\int_{x_2}^{x_1} \frac{ 1 }{(1-2x)^2} \,dx= \frac{x_1-x_2}{ (1-2x_1)(1-2x_2) }. 
\end{equation}

\subsection{Post-critical regime}

Recall from Theorem \ref{Thm_2D eq msr} (a) that $S_W = \overline{ \mathbb{D}(0,r_1) } \setminus \mathbb{D }(\mathcal{A}, r_0 )$. 
By \eqref{def of A center}, \eqref{def of r0} and \eqref{def of r1}, as $\rho \to 0$ we have 
\begin{align}\label{lol10}
\mathcal{A} = \frac{a}{(1-a^2)c} \rho +O(\rho^2), \qquad r_0= 1+ \frac{ c-1-a^{2}(1+c) }{ 2(1-a^{2})c }\rho+O(\rho^2), \qquad r_1= 1+\frac{c}{2(1+c)} \rho +O(\rho^2). 
\end{align}
The last expansion in \eqref{lol10} directly yields $x_{1}=\frac{c}{2(1+c)}$. 
Let $\theta \in (-\pi,\pi]$ be fixed and let $x$ be such that $x e^{i \theta} \in  \partial \mathbb{D}(\mathcal{A},r_0)$. Substituting the expansion $x=1+x_{2} \rho + \bigO(\rho^{2})$ as $\rho \to 0$ into
$$
x^2+\mathcal{A}^2- 2 x \mathcal{A} \cos \theta =r_0^2,
$$
we obtain $x_{2} = \frac{ 1+a^2-c+a^2c-2a \cos \theta }{ 2c(a^2-1) }$.  
Then by substituting $x_{1}$ and $x_{2}$ in \eqref{def of Upsilon}, we obtain \eqref{def of eq msr 1D post}.

\subsection{Pre-critical regime} \label{Subsection_1D msr precritical}

Throughout this subsection, we assume that $\frac{1}{1+2c}<a\leq1$.  
Recall that $f\equiv f_\rho$ is the conformal map from $\C\setminus\overline{\mathbb D}$ onto $\C\setminus S_W$ and is given by \eqref{def of f pre trunc}.

We state below the formal expansions of $\RRR$, $\AAA$, and
$\QQQ$ as $\rho\to0$; their derivation is given in Appendix~\ref{subsec:exp param weak pre crit}. Writing $\widehat{\RRR}\in\R$ for an undetermined first-order coefficient, we have
\begin{equation} \label{small rho expansion of QQQ}
\begin{split}
    \RRR=\QQ+\widehat{\RRR}\rho+O(\rho^2),
    \qquad 
    \AAA
    =\frac{4a(c+1)}{(2c+1-a)^2}\rho+O(\rho^2),
    \qquad 
    \QQQ
    =\QQ+
    \left(
        \widehat{\RRR}
        +\frac{1-2c-a}{2(1+2c-a)}\QQ
    \right)\rho
    +O(\rho^2),
\end{split}
\end{equation}
where
\begin{equation} \label{def of QQ}
    \QQ=
    \frac{a+1}{2\sqrt{a}}
    \frac{\sqrt{2c+1}}{c+1}.
\end{equation}
In particular (see also Appendix~\ref{subsec:exp param weak pre crit}),
\begin{equation} \label{f pre rho0}
    f_0(z)
    :=\lim_{\rho\to0}f_\rho(z)
    =\frac{z(\QQ z-1)}{z-\QQ},
\end{equation}
locally uniformly for $z\in\C\setminus\{\QQ\}$.

\begin{rem}[Geometry of the map $f_0$]
Notice that, for $|z|=1$,
\[
\begin{split}
    |f_0(z)|
    &=\left|\frac{\QQ z-1}{z-\QQ}\right|
    =\left|\frac{\QQ\overline z-1}{z-\QQ}\right|
    =\left|\frac{\QQ/z-1}{z-\QQ}\right|
    =1.
\end{split}
\]
Thus $f_0$ maps the unit circle onto a subset of the unit circle. Moreover,
\[
    f_0'(z)
    =\frac{\QQ(z^2-2\QQ z+1)}{(z-\QQ)^2},
\]
so the critical points of $f_0$ are $  \QQ\pm i\sqrt{1-\QQ^2}$, both of which lie on the unit circle. 
The corresponding critical values are
\[
\begin{split}
    f_0\left(\QQ\pm i\sqrt{1-\QQ^2}\right) =2\QQ^2-1
      \pm 2i\QQ\sqrt{1-\QQ^2}  =\left(\QQ\pm i\sqrt{1-\QQ^2}\right)^2.
\end{split}
\]
Since $f_0(1)=-1$, it follows that
\[
    \{f_0(z):|z|=1\}
    =
    \left\{
        e^{i\theta}:
        2\arccos(\QQ)\leq|\theta|\leq\pi
    \right\}.
\]
By \eqref{def of QQ}, we have $2\arccos(\QQ)=\theta_*$, where $\theta_*$ is given by \eqref{def of theta star support}.

Let $e^{i\theta_1}$ and $e^{i\theta_2}$ be two points of the unit circle such that the line through them passes through $1/\QQ$. For $|z|=1$, we have
\[
    \arg f_0(z)
    =\arg\left(\frac{z(z-1/\QQ)}{z-\QQ}\right)
    =2\arg(z-1/\QQ)+\pi \pmod{2\pi},
\]
where we used $z-\QQ=-\QQ z\,\overline{z-1/\QQ}$.
The arguments of $e^{i\theta_1}-1/\QQ$ and
$e^{i\theta_2}-1/\QQ$ are equal, so
$f_0(e^{i\theta_1})=f_0(e^{i\theta_2})$; see Figure~\ref{Fig_conformal rho0}.

\begin{figure}[t]
    \centering
    \begin{tikzpicture}
        \draw[->] (-3.5,0) -- (4.5,0);
        \draw[green] (0,0) circle(2);
        \draw[dashed] (1,0) -- (1,{sqrt(3)});

        \filldraw (0,0) circle(2pt) node[below] {$0$};
        \filldraw (1,0) circle(2pt) node[below] {$\QQ$};
        \filldraw (4,0) circle(2pt) node[below] {$1/\QQ$};

        \draw[dotted] (4,0) -- (-1,{sqrt(3)});

        \filldraw[red] (1,{sqrt(3)}) circle(2pt);
        \filldraw[red] (1,{-sqrt(3)}) circle(2pt);

        \filldraw[blue] (-1,{sqrt(3)}) circle(2pt)
            node[above] {$e^{i\theta_1}$};
        \filldraw[blue] ({13/7},{3*sqrt(3)/7}) circle(2pt)
            node[above right] {$e^{i\theta_2}$};

        \draw[->] (4.5,1.5)
            arc[start angle=120,end angle=60,radius=2];
        \node[above] at (5.65,1.75) {$f_0$};

        \draw[->] (6.5,0) -- (13.5,0);
        \filldraw (10,0) circle(2pt) node[below] {$0$};

        \draw[green] (9,{sqrt(3)})
            arc[start angle=120,end angle=240,radius=2];

        \filldraw[red] (9,{sqrt(3)}) circle(2pt);
        \filldraw[red] (9,{-sqrt(3)}) circle(2pt);
        \filldraw[blue] ({59/7},{5*sqrt(3)/7}) circle(2pt);
    \end{tikzpicture}
    \caption{Illustration of the limiting map $f_0$ for $\QQ=1/2$.}
    \label{Fig_conformal rho0}
\end{figure}
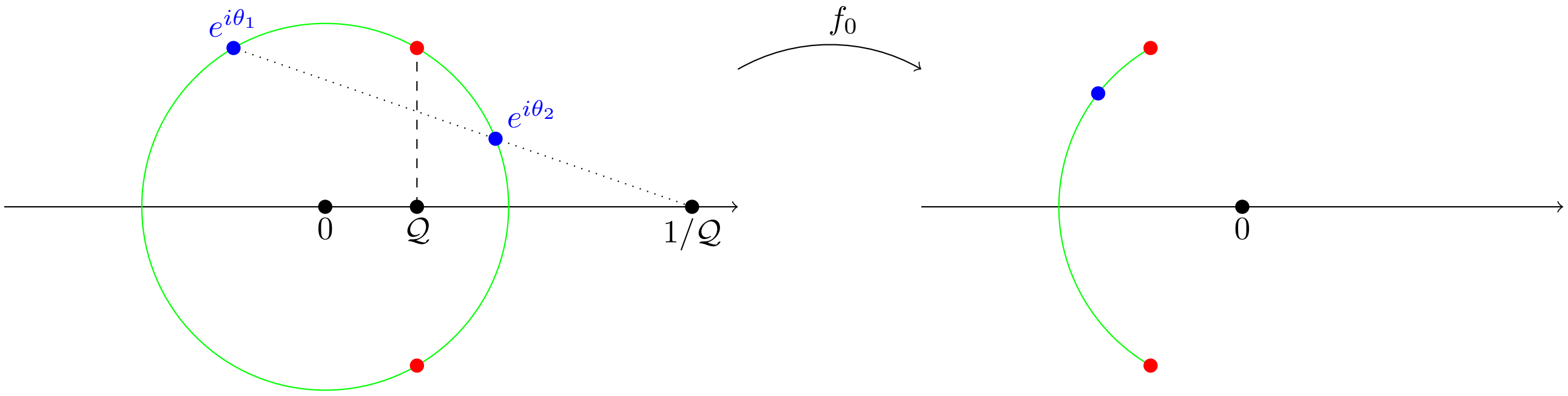
\end{rem}

The limiting map determines the support of the limiting measure, but to recover its density we must also determine the radial thickness of the planar droplet.
For fixed $\theta$ with $\theta_*<|\theta|\leq\pi$, recall from \eqref{def of x1 x2} that the two intersections of $\partial S_W$ with the ray
$\{re^{i\theta}:r>0\}$ have moduli
$r_{j,\rho}(\theta)=1+\rho x_j(\theta)+O(\rho^2)$, $j=1,2$. 
By \eqref{density weak scaling}, the planar density is of order
$\rho^{-1}$, while \eqref{def of x1 x2} gives the radial thickness
\begin{align*}
r_{1,\rho}(\theta)-r_{2,\rho}(\theta)
    =\rho\bigl(x_1(\theta)-x_2(\theta)\bigr)+O(\rho^2).
\end{align*}
Integrating the planar density in the radial direction gives the limiting density $\Upsilon(\theta)/\pi$ in \eqref{def of Upsilon}. It therefore remains to determine $x_1(\theta)$ and $x_2(\theta)$.

\begin{rem}\label{Rem_Rhat independence}
Before computing $x_1$ and $x_2$, we explain why they do not
depend on the undetermined coefficient $\widehat{\RRR}$.
Although $\widehat{\RRR}$ enters the first-order expansion of $f_\rho$, its
effect is purely tangential: it changes the argument of boundary points, but not their modulus to first order. 
Indeed, substituting \eqref{small rho expansion of QQQ} into $f_\rho$ (defined in \eqref{def of f pre trunc}) and using \eqref{Kappa RAQ}, we obtain, as $\rho\to0$,
\begin{align}\label{f rho first order expansion}
f_\rho(z)
=
f_0(z)
+
\rho\Bigl(h_{\widehat{\RRR}}(z)+h_1(z)\Bigr)
+
O(\rho^2),
\qquad
h_{\widehat{\RRR}}(z)
=
\widehat{\RRR}\,
\frac{z(z^2-1)}{(z-\QQ)^2},
\end{align}
where $h_1(z)$ is independent of $\widehat{\RRR}$. To examine the contribution of $h_{\widehat{\RRR}}$, fix $z\in\partial\D$ and consider $f_0(z)+\rho h_{\widehat{\RRR}}(z)$. Since $|f_0(z)|=1$, as $\rho\to0$,
\begin{align*}
& \bigl|f_0(z)+\rho h_{\widehat{\RRR}}(z)\bigr|
    =
    1+\rho\,\re\left(
        \frac{h_{\widehat{\RRR}}(z)}{f_0(z)}
    \right)+O(\rho^2),
    \\
    & \arg\bigl(f_0(z)+\rho h_{\widehat{\RRR}}(z)\bigr)
    =
    \arg f_0(z)+\rho\,\im\left(
        \frac{h_{\widehat{\RRR}}(z)}{f_0(z)}
    \right)+O(\rho^2).
\end{align*}
For $|z|=1$, we have
\[
    \frac{1}{f_0(z)}
    \frac{z(z^2-1)}{(z-\QQ)^2}
    =
    \frac{z^2-1}{(z-\QQ)(\QQ z-1)} =
    -\frac{2i\,\im z}{|z-\QQ|^2}
    \in i\R.
\]
Since $\widehat{\RRR}\in\R$, it follows that
\begin{align}\label{Rhat zero radial contribution}
\re\left(\frac{h_{\widehat{\RRR}}(z)}{f_0(z)}\right)=0.
\end{align}
Using \eqref{Rhat zero radial contribution} and $|f_0(z)|=1$
in \eqref{f rho first order expansion}, we obtain, as
$\rho\to0$,
\begin{align}\label{f rho modulus expansion}
|f_\rho(z)|
    =1+\rho\,\re\left(\frac{h_1(z)}{f_0(z)}\right)
    +O(\rho^2),
    \qquad |z|=1.
\end{align}
Let $\zeta_1,\zeta_2\in\T$ be the two preimages of
$e^{i\theta}$ under $f_0$, and let $\zeta_{1,\rho},\zeta_{2,\rho}$ be the corresponding preimages for $f_\rho$. Since $f_0'(\zeta_j)\ne0$, the expansions
\eqref{f rho first order expansion} and \eqref{def of x1 x2}
give $\zeta_{j,\rho}=\zeta_j+O(\rho)$ as
$\rho\to0$. Applying \eqref{f rho modulus expansion} with
$z=\zeta_{j,\rho}$, we obtain, as $\rho\to0$,
\begin{align*}
r_{j,\rho}(\theta) =|f_\rho(\zeta_{j,\rho})|  =1+\rho\,\re\left(
        \frac{h_1(\zeta_{j,\rho})}{f_0(\zeta_{j,\rho})}
    \right)+O(\rho^2)  =1+\rho\,\re\left(
        \frac{h_1(\zeta_j)}{f_0(\zeta_j)}
    \right)+O(\rho^2).
\end{align*}
Comparing with \eqref{def of x1 x2}, we conclude that
\[
    x_j(\theta)
    =\re\left(\frac{h_1(\zeta_j)}{f_0(\zeta_j)}\right),
    \qquad j=1,2.
\]
Since both $f_0$ and $h_1$ are independent of
$\widehat{\RRR}$, so are $x_1$, $x_2$, and the limiting
density in \eqref{def of Upsilon}.
\end{rem}

We now use the two inverse branches of $f_\rho$. For $\rho>0$, solving $f_\rho(z)=w$ using \eqref{def of f pre trunc} and \eqref{Kappa RAQ} gives
\begin{equation}\label{def of gpm}
g_\pm(w)= \frac{ \QQQ w+ \RRR + \AAA\QQQ^3-\AAA\QQQ \mp \sqrt{ ( \QQQ w+ \RRR + \AAA\QQQ^3-\AAA \QQQ )^2-4 \QQQ^3 \RRR w }  }{ 2 \QQQ \RRR }. 
\end{equation}
Since $f_\rho(\T)=\partial S_W$, a point $re^{i\theta}$
lies on $\partial S_W$ if and only if $|g_+(re^{i\theta})|=1$ or $|g_-(re^{i\theta})|=1$.
Using \eqref{def of x1 x2}, we therefore determine
$x_1(\theta)$ and $x_2(\theta)$ from
\begin{equation}\label{radial boundary condition}
    \left|g_\pm\bigl((1+\rho x)e^{i\theta}\bigr)\right|
    =1+O(\rho^2),
    \qquad \rho\to0.
\end{equation}
We formally write
\begin{equation}\label{gpm first order expansion}
g_\pm(w) = g_{\pm,0}(w) + g_{\pm,1}(w) \rho +O(\rho^2),\qquad \rho\to0.
\end{equation}
Substituting \eqref{small rho expansion of QQQ} into \eqref{def of gpm}, we obtain
\begin{equation}\label{def of g0pm}
 g_{\pm,0}(w)= \frac{ w+1 \mp \sqrt{ (w+1)^2-4\QQ^2w }  }{ 2 \QQ }. 
\end{equation}

We now give the formulas for $x_1(\theta)$ and $x_2(\theta)$;
their derivation from \eqref{radial boundary condition} is
given in Appendix~\ref{subsec:radial density weak pre crit}. Since $g_{-,0}(w)g_{+,0}(w)=w$, we define $\varphi\in\R$ by
\begin{equation}\label{def of varphi}
	g_{+,0}(e^{i\theta}) = e^{ i \varphi }, \qquad g_{-,0}(e^{i\theta}) = e^{ i (\theta-\varphi) }.
\end{equation} 
Set
\begin{align}\label{def of ABC weak pre}
	A= \frac{1}{\QQ}, \qquad B=  \frac{1-2c-a }{ 2(1+2c-a) }, \qquad C= A \Big( B+ \frac{c}{c+1} \Big), 
\end{align}
so that
\begin{align}\label{x2 explicit}
\begin{split}
x_2  =  \frac{  AB \cos( \theta-\varphi)  - AC \cos\theta + C \cos \varphi-B      }{ A^2-2A \cos(\theta-\varphi)+1 } 
 =    \frac{  (A^2-1) B    + A \frac{ c}{c+1} \cos \varphi -A^2 \frac{c}{c+1} \cos \theta  }{ 1-A^2-2A^2 \cos\theta +2 A  \cos \varphi  },
\end{split}
\end{align}
and
\begin{align}\label{x1 explicit}
\begin{split}
x_1  & = \frac{  AB \cos(  \varphi)  - AC \cos\theta + C \cos (\theta-\varphi)-B      }{ A^2-2A \cos \varphi+1 }
 =   \frac{  (A^2-1) B   - A \frac{ c}{c+1} \cos \varphi + A^2 \frac{c}{c+1}     }{ A^2-2A \cos \varphi+1 }. 
\end{split}
\end{align} 
Substituting these expressions into \eqref{def of Upsilon} yields
\begin{align}\label{Upsilon in terms of varphi}
	\begin{split}
		\Upsilon(\theta) & =	 \frac{  ( a(2c+1)-1 )^2 }{ a+1    }  \frac{a   (c+1) -   \sqrt{a} (a+1)   \sqrt{2c+1}   \cos \varphi+    a   (c+1)   \cos\theta}{(a-1)(a(2c+1)-1)       +   2\sqrt{a} (a+1)    \sqrt{2c+1} \cos \varphi -    4a   (c+1)   \cos \theta}
		\\
		&\quad \times \frac{1 }{     1+a(2c+1)    -      2\sqrt{a}     \sqrt{2c+1}   \cos \varphi   } 
	\end{split}
\end{align}
Finally, \eqref{def of varphi} and \eqref{def of g0pm} give
\begin{equation}\label{eiphi in terms of theta}
	e^{i\varphi} =  \frac{ e^{i\theta}+1 - \sqrt{ (e^{i\theta}+1)^2-4\QQ^2e^{i\theta} }  }{ 2 \QQ }. 
\end{equation} 
Taking real parts in \eqref{eiphi in terms of theta}, with the square root chosen to have nonnegative real part, we obtain
\begin{align}\label{cos varphi in terms of theta}
	\cos \varphi &= \frac{1}{\QQ} \Big(  \cos^2 (\theta/2) -    \sqrt{ \QQ^2- \cos^2(\theta/2) } |\sin (\theta/2)| \Big), \qquad |\theta| \in [2\arccos \mathcal{Q},\pi]. 
\end{align}
Substituting \eqref{cos varphi in terms of theta} into \eqref{Upsilon in terms of varphi} and using \eqref{def of Upsilon}, we obtain the density
\eqref{def of eq msr 1D pre}.

\section{Proof of Theorem~\ref{Thm_2D eq msr}}  \label{Section_EL strong}

In this section, we prove Theorem~\ref{Thm_2D eq msr} by verifying
the Euler--Lagrange variational conditions \eqref{var eq_2D} and
\eqref{var ineq_2D} for the candidate measures obtained in
Section~\ref{Section_conformal strong}. We first check that the
candidate droplets are contained in $\mathbb D(0,\sqrt{1+\rho})$
and that the corresponding measures have total mass one.
In the pre-critical regime, we also verify that the map
\eqref{def of f pre trunc main} is injective on the unit circle,
so that its image is a Jordan curve.

\begin{lem}[Properties of the candidate droplets]
\label{Lem_admissibility strong}
Assume that \(\rho>0\), \(c>0\), and
\(0<a<\sqrt{1+\rho}\).

\begin{itemize}
    \item[(a)]
    If \(0<a\leq a_{\rm cri}\), then
    \begin{equation} \label{eq for disc inclusions}
        \D(\mathcal A,r_0)\subset\D(0,r_1)
        \subset\D(0,\sqrt{1+\rho}),
        \qquad
        a\in\D(\mathcal A,r_0).
    \end{equation} 
    Moreover, the measure in \eqref{def of mu W 2D main}, with
    \(S_W\) given by \eqref{droplet post trunc}, has total mass one.

    \smallskip

    \item[(b)]
    Suppose that \(Y\) and \(\QQQ\) satisfy $\PP(Y)=0$ and \eqref{c in terms of R Q A}, with $Y\in ( \frac{1-\QQQ}{2}, \frac{1+\QQQ}{2} )$ and $0<\QQQ< \min\{1,a^{-1}\}$.  
    Let \(X\), \(\RRR\), \(\AAA\), and \(\mathcal K\) be given by
    \eqref{relation X and Y}, \eqref{RRR in terms of X},
    \eqref{AAA in terms of X}, and
    \eqref{def of f pre trunc main}, respectively. Then $\RRR>0, \AAA>0$, and  $\mathcal{K}>0$.  
    
    In addition, the rational function \(f\) in
    \eqref{def of f pre trunc main} is univalent on
    \(\{w:|w|>1\}\), extends injectively to \(\T\), and maps
    \(\T\) into \(\D(0,\sqrt{1+\rho})\). Consequently,
    \(f(\T)\) is a Jordan curve, and the closed region
$S_W$ bounded by this curve is contained in
$\D(0,\sqrt{1+\rho})$. The measure in \eqref{def of mu W 2D main} has total mass one. Moreover, $a\notin S_W$.
\end{itemize}
\end{lem}

The proof of Lemma~\ref{Lem_admissibility strong} is provided in Appendix~\ref{Appendix_admissble droplet}. 

We now verify the Euler--Lagrange conditions \eqref{var eq_2D} and \eqref{var ineq_2D} for the candidate measure $\mu_W$. For this purpose, define
\begin{equation}
    H(z)
    :=
    \int_{\C}
    \log\frac{1}{|\zeta-z|^2}\,d\mu_W(\zeta)
    -
    \rho\log
    \left(
        1-\frac{|z|^2}{1+\rho}
    \right)
    -
    2c\log|z-a|,
    \qquad
    |z|<\sqrt{1+\rho}.
\end{equation}
By \eqref{def of potential W strong}, to verify \eqref{var eq_2D} and \eqref{var ineq_2D}, it suffices to show that
\begin{equation} \label{var conditions in terms of H}
    H(z)
    \begin{cases}
        =\mathsf C_W,
        &\textup{if }z\in S_W,
        \smallskip
        \\
        \geq\mathsf C_W,
        &\textup{if }
        z\in
        S_W^c\cap\D(0,\sqrt{1+\rho}),
    \end{cases}
\end{equation}
for some constant \(\mathsf C_W\). (Outside the hard-wall disc $\D(0,\sqrt{1+\rho})$, the variational inequality is automatic because \(W=+\infty\).)

To verify \eqref{var conditions in terms of H}, we will use the
following lemma.

\begin{lem}\label{Lem_EL reduction strong}
To prove the variational equality in \eqref{var conditions in terms of H}, it suffices to show that
    \begin{equation} \label{var eq modified}
        \frac{1}{2\pi i}
        \int_{\partial S_W}
        \frac{1}{z-\zeta}
        \frac{\rho\,\overline{\zeta}}
        {1+\rho-|\zeta|^2}\,d\zeta
        =
        -\frac{c}{z-a},
        \qquad
        z\in S_W^{\circ}.
    \end{equation}
Moreover, once the variational equality has been established, the
variational inequality follows if $H$ has no local minimum in
$\D(0,\sqrt{1+\rho})\setminus(S_W\cup\{a\})$ whose value is strictly
smaller than $\mathsf C_W$. In particular, it is sufficient to prove
either that
    \begin{equation} \label{var ineq modified}
        \frac{1}{2\pi i}
        \int_{\partial S_W}
        \frac{1}{z-\zeta}
        \frac{\rho\,\overline{\zeta}}
        {1+\rho-|\zeta|^2}\,d\zeta
        \neq
        \frac{\rho\,\overline z}
        {1+\rho-|z|^2}
        -
        \frac{c}{z-a}
    \end{equation}
    for every $ z\in \D(0,\sqrt{1+\rho})\setminus S_W$ with $z \neq a$, or, more generally, that every point at which equality holds in
    \eqref{var ineq modified} is not a local minimum of \(H\). 

Here and below, \(\partial S_W\) is positively oriented with respect
to \(S_W\). Thus, in the post-critical regime, the outer circular
boundary is oriented counterclockwise and the inner circular boundary
is oriented clockwise.
\end{lem}

\begin{proof}
By \eqref{def of mu W 2D main}, for \(z\in S_W^{\circ}\), the equality \(\partial_zH(z)=0\) is equivalent to 
\begin{equation}\label{lol16}
    \int_{S_W}
    \frac{1}{z-\zeta}
    \frac{\rho(1+\rho)}
    {(1+\rho-|\zeta|^2)^2}\,dA(\zeta)
    =
    \frac{\rho\,\overline z}{1+\rho-|z|^2}
    -
    \frac{c}{z-a}.
\end{equation}
For \(|z|<\sqrt{1+\rho}\), $z\notin \partial S_{W}$, Green's formula gives
\begin{equation} \label{Green thm Cauchy var eq}
\begin{split}
    &\int_{S_W}
    \frac{1}{z-\zeta}
    \frac{\rho(1+\rho)}
    {(1+\rho-|\zeta|^2)^2}\,dA(\zeta) 
    =
    \frac{1}{2\pi i}
    \int_{\partial S_W}
    \frac{1}{z-\zeta}
    \frac{\rho\,\overline{\zeta}}
    {1+\rho-|\zeta|^2}\,d\zeta
    +
    \frac{\rho\,\overline z}
    {1+\rho-|z|^2}
    \mathbbm{1}_{\{z\in S_W\}}.
\end{split}
\end{equation}
Substituting \eqref{Green thm Cauchy var eq} into
\eqref{lol16}, we see that \eqref{var eq modified} is equivalent to
\(\partial_zH(z)=0\) on \(S_W^{\circ}\). Since \(H\) is real-valued and
\(S_W^{\circ}\) is connected, \(H\) is constant on \(S_W^{\circ}\). 
By continuity, $H$ has the same constant value on $S_W=\overline{S_W^\circ}$, proving the variational equality in \eqref{var conditions in terms of H}.

Suppose now that the variational inequality in \eqref{var conditions in terms of H} fails. 
Then there exists \(z_0 \in\D(0,\sqrt{1+\rho})\setminus(S_W\cup\{a\})\) such that $ H(z_0)<\mathsf C_W.$  
On the other hand, $ H=\mathsf C_W$ on $\partial S_W$, while $H(z) \to +\infty$ as $|z| \to \sqrt{1+\rho}$ or $z \to a$.  
Consequently, the sublevel set $K:=\{z\in\D(0,\sqrt{1+\rho}):H(z)\leq H(z_0)\}$ is compact and contained in $ \D(0,\sqrt{1+\rho})\setminus(S_W\cup\{a\}).$ 
The function \(H\) attains its minimum on $K$ at some point $z_{\min}$. 
Since $H>H(z_0)$ outside $K$, this point is also a local minimum of $H$, with
$H(z_{\min})\leq H(z_0)<\mathsf C_W$. The function $H$ is differentiable near $z_{\min}$, so $\partial_zH(z_{\min})=0$. 
Since $z_{\min}\notin S_W$, the indicator in \eqref{Green thm Cauchy var eq} vanishes at $z_{\min}$. Hence \(\partial_zH=0\) means that the two sides of
\eqref{var ineq modified} are equal at $z_{\min}$. This proves the assertion.
\end{proof}

In the post-critical regime, we shall prove the stronger condition
\eqref{var ineq modified} by contradiction. Namely, we suppose that
\begin{equation} \label{var ineq modified contradiction}
    \frac{1}{2\pi i}
    \int_{\partial S_W}
    \frac{1}{z-\zeta}
    \frac{\rho\,\overline{\zeta}}
    {1+\rho-|\zeta|^2}\,d\zeta
    =
    \frac{\rho\,\overline z}
    {1+\rho-|z|^2}
    -
    \frac{c}{z-a}
\end{equation}
for some $z\in  \D(0,\sqrt{1+\rho})\setminus S_W$ with $z \neq a$, and show that such a point cannot exist. In the pre-critical regime, \eqref{var ineq modified contradiction} may have solutions in $\D(0,\sqrt{1+\rho})\setminus (S_W\cup \{a\})$; in that case, we shall instead show that $H$ does not have a local minimum at any of these points.

\subsection{Proof of Theorem~\ref{Thm_2D eq msr} (a): the post-critical regime}

In this subsection, we verify that \eqref{var eq modified} and
\eqref{var ineq modified} hold with $  S_W=\overline{\mathbb{D}(0,r_1)}\setminus    \mathbb{D}(\mathcal{A},r_0)$, as in \eqref{droplet post trunc}, whenever $a\leq a_{\rm cri}$.

\subsubsection{Variational equality \eqref{var eq modified}}

Suppose first that $z \in S_W^{ \circ }$. Since
$|z|<r_1$, we have
\begin{align} \label{var eq modified post 1}
    \frac{1}{2\pi i} \int_{|\zeta|=r_1}
    \frac{1}{z-\zeta}
    \frac{\rho\,\overline\zeta}{1+\rho-|\zeta|^2}\,d\zeta
    &=
    \frac{\rho r_1^2}{1+\rho-r_1^2}
    \frac{1}{2\pi i}\int_{|\zeta|=r_1}
    \frac{1}{z-\zeta}\frac{d\zeta}{\zeta}
    =0.
\end{align}
Indeed, for $z\neq0$, the residues of the last integrand at $\zeta=0$ and
$\zeta=z$ cancel. For $z=0$, the integrand is $-1/\zeta^2$, whose residue is zero.

On the other hand, let $g(w)=\mathcal A+r_0w$ and orient the circle
$\{|\zeta-\mathcal A|=r_0\}$ counterclockwise. Then
\begin{align}
\begin{split} 
&\quad \int_{|\zeta-\mathcal{A}|=r_0}
  \frac{1}{z-\zeta}
  \frac{\rho\,\overline{\zeta}}{1+\rho-|\zeta|^2}\,d\zeta
  =r_0\rho\int_{|w|=1}
  \frac{1}{z-g(w)}
  \frac{\overline{g(w)}}{1+\rho-g(w)\overline{g(w)}}\,dw 
\\
&  =r_0\rho\int_{|w|=1}
  \frac{1}{z-g(w)}
  \frac{g(1/w)}{1+\rho-g(w)g(1/w)}\,dw
  =-\frac{\rho}{\mathcal A}\int_{|w|=1}
  \frac{1}{z-\mathcal A-r_0w}
  \frac{\mathcal Aw+r_0}{(w-w_+)(w-w_-)}\,dw, \label{inner circle integral post} 
  \end{split}
\end{align} 
where
\begin{align*}
    w_\pm
    &=
    \frac{-(\mathcal A^2+r_0^2-\rho-1)
    \pm\sqrt{(\mathcal A^2+r_0^2-\rho-1)^2
    -4\mathcal A^2r_0^2}}
    {2\mathcal A r_0}.
\end{align*}
The two roots satisfy $w_+w_-=1$. Moreover, by
\eqref{def of A center} and \eqref{def of r0},
\[
    w_- = \frac{a-\mathcal A}{r_0},
    \qquad
    w_+-w_-
    =
    \frac{1}{\mathcal A r_0}
    \frac{\rho(\rho+1)(1+\rho-a^2)}
    {\rho(\rho+1)+c(1+\rho-a^2)}.
\]
By Lemma~\ref{Lem_admissibility strong}, we have
$a\in\D(\mathcal A,r_0)$, so $|w_-|<1$ and therefore $|w_+|>1$.
Furthermore, since $z\in S_W^{\circ}$, the pole
$w=(z-\mathcal A)/r_0$ lies outside the unit circle. Thus only the pole
at $w=w_-$ contributes, and a residue computation (using also \eqref{def of A center} and \eqref{def of r0}) yields
\begin{align} \label{var eq modified post 2}
\begin{split}
    \frac{1}{2\pi i}
    \int_{|\zeta-\mathcal A|=r_0}
    \frac{1}{z-\zeta}
    \frac{\rho\,\overline\zeta}{1+\rho-|\zeta|^2}\,d\zeta
    &=-\frac{\rho}{\mathcal A}
    \frac{\mathcal Aw_-+r_0}{w_--w_+}
    \frac{1}{z-\mathcal A-r_0w_-}
   =\frac{c}{z-a}.
\end{split}
\end{align}
The positively oriented boundary of $S_W$ consists of the outer
circle oriented counterclockwise and the inner circle oriented
clockwise. Therefore, by \eqref{var eq modified post 2}, the
inner-circle contribution to the integral in
\eqref{var eq modified} is $-c/(z-a)$. Together with
\eqref{var eq modified post 1}, this proves
\eqref{var eq modified}. 

\subsubsection{Variational inequality \eqref{var ineq modified}}

We prove the stronger condition \eqref{var ineq modified} by
contradiction. Suppose that \eqref{var ineq modified contradiction}
holds for some $z\in\mathbb{D}(0,\sqrt{1+\rho})\setminus S_W$ with $z \neq a$.  
Since $z\notin S_W$, \eqref{droplet post trunc} implies
that either $|z|>r_1$ or
$|z-\mathcal A|<r_0$.

We first consider the case $r_1<|z|<\sqrt{1+\rho}$. 
Repeating the calculation in \eqref{var eq modified post 1}, with only the pole $\zeta=0$ inside the contour, and using \eqref{var eq modified post 2} for the inner circle, we obtain
\begin{equation}
    \frac{1}{2\pi i}\int_{\partial S_W}
    \frac{1}{z-\zeta}
    \frac{\rho\,\overline\zeta}{1+\rho-|\zeta|^2}\,d\zeta
    =
    \frac{\rho r_1^2}{1+\rho-r_1^2}\frac1z
    -\frac{c}{z-a}.
\end{equation}
If \eqref{var ineq modified contradiction} were true, cancelling
the common term $-c/(z-a)$ and dividing by $\rho$ would give
\[
    \frac{r_1^2}{z(1+\rho-r_1^2)}
    =
    \frac{\overline z}{1+\rho-|z|^2}.
\]
Multiplying by $z$ and clearing denominators yields $|z|^2=r_1^2,$ contradicting $|z|>r_1$.

We next consider the case $|z-\mathcal A|<r_0$. In this case, the pole $w=(z-\mathcal A)/r_0$ now lies inside the unit circle. The residue of the integrand in \eqref{inner circle integral post} at this point is (using \eqref{def of A center} and \eqref{def of r0})
\begin{align}\label{res at z-A/r0}
\underset{w=(z-\mathcal A)/r_0}{\Res}
    \left[
        r_0\rho\frac{1}{z-g(w)}
        \frac{g(1/w)}{1+\rho-g(w)g(1/w)}
    \right]
    =
    -\frac{a\rho}{1+\rho-az}
    -c\frac{1+\rho-a^2}
    {(1+\rho-az)(z-a)}.
\end{align}
Since $|z|<r_1$, the outer-circle integral vanishes by \eqref{var eq modified post 1}. For the counterclockwise inner-circle integral in \eqref{inner circle integral post}, we add the residue at $w=w_-$ computed in \eqref{var eq modified post 2} to the residue at $w=(z-\mathcal A)/r_0$ just computed in \eqref{res at z-A/r0}. Reversing the orientation of the inner circle, we obtain
\begin{equation}
    \frac{1}{2\pi i}\int_{\partial S_W}
    \frac{1}{z-\zeta}
    \frac{\rho\,\overline\zeta}{1+\rho-|\zeta|^2}\,d\zeta
    =
    \frac{a\rho}{1+\rho-az}
    +c\frac{1+\rho-a^2}
    {(1+\rho-az)(z-a)}
    -\frac{c}{z-a}.
\end{equation}
Consequently, \eqref{var ineq modified contradiction} is equivalent
to
\[
    c(1+\rho-|z|^2)(1+\rho-a^2)
    =\rho(1+\rho)|z-a|^2.
\]
Using \eqref{def of A center} and \eqref{def of r0}, the difference
between the two sides factorises as
\[
\begin{split}
c(1+\rho-a^2)(1+\rho-|z|^2)
    -\rho(1+\rho)|z-a|^2
    =
    \bigl[\rho(1+\rho)+c(1+\rho-a^2)\bigr]
    \bigl(r_0^2-|z-\mathcal A|^2\bigr).
\end{split}
\]
Since $\rho(1+\rho)+c(1+\rho-a^2)>0$, the critical-point
equation \eqref{var ineq modified contradiction} can hold
only when $|z-\mathcal A|=r_0$, contradicting the assumption
that $|z-\mathcal A|<r_0$.

We have therefore proved that \eqref{var ineq modified} holds for every
$z\in\mathbb D(0,\sqrt{1+\rho})\setminus(S_W\cup\{a\})$.
By Lemma~\ref{Lem_EL reduction strong}, this establishes
the variational inequality in \eqref{var conditions in terms of H}
and completes the proof of Theorem~\ref{Thm_2D eq msr}~(a).

\subsection{Proof of Theorem~\ref{Thm_2D eq msr} (b): the pre-critical regime}

We now consider the pre-critical regime $a>a_{\rm cri}$. 
By Lemma~\ref{Lem_admissibility strong}, the rational function $f_\rho$ in \eqref{def of f pre trunc main} is a conformal map from $\C\setminus\overline{\mathbb D}$ onto $S_W^c$. To lighten the notation, we write $f\equiv f_\rho$ throughout this subsection. The condition $a>a_{\rm cri}$ guarantees the existence of the parameters defining $f$, by Proposition~\ref{Prop_admissible pre parameters}.

Let $w_\pm$ be the solutions of $f(w)=z$ (recall \eqref{def of f pre trunc main}), given by
\begin{equation} \label{def of w pm pre}
    w_\pm
    =
    \frac{
        \RRR\QQQ^2+\mathcal K+\QQQ z
        \pm
        \sqrt{
            (\RRR\QQQ^2+\mathcal K+\QQQ z)^2
            -4\RRR\QQQ^3z
        }
    }{
        2\RRR\QQQ
    }.
\end{equation}
No global choice of the square-root branch will be needed. 
Equivalently, $w_+$ and $w_-$ are the two roots of
\begin{equation} \label{w pm qua eq_pre}
    \RRR\QQQ w^2
    -(\RRR\QQQ^2+\mathcal K+\QQQ z)w
    +\QQQ^2z=0.
\end{equation}
Thus, by \eqref{def of f pre trunc main},
\begin{align}\label{factorization of f(z)-z}
    f(w)-z
    =\RRR\frac{(w-w_+)(w-w_-)}{w-\QQQ}.
\end{align}

We shall use the following notation. If
$z\in S_W^{\circ}$, then both roots $w_\pm$ lie in $\D$. Indeed, by Lemma~\ref{Lem_admissibility strong}, $f$ maps $\{w:|w|>1\}$ onto $S_W^c$ and $\T$ onto $\partial S_W$. If $z\in S_W^c$, we label the roots so that 
\[
    w_+=f^{-1}(z)\in \C\setminus\overline{\mathbb D},
    \qquad
    w_-\in\D.
\]
The exterior root $w_{+}$ is simple and unique by Lemma~\ref{Lem_admissibility strong}, so $w_{-} \in \overline{\D}$. Actually $w_{-}\notin \T$, because $f(\T)=\partial S_W$ and $z\in S_W^c$. Thus $w_{-}\in \D$. 
We first assume that $w_+\neq w_-$. At the finitely many values of $z$ for which the roots coincide, the identities below follow by continuity. 
Note that 
\begin{equation} \label{w pm sum prod pre}
    w_++w_-
    =\frac{1}{\RRR}
    \left(
        \RRR\QQQ+\frac{\mathcal K}{\QQQ}+z
    \right),
    \qquad
    w_+w_-=\frac{\QQQ}{\RRR}z.
\end{equation}

For $z\notin\partial S_W$, the change of variables $\zeta=f(w)$ and the identity $\overline{f(w)}=f(1/w)$ for $|w|=1$ give
\begin{align}\label{lol17}
    \int_{\partial S_W}
    \frac{1}{z-\zeta}
    \frac{\rho\,\overline\zeta}{1+\rho-|\zeta|^2}\,d\zeta
    &=
    \int_{|w|=1}
    \frac{1}{z-f(w)}
    \frac{\rho\,\overline{f(w)}f'(w)}
    {1+\rho-f(w)\overline{f(w)}}\,dw =\int_{|w|=1}G_z(w)\,dw,
\end{align}
where
\begin{equation}\label{def of Gz}
    G_z(w)
    :=\frac{F(w)}{z-f(w)}
    =-\frac{1}{\RRR}
    \frac{(w-\QQQ)F(w)}{(w-w_+)(w-w_-)}.
\end{equation}
Here $F$ is defined in \eqref{def of F capital}, and the second equality in \eqref{def of Gz} follows from \eqref{factorization of f(z)-z}. 
By Lemma~\ref{Lem_F pole residue}, the pole of $F$ at $w=\QQQ$ is simple; it is therefore cancelled by the factor $w-\QQQ$ in \eqref{def of Gz}. Consequently, the only possible poles of $G_z$ in $\D$ are $0$, $1/z_*$, and those among $w_\pm$ that lie in $\D$. 
If some of these points coincide, the resulting integral identities below follow by continuity in $z$.

By \eqref{def of f pre trunc main} and
\eqref{evaluations of zstar}, we have
\[
    f(0)=0,
    \qquad
    f(1/z_*)=\frac{1+\rho}{a}.
\]
Using \eqref{def of Gz}, Lemma~\ref{Lem_F pole residue}, and \eqref{mass one Res eqn}, we obtain, for $z\notin\{0,(1+\rho)/a\}$,
\begin{align} \label{Residue of Gz at 0 zstar}
    \underset{w=0}{\Res}\,G_z(w)
    &=\frac{1}{z}\underset{w=0}{\Res}\,F(w) =\frac{1+c}{z},
    \qquad
    \underset{w=1/z_*}{\Res}\,G_z(w)
    =
    \frac{
        1-\rho-\underset{w=0}{\Res}\,F(w)
    }{
        z-\frac{1+\rho}{a}
    }=-\frac{\rho+c}{z-\frac{1+\rho}{a}}.
\end{align}
On the other hand, differentiating
\eqref{factorization of f(z)-z} at $w=w_\pm$ gives
\[
    f'(w_+)
    =\RRR\frac{w_+-w_-}{w_+-\QQQ},
    \qquad
    f'(w_-)
    =-\RRR\frac{w_+-w_-}{w_--\QQQ}.
\]
Using \eqref{def of Gz}, \eqref{def of F capital}, and $f(w_\pm)=z$, we obtain
\begin{align} \label{Residue of Gz at wpm}
\begin{split}
    \underset{w=w_+}{\Res}\,G_z(w)
    &=-\frac{F(w_+)}{f'(w_+)}=
    \frac{\rho}{z}
    \left(
        1-\frac{1+\rho}{1+\rho-zf(1/w_+)}
    \right),
    \\
    \underset{w=w_-}{\Res}\,G_z(w)
    & =-\frac{F(w_-)}{f'(w_-)}=
    \frac{\rho}{z}
    \left(
        1-\frac{1+\rho}{1+\rho-zf(1/w_-)}
    \right).
\end{split}
\end{align}

\subsubsection{Variational equality \eqref{var eq modified}}

Suppose that $z\in S_W^\circ$. Then $w_\pm\in\D$, and
$1/z_*\in\D$ by Lemma~\ref{Lem_admissibility strong}. Therefore, by
\eqref{lol17} and the residue theorem, the desired identity
\eqref{var eq modified} is equivalent to
\begin{equation} \label{var equality pre v1}
    \underset{w=0}{\Res}\,G_z(w)
    +\underset{w=1/z_*}{\Res}\,G_z(w)
    +\underset{w=w_+}{\Res}\,G_z(w)
    +\underset{w=w_-}{\Res}\,G_z(w)
    =-\frac{c}{z-a}.
\end{equation}
Using \eqref{Residue of Gz at 0 zstar} and
\eqref{Residue of Gz at wpm}, this identity can be rewritten as
\begin{equation} \label{var equality pre v2}
    \rho(1+\rho)\mathcal F(z)
    =
    1+\rho+c
   - \frac{1+\rho}{a}
    \frac{
        \rho+c
    }{
        z-\frac{1+\rho}{a}
    }
    +\frac{ac}{z-a},
\end{equation}
where
\begin{equation} \label{def of mathcal F}
 \mathcal F(z)
    :=
    \frac{
        2(1+\rho)
        -z\bigl(f(1/w_+)+f(1/w_-)\bigr)
    }{
        (1+\rho)^2
        -(1+\rho)z\bigl(f(1/w_+)+f(1/w_-)\bigr)
        +z^2f(1/w_+)f(1/w_-)
    }.
\end{equation} 
The function $\mathcal F$ can also be written as
\begin{align}\label{Fcal as a sum}
\mathcal F(z)
    =
    \frac{1}{1+\rho-zf(1/w_+)}
    +
    \frac{1}{1+\rho-zf(1/w_-)}.
\end{align}
We will prove \eqref{var equality pre v2} by showing that both sides
have the same simple poles and residues, and the same limit
as $z\to\infty$.

\begin{lem}\label{Lem_mathcal F partial fractions}
The rational function $\mathcal F$ in \eqref{def of mathcal F} satisfies
\begin{equation*}
\begin{split}  
    \mathcal F(z)
    &=
    \frac{1+\rho+c}{\rho(1+\rho)}
    +\frac{ac}{\rho(1+\rho)}\frac{1}{z-a} 
 - \frac{
        \rho+c
    }{a\rho}
    \frac{1}{z-\frac{1+\rho}{a}}.
\end{split}
\end{equation*}
\end{lem}

\begin{proof}
The expression in \eqref{def of mathcal F} is a symmetric
rational function of $w_+$ and $w_-$. It is therefore rational
in their sum and product, and hence in $z$ by
\eqref{w pm sum prod pre}.
Since $f(w_\pm)=z$, we have
\[
    \mathcal F(z)
    =
    \frac{1}{1+\rho-f(w_+)f(1/w_+)}
    +
    \frac{1}{1+\rho-f(w_-)f(1/w_-)}.
\]
Thus $\mathcal F(z)$ can have a finite pole only if there
exists $w$ satisfying
\[
    f(w)=z,
    \qquad
    1+\rho-f(w)f(1/w)=0.
\]
By the factorisation \eqref{factorisation denominator F},
the second equation implies that $w=z_*$ or $w=z_*^{-1}$. 
By \eqref{evaluations of zstar}, we have
\[
    f(z_*)=a,
    \qquad
    f(1/z_*)=\frac{1+\rho}{a}.
\]
Therefore, the only possible finite poles of $\mathcal F$ are $a$ and $(1+\rho)/a$.

Let $w_0\in\{z_*,z_*^{-1}\}$. Since $w_0$ is real and $f'(w_0)=\RRR+\mathcal K/(w_0-\QQQ)^2>0$, there is a local inverse branch $w=w(z)$ near $z=f(w_0)$
satisfying $w(f(w_0))=w_0$.
Changing variables $z=f(w)$ and using \eqref{def of F capital}, we obtain
\begin{align}\label{residue change variables mathcal F}
\underset{z=f(w_0)}{\Res}
    \frac{1}{1+\rho-zf(1/w(z))} =
    \underset{w=w_0}{\Res}
    \frac{f'(w)}{1+\rho-f(w)f(1/w)}
    =
    \frac{
        \underset{w=w_0}{\Res}\,F(w)
    }{
        \rho f(1/w_0)
    }.
\end{align}
At $z=(1+\rho)/a$, the term $1/(1+\rho-zf(1/w_+))$ is holomorphic (the conformal inverse $w_+$ is holomorphic here, and \eqref{factorisation denominator F} excludes a pole: $w_+\neq z_*^{-1}$ since $|w_+|>1$, and $w_+\neq z_*$ since $f(z_*)=a\neq(1+\rho)/a$). Therefore, by \eqref{Fcal as a sum}, the
residue of $\mathcal F$ equals that of $1/(1+\rho-zf(1/w_-))$, with $w_-=1/z_*$. Using \eqref{mass one Res eqn}, we therefore obtain
\[
    \underset{z=\frac{1+\rho}{a}}{\Res}\,\mathcal F(z)
    =
    \frac{\underset{w=1/z_*}{\Res}\,F(w)}{a\rho}
    =
    -\frac{\rho+c}{a\rho}. 
\]
To compute the other residue, we first note that
\[
    \underset{w=z_*}{\Res}\,F(w)=c.
\]
Indeed, by \eqref{def of F capital} and \eqref{factorisation denominator F}, one readily checks that the finite poles of $F$ are $0$, $\QQQ$, $1/z_*$, and $z_*$. At $w=1/\QQQ$, the pole of $f(1/w)$ cancels between the numerator and denominator in \eqref{def of F capital}, since $f(1/\QQQ)=\AAA>0$.
By \eqref{residue of F at 0}, \eqref{residue of F at QQQ},
and \eqref{mass one Res eqn}, we have
\[
    \underset{w=0}{\Res}\,F(w)
    +\underset{w=\QQQ}{\Res}\,F(w)
    +\underset{w=1/z_*}{\Res}\,F(w)
    =(1+c)+\rho-(\rho+c)=1.
\] 
Using \eqref{def of f pre trunc main} and \eqref{f'(0) relation} in
\eqref{def of F capital}, we obtain
\[
    F(w)
    =\frac{\rho\RRR f'(0)}{1+\rho-\RRR f'(0)}
      \frac1w+O(w^{-2})
    =\frac{1+c}{w}+O(w^{-2}),
    \qquad w\to\infty.
\]
The residue theorem on the Riemann sphere therefore implies
that the sum of all finite residues of $F$ is $1+c$.
Since the residues at $0$, $\QQQ$, and $1/z_*$
sum to $1$, we conclude that
\[
    \underset{w=z_*}{\Res}\,F(w)=c.
\]
At $z=a$, the pole in \eqref{Fcal as a sum} comes from
the term with $w_+=z_*$. Changing variables $z=f(w)$
and using \eqref{evaluations of zstar}, we obtain
\[
    \underset{z=a}{\Res}\,\mathcal F(z)
    =
    \frac{\underset{w=z_*}{\Res}\,F(w)}
         {\rho f(1/z_*)}
    =
    \frac{ac}{\rho(1+\rho)}.
\]
It remains to determine the value of $\mathcal F(z)$ at infinity. 
As $z\to\infty$, by \eqref{w pm sum prod pre}, $w_+=z/\RRR+O(1)$ and $w_-\to\QQQ$. Hence
$zf(1/w_+)\to\RRR f'(0)$, while
$f(1/w_-)\to f(1/\QQQ)=\AAA>0$.
The second term in \eqref{Fcal as a sum} therefore tends
to zero, and
\[
    \mathcal F(z)
    \to 
    \frac{1}{1+\rho-\RRR f'(0)}
    =\frac{1+\rho+c}{\rho(1+\rho)},
\]
where we again used \eqref{f'(0) relation}. The claim follows.
\end{proof}

Lemma~\ref{Lem_mathcal F partial fractions} is precisely
\eqref{var equality pre v2}. This proves
\eqref{var equality pre v1}, and therefore the variational equality
\eqref{var eq modified}.  

\subsubsection{Variational inequality \eqref{var ineq modified}}

Let $ z\in \D(0,\sqrt{1+\rho})\setminus(S_W\cup\{a\}),$  and label the roots so that $|w_+|>1$ and $|w_-|<1$. The contour in
\eqref{lol17} encloses the pole at $w_-$ but not the pole at $w_+$. 
Since \eqref{var equality pre v1} holds as a rational identity in $z$, it also applies here.
Using \eqref{Green thm Cauchy var eq}, we obtain
\begin{align*}
    \partial_zH(z)
    &=
    \underset{w=w_+}{\Res}\,G_z(w)
    +\frac{\rho\,\overline z}{1+\rho-|z|^2}
=
    \frac{
        \rho(1+\rho)
        \bigl(\overline z-f(1/w_+)\bigr)
    }{
        (1+\rho-|z|^2)
        \bigl(1+\rho-zf(1/w_+)\bigr)
    }, 
\end{align*}
where the last equality follows from \eqref{Residue of Gz at wpm}.
By \eqref{factorisation denominator F}, the second factor in the denominator can vanish for the exterior root
only when $w_+=z_*$, which corresponds to $z=a$ by \eqref{evaluations of zstar}. Since $z\neq a$, this factor cannot vanish. 
Since $f$ has real coefficients and $f(w_\pm)=z$, the two
solutions of $f(w)=\overline z$ are $\overline{w_+}$ and
$\overline{w_-}$. Thus $f(1/w_+)=\overline z$ implies that
$1/w_+=\overline{w_+}$ or $1/w_+=\overline{w_-}$.
The first equality would give $|w_+|^2=1$, contradicting
$|w_+|>1$. Hence $1/w_+=\overline{w_-}$. 

Write $w_-=r e^{i\varphi}$, where $0<r<1$. Then $ w_+=r^{-1}e^{i\varphi}.$ 
The product relation in \eqref{w pm sum prod pre} gives $  z=\frac{\RRR}{\QQQ}e^{2i\varphi}. $ 
Substituting this into the sum relation in
\eqref{w pm sum prod pre} and multiplying by $e^{-i\varphi}$, we
obtain 
\begin{equation}
  r+r^{-1}
    =
    \frac{
        (\RRR\QQQ^2+\mathcal K)e^{-i\varphi}
        +\RRR e^{i\varphi}
    }{
        \RRR\QQQ
    }. \label{eq for r+1/r}
\end{equation} 
Taking imaginary parts and using
$\mathcal K=(\RRR-\AAA\QQQ)(1-\QQQ^2)$ (from \eqref{Kappa RAQ}) gives $  \AAA\QQQ(1-\QQQ^2)\sin\varphi=0. $ 
Since $\AAA>0$ by Lemma~\ref{Lem_admissibility strong}, we have
$\sin\varphi=0$. The choice $\varphi=\pi$ is impossible, because the
right-hand side of \eqref{eq for r+1/r} then becomes negative, whereas $r+r^{-1}>0$. Therefore $\varphi=0$, so $H$ can have a critical point outside
$S_W$ only at $z_0=\RRR/\QQQ$. 

If $z_0 \notin \D(0,\sqrt{1+\rho})\setminus S_W$, then $H$ has no critical points in this set. 
Suppose now that $z_0 \in \D(0,\sqrt{1+\rho})\setminus S_W$. 
Let $u>1$ be the exterior solution of $f(u)=z_0$. 
By \eqref{w pm sum prod pre}, the product of the two roots is $(\QQQ/\RRR)z_0=1$, so the other solution is $u^{-1}$. 
From $f(u)=f(u^{-1})$ and \eqref{def of f pre trunc main}, we obtain 
\begin{equation} \label{eq for K/R}
 \frac{\mathcal K}{\RRR}
    =\frac{(u-\QQQ)(\QQQ u-1)}{u}.
\end{equation} 
Since $\mathcal K/\RRR>0$ and $u>1>\QQQ$, it follows that $\QQQ u>1$.

Let $f^{-1}$ denote the exterior inverse of $f$ near $z_0$, and set
\[
    \phi(z)
    :=f\left(\frac{1}{f^{-1}(z)}\right).
\]
Then $\phi(z_0)=z_0$. Using \eqref{eq for K/R}, we obtain 
\[
    f'(u)
    =\RRR\frac{u^2-1}{u(u-\QQQ)},
    \qquad
    f'(u^{-1})
    =\RRR\frac{u^2-1}{\QQQ u-1}, \qquad  \phi'(z_0)
    =-\frac{f'(u^{-1})}{u^2f'(u)}
    =-\frac{u-\QQQ}{u(\QQQ u-1)}. 
\] 
Furthermore, by \eqref{w pm sum prod pre} and $  \RRR-(\RRR\QQQ^2+\mathcal K)    =\AAA\QQQ(1-\QQQ^2)>0$, we have 
\[
    u+u^{-1}
    =\frac{\RRR\QQQ^2+\mathcal K+\RRR}{\RRR\QQQ}
    <\frac{2}{\QQQ}. 
\] 
It follows that
\[
    1+\phi'(z_0)
    =\frac{\QQQ u^2-2u+\QQQ}{u(\QQQ u-1)}<0,
    \qquad
    1-\phi'(z_0)>0.
\]

Writing $z=z_0+h$ in the formula for $\partial_zH$, we find
\[
    \partial_zH(z_0+h)
    =
    \frac{\rho(1+\rho)}{(1+\rho-z_0^2)^2}
    \bigl(\overline h-\phi'(z_0)h\bigr)
    +O(|h|^2), \qquad \mbox{as } h \to 0.
\]
Since $\phi'(z_0)$ is real, the two eigenvalues of the Hessian of $H$
at $z_0$ are
\[
    \frac{2\rho(1+\rho)}{(1+\rho-z_0^2)^2}
    \bigl(1-\phi'(z_0)\bigr)>0, \qquad    \frac{2\rho(1+\rho)}{(1+\rho-z_0^2)^2}
    \bigl(1+\phi'(z_0)\bigr)<0.
\] 
Thus $z_0$ is a saddle point and, in particular, is not a local
minimum of $H$. Lemma~\ref{Lem_EL reduction strong} therefore yields
the variational inequality. This completes the proof in the
pre-critical regime.

\appendix

\section{Algebraic details for the pre-critical regime}
\label{Appendix_lemmas}

Throughout this appendix, we assume that $\rho>0$, $0<a<\sqrt{1+\rho}$ and $0<\QQQ<1$ and use the notation
\begin{equation} \label{def of delta mathsf s t in appendix}
    \delta:=1+\rho-a^2, \qquad
    \mathsf{s}:=\frac{1-\QQQ}{2}\delta, \qquad
    \mathsf{t}:=\frac{1+\QQQ}{2}\delta.
\end{equation} 
In particular, \(\delta>0\) and \(0<\mathsf{s}<\mathsf{t}\).

\begin{lem}[The admissible zero of \(\PP\)]
\label{Lem_poly PP unique root}
Let \(\PP\) be the polynomial defined in \eqref{def of polynomial PP}. Then $\PP$ has a zero in $(\frac{1-\QQQ}{2},\frac{1+\QQQ}{2})$ if and only if $a \QQQ<1$. Whenever this zero exists, it is unique and simple.
\end{lem}

\begin{proof}
For $u>0$, consider the change of variables
\[
    y=
    \frac{(1-\QQQ)+(1+\QQQ)u}{2(1+u)}.
\]
This gives a bijection from $(0,\infty)$ onto
$(\frac{1-\QQQ}{2},\frac{1+\QQQ}{2})$. Write
\begin{equation}
\widehat{\PP}(u):= \frac{\delta}{\QQQ^2}(1+u)^6
    \PP\left(
        \frac{(1-\QQQ)+(1+\QQQ)u}{2(1+u)}
    \right)
    =
    \sum_{j=0}^{6}b_j u^j.
\end{equation} 
A direct expansion gives
\begin{equation}
 b_0
    =
    a^2(1-\QQQ)^2(a^2\QQQ^2-1),
    \qquad
    b_1
    =
    2a^2(1-\QQQ)
    \bigl(\QQQ^2(1+\rho+a^2)-2\bigr),
\end{equation} 
and
\begin{equation}
\label{b2 appendix}
\begin{split}
    \QQQ^2 b_2
    &=
    \bigl(\QQQ^2(1+\rho+a^2)-2\bigr)
    \bigl(1+(5-2\QQQ^2)a^2\QQQ^2\bigr)
    +2(1-a^4\QQQ^4)
    +a^2\QQQ^2(1-\QQQ^2)
    (3-\QQQ^2-2a^2\QQQ^2).
\end{split}
\end{equation}
Moreover,
\begin{align*}
&b_3=  4(1+\QQQ)  \bigl(\delta+a^2\QQQ^2\rho\bigr), \qquad b_4= (1+\QQQ)^2
    \bigl[
        6\delta+4a^2
        +a^2(1-\QQQ^2+\rho\QQQ^2)
    \bigr],
\\
&   b_5=4(1+\rho)(1+\QQQ)^3,\qquad \qquad b_6 =(1+\rho)(1+\QQQ)^4.
\end{align*} 
In particular, $b_3,b_4,b_5,b_6>0$.

Suppose first that $a\QQQ\geq1$. Since $\delta>0$,
\[
    \QQQ^2(1+\rho+a^2)-2
    =
    \delta\QQQ^2+2a^2\QQQ^2-2
    >
    2a^2\QQQ^2-2.
\]
It therefore follows from \eqref{b2 appendix} that
\[
\begin{split}
    \QQQ^2 b_2
    &>
    (2a^2\QQQ^2-2)
    \bigl(1+(5-2\QQQ^2)a^2\QQQ^2\bigr)
    +2(1-a^4\QQQ^4) 
    +a^2\QQQ^2(1-\QQQ^2)
    (3-\QQQ^2-2a^2\QQQ^2)
    \\
    &=
    a^2\QQQ^2
    \bigl(
        \QQQ^4-2a^2\QQQ^4+6a^2\QQQ^2-5
    \bigr) \geq
    a^2\QQQ^2(1-\QQQ^2)^2
    >0.
\end{split}
\]
Hence $b_0\geq0$ and $b_j>0$ for $j=1,\ldots,6$. Thus $\widehat{\PP}(u)>0$ for every $u>0$, so $\PP$ has no zero in $(\frac{1-\QQQ}{2},\frac{1+\QQQ}{2})$.

Suppose now that $a\QQQ<1$. Then $b_0<0$. If $\QQQ^2(1+\rho+a^2)-2>0,$ then $b_1>0$ and all three terms on the right-hand side of \eqref{b2 appendix} are positive. Indeed, $a\QQQ<1$ gives $1-a^4\QQQ^4>0$, and $   3-\QQQ^2-2a^2\QQQ^2
    >
    1-\QQQ^2
    >0.$ 
Thus $b_2>0$. 
If instead $\QQQ^2(1+\rho+a^2)-2\leq0,$  then $b_1\leq0$, while the sign of $b_2$ is of no importance. Indeed, in either case, after any zero coefficients are omitted, the sequence $ b_0,b_1,b_2,b_3,b_4,b_5,b_6$ has exactly one sign change. Descartes' rule of signs therefore shows that $\widehat{\PP}$ has exactly one positive zero, counted with multiplicity. The above change of variables then implies that $\PP$ has exactly one zero in $(\frac{1-\QQQ}{2},\frac{1+\QQQ}{2})$, counted with multiplicity. In particular, this zero is simple.
\end{proof}

We next show that $\mathcal K>0$ if and only if $\mathsf s<X<\mathsf t$. This will be used below and in Appendix~\ref{Appendix_admissble droplet}.

\begin{lem}[Positivity of \(\mathcal K\)]
\label{Lem_mathcal K positive}
Let \(X>0\) with \(X\neq\mathsf{t}\), and let \(\RRR\) and \(\AAA\) be given by \eqref{RRR in terms of X} and \eqref{AAA in terms of X}. Then
\begin{equation}
\label{factorisation K appendix}
\begin{split}
\mathcal K
&=-\frac{\QQQ(1+\QQQ)}{2aX}
  \frac{X^2-\mathsf{s}^2}{X^2-\mathsf{t}^2}
  \Bigl[
      X^2+(1+\QQQ)(1+\rho+a^2)X+\mathsf{t}^2
  \Bigr].
\end{split}
\end{equation}
Consequently, $\mathcal K>0$ if and only if $ \mathsf{s}<X<\mathsf{t}.$  
In particular, if \(X=\delta Y\) and $Y \in (\frac{1-\QQQ}{2},\frac{1+\QQQ}{2})$, then \(\mathcal K>0\).
\end{lem}

\begin{proof}
Substituting \eqref{RRR in terms of X} and \eqref{AAA in terms of X} into $ \mathcal K=(\RRR-\AAA\QQQ)(1-\QQQ^2)$ and simplifying gives \eqref{factorisation K appendix}. The last factor is strictly positive, since
\[
    X^2+(1+\QQQ)(1+\rho+a^2)X+\mathsf{t}^2
    =(X+\mathsf{t})^2+2a^2(1+\QQQ)X>0.
\]
Since $\frac{\QQQ(1+\QQQ)}{2aX}>0$, \(\mathcal K>0\) precisely when $ (X^2-\mathsf{s}^2)(X^2-\mathsf{t}^2)<0,$ which, for \(X>0\), is equivalent to \(\mathsf{s}<X<\mathsf{t}\).
\end{proof}

\subsection{Existence and uniqueness of the pre-critical parameters}

We now turn from the equation $\PP(Y)=0$ for fixed $\QQQ$
to the full system consisting of this equation and
\eqref{c in terms of R Q A}. Here, $\PP$ is defined by \eqref{def of polynomial PP}.
By Lemma~\ref{Lem_poly PP unique root}, for each
$\QQQ\in(0,\min\{1,a^{-1}\})$, the equation $\PP(Y)=0$
has a unique solution
$Y=Y(\QQQ)\in(\frac{1-\QQQ}{2},\frac{1+\QQQ}{2})$.
It remains to show that \eqref{c in terms of R Q A} determines
$\QQQ$ uniquely.

\begin{prop}[Existence and uniqueness of the admissible pre-critical parameters]
\label{Prop_admissible pre parameters}
Let $\rho>0$, $c>0$, and
$a_{\rm cri}<a<\sqrt{1+\rho}$. Then there exists a unique pair
$(Y,\QQQ)$ satisfying $\PP(Y)=0$  and \eqref{c in terms of R Q A} with $Y\in ( \frac{1-\QQQ}{2}, \frac{1+\QQQ}{2} )$ and $0<\QQQ< \min\{1,a^{-1}\}$.
\end{prop}
\begin{proof}
Fix $\QQQ\in(0,\min\{1,a^{-1}\})$. By
Lemma~\ref{Lem_poly PP unique root}, the polynomial $\PP$ has a unique
simple zero $Y=Y(\QQQ)$ in $(\frac{1-\QQQ}{2},\frac{1+\QQQ}{2})$. Since the
coefficients of $\PP$ depend continuously on $\QQQ$, the simplicity
and uniqueness of this zero imply that $Y(\QQQ)$ is continuous.

Set $X=\delta Y(\QQQ)$, and define $\RRR$, $\AAA$ and $\mathcal K$ by \eqref{RRR in terms of X}, \eqref{AAA in terms of X} and \eqref{Kappa RAQ}, respectively. Define
\[
    \Lambda(\QQQ)
    :=
    \RRR^2+\frac{\RRR\mathcal K}{\QQQ^2}
    =
    \frac{
        \RRR\bigl(\RRR-\AAA\QQQ(1-\QQQ^2)\bigr)
    }{\QQQ^2}.
\]
Then $\Lambda$ is continuous, and \eqref{c in terms of R Q A} is
equivalent to
\begin{equation}\label{Lambda equation appendix}
    \Lambda(\QQQ)
    =
    \frac{(1+c)(1+\rho)}{\rho+c+1}=r_{1}^{2},
\end{equation}
where $r_{1}$ is defined in \eqref{def of r1}.

We first determine the limit as $\QQQ\to0$. The bounds $Y\in ( \frac{1-\QQQ}{2}, \frac{1+\QQQ}{2} )$ imply that $Y(\QQQ)\to1/2$. Expanding
$\PP(Y(\QQQ))=0$ as $\QQQ\to0$ and using \eqref{RRR in terms of X} and \eqref{factorisation K appendix}, we obtain
\[
\begin{split}
    Y(\QQQ)
    &=
    \frac12
    -
    \frac{\sqrt{1+\rho}-a}
         {2(\sqrt{1+\rho}+a)}\,\QQQ
    +O(\QQQ^2),  \qquad
    \RRR
    =
    \sqrt{1+\rho}\,\QQQ+O(\QQQ^2),
    \qquad
    \mathcal K
    =
    \sqrt{1+\rho}\,\QQQ+O(\QQQ^2).
\end{split}
\]
Consequently, $\Lambda(\QQQ)\to1+\rho$ as $\QQQ\to0$.

Suppose first that $0<a\leq1$. Set
$\varepsilon=1-\QQQ$. 
Expanding $\PP(Y(\QQQ))=0$ as $\varepsilon\to0$ and using \eqref{RRR in terms of X} and \eqref{factorisation K appendix}, we obtain that $Y(\QQQ)=O(\varepsilon)$,
$\mathcal K\to0$, and $\RRR\to r_*$, where $r_*$ is the unique positive solution of $\Phi_a(r)=0$, where 
\[ 
\Phi_a(r):=    (1-a^2)r^2+2a\rho r
    -\bigl(1+\rho+a^2(\rho-1)\bigr).
\]
It follows that $ \lim_{\QQQ\to 1}\Lambda(\QQQ)=r_*^2.$ 

Since $0<a\leq1$, the function $\Phi_a$ is strictly increasing
for $r>0$. Using \eqref{def of r1} and the definition of $a_{\rm cri}$
in \eqref{def of c cri}, a direct factorisation gives
\[
\begin{split}
    \Phi_a(r_1)
    &=
    \frac{\rho(2c+\rho+1)}{\rho+c+1}
    (a-a_{\rm cri})
    \left(
        \frac{\sqrt{\rho+1}}
        {\sqrt{(\rho+c+1)(c+1)}-\sqrt{(\rho+c)c}}
        -a
    \right).
\end{split}
\]
The last factor is strictly positive for
$0<a\leq1$. Indeed, its first term is strictly larger than $1$, since
\[
    \sqrt{(\rho+c+1)(c+1)}-\sqrt{(\rho+c)c}
    <\sqrt{\rho+1};
\]
after squaring twice, this inequality reduces to
\[
    (c+1)(c+\rho)>c(c+\rho+1),
\]
whose two sides differ by $\rho>0$. Since $a>a_{\rm cri}$ by
assumption, we conclude that $\Phi_a(r_1)>0$. As
$\Phi_a(r_*)=0$ and $\Phi_a$ is strictly increasing, this implies
$r_*<r_1$. Therefore,
\[
    \lim_{\QQQ\to 1}\Lambda(\QQQ) =r_*^2
    < r_1^2=\frac{(1+c)(1+\rho)}{\rho+c+1}
    < 1+\rho =
    \lim_{\QQQ\to0}\Lambda(\QQQ).
\]
Since $\Lambda$ is continuous, \eqref{Lambda equation appendix} has at least one solution $\QQQ\in(0,1)$.

Suppose next that $1<a<\sqrt{1+\rho}$. As $\QQQ\to a^{-1}$, the analysis in the proof of Lemma~\ref{Lem_poly PP unique root} gives
\[
    Y(\QQQ)\to\frac{a-1}{2a}.
\]
Substitution into \eqref{RRR in terms of X} and
\eqref{factorisation K appendix} then yields
$\RRR\to1$, $\mathcal K\to0$, and hence
$\Lambda(\QQQ)\to1$. Since
\[
  \lim_{\QQQ\to a^{-1}}\Lambda(\QQQ)= 1< r_1^2=\frac{(1+c)(1+\rho)}{\rho+c+1}
    <1+\rho =\lim_{\QQQ\to0}\Lambda(\QQQ),
\]
continuity of $\Lambda$ again implies that
\eqref{Lambda equation appendix} has at least one solution, now with $\QQQ\in(0,a^{-1})$. This proves existence.

It remains to prove uniqueness. Suppose that two admissible pairs
$(Y_1,\QQQ_1)$ and $(Y_2,\QQQ_2)$ satisfy the statement of the proposition, and
let $f_1,f_2$ be the corresponding maps defined by
\eqref{def of f pre trunc}. By
Lemma~\ref{Lem_admissibility strong} and the Euler--Lagrange
verification in Section~\ref{Section_EL strong}, each pair produces
an equilibrium measure for the same potential $W$. The proof of this
uses only the equations and inequalities satisfied by each pair; it does not assume that the pair is unique.

The equilibrium measure associated with $W$ is unique. Hence the two
measures, and therefore their droplets, coincide. The maps $f_1$ and
$f_2$ are the normalised exterior conformal maps of the same domain,
so they also coincide. Since $\mathcal K_j>0$, the unique pole of
$f_j$ in $\D$ is $\QQQ_j$. We conclude that
$\QQQ_1=\QQQ_2$, and Lemma~\ref{Lem_poly PP unique root} then gives
$Y_1=Y_2$.
\end{proof}

\subsection{Reduction of the mass-one condition}

We now prove Lemma~\ref{Lem_alg eq for X} by showing that the
mass-one condition \eqref{mass one Res eqn} is equivalent to
$\PP(X/\delta)=0$.

\begin{proof}[Proof of Lemma~\ref{Lem_alg eq for X}]\label{proof lemma appendix}
For brevity, set
\begin{equation} \label{Delta positive appendix}
   \mathsf{H}
    := \RRR-\AAA\QQQ(1-\QQQ^2), \qquad   \Delta
    :
    =\QQQ^2(1+\rho)-\RRR\mathsf{H}   =\QQQ^2\frac{\rho (1+\rho)}{\rho+c+1} >0 .
\end{equation} 
Here, we have used \eqref{f'(0) relation}.  
Thus, the denominator $\Delta$ in \eqref{residue of F at 0}
and \eqref{residue of F at zstar} is nonzero.

Substituting \eqref{residue of F at zstar} into \eqref{mass one Res eqn}, multiplying by \(\Delta\), and using
$\rho\RRR\mathsf H=(1+c)\Delta$ from
\eqref{residue of F at 0}, we obtain
\[
\begin{split}
\rho\RRR\mathsf{H}
+a\rho\QQQ
 \frac{z_*-\QQQ}{z_*^2-1}
 \frac{
     \mathsf{H}z_*^2-2\QQQ\RRR z_*+\RRR
 }{
     z_*(1-\QQQ z_*)
 } 
=(1-\rho)\Delta.
\end{split}
\]
After rearrangement, this is equivalent to
\begin{equation}
\label{mass relation cleared appendix}
\begin{split}
a\rho\QQQ(z_*-\QQQ)
 \bigl(
     \mathsf{H}z_*^2-2\QQQ\RRR z_*+\RRR
 \bigr) 
=(z_*^2-1)z_*(\QQQ z_*-1)
 \bigl(
     \RRR\mathsf{H}-\QQQ^2(1-\rho^2)
 \bigr).
\end{split}
\end{equation}

We now substitute the parametrisation in terms of \(X\). Define
\begin{align}\label{def of U}
U:=X^2+\QQQ(1+\rho+a^2)X
      -\frac{1-\QQQ^2}{4}\delta^2.
\end{align}
Then \eqref{RRR in terms of X} and \eqref{AAA in terms of X} can be written as
\begin{equation}
\label{R compact appendix}
    \RRR=\frac{U}{2aX}, \qquad   \AAA
    =
    \frac{
        U^2-4(1+\rho)a^2\QQQ^2X^2
    }{
        2aX\QQQ(1-\QQQ)(X^2-\mathsf{t}^2)
    }.
\end{equation} 
Moreover, \eqref{zstar in terms of X} becomes
\begin{equation}
\label{zstar compact appendix}
    z_*=\frac{X+\mathsf{t}}{\mathsf{t}-X}.
\end{equation}
Since \(\mathsf{s}<X<\mathsf{t}\), all denominators below are non-zero, and
\begin{equation}
\label{zstar minus Q appendix}
    z_*-\QQQ
    =\frac{(1+\QQQ)(X+\mathsf{s})}{\mathsf{t}-X}, \qquad  \QQQ z_*-1
    =\frac{(1+\QQQ)(X-\mathsf{s})}{\mathsf{t}-X}, \qquad  z_*^2-1
    =\frac{4X\mathsf{t}}{(\mathsf{t}-X)^2}. 
\end{equation}  

Let \(\mathfrak E(X)\) denote the left-hand side of
\eqref{mass relation cleared appendix} minus its right-hand side.
Substituting \eqref{R compact appendix}--\eqref{zstar minus Q appendix} and simplifying, we obtain
\begin{equation}
\label{mass polynomial factorisation appendix}
\begin{split}
    \mathfrak E(X)
    &=
    -\frac{
        \delta^6\QQQ(1+\rho)(1+\QQQ)^2
        (X-\mathsf{s})
    }{
        2a^2X(\mathsf{t}-X)^5
    }
    \PP\left(\frac{X}{\delta}\right).
\end{split}
\end{equation}
The prefactor in \eqref{mass polynomial factorisation appendix} is non-zero for
\(\mathsf{s}<X<\mathsf{t}\). Consequently, \eqref{mass relation cleared appendix} is equivalent to $\PP(\frac{X}{1+\rho-a^2})=0,$ which is the desired conclusion.
\end{proof}

\section{Proof of Lemma~\ref{Lem_admissibility strong}}
\label{Appendix_admissble droplet}

We first consider the post-critical regime. By \eqref{def of A center}--\eqref{def of r1}, we have 
\[
 r_1^2<1+\rho , \qquad     r_0^2-(a-\mathcal A)^2
    =
    \frac{
        c(1+\rho-a^2)^2
    }{
        \rho(1+\rho)+c(1+\rho-a^2)
    }>0.
\]
Thus \(a\in\D(\mathcal A,r_0)\). 
A direct computation also shows that $\mathcal A+r_0\leq r_1$ if and only if $ a\leq a_{\rm cri}.$ Thus \eqref{eq for disc inclusions} follows.  
It remains to verify the mass-one condition. 
Using \eqref{def of A center}--\eqref{def of r1}, we obtain
\[
    \int_{|z|<r_1}
    \frac{\rho(1+\rho)}
    {(1+\rho-|z|^2)^2}\,dA(z)
    =c+1, \qquad    \int_{|z-\mathcal A|<r_0}
    \frac{\rho(1+\rho)}
    {(1+\rho-|z|^2)^2}\,dA(z)
    =c.
\] 
Since \(S_W =\overline{ \mathbb{D}(0,r_1) } \setminus \mathbb{D }(\mathcal{A}, r_0 )\), the total mass of the candidate measure in \eqref{def of mu W 2D main} is equal to one.

We next consider the pre-critical regime. We use the notation of
Appendix~\ref{Appendix_lemmas}. Since $ X=(1+\rho-a^2)Y$ and
$\frac{1-\QQQ}{2}<Y<\frac{1+\QQQ}{2}$, we have $ \mathsf s<X<\mathsf t.$ 
By \eqref{R compact appendix}, we have $ \RRR=\frac{U}{2aX}.$ 
Moreover, using \eqref{def of U} and \eqref{def of delta mathsf s t in appendix}, we obtain
\[
    U'(X)=2X+\QQQ(1+\rho+a^2)>0, \qquad    U(\mathsf s)
    =a^2\QQQ(1-\QQQ)(1+\rho-a^2)>0.
\] 
It follows that \(U>0\) for \(\mathsf s<X<\mathsf t\), and hence
\(\RRR>0\).

We next prove that \(\AAA>0\). 
Substituting $U=2aX\RRR$ into the expression for $\AAA$ in \eqref{R compact appendix}, we obtain
\begin{align}\label{new expr for A}
\AAA
    =
    \frac{
        2aX\bigl(\RRR^2-(1+\rho)\QQQ^2\bigr)
    }{
        \QQQ(1-\QQQ)(X^2-\mathsf t^2)
    }.
\end{align}
Suppose, for contradiction, that \(\AAA\leq0\). 
Since $\RRR>0$ and $0<\QQQ<1$, \eqref{f'(0) relation} gives
\[
    \RRR^2
    \leq \RRR\bigl(\RRR-\AAA\QQQ(1-\QQQ^2)\bigr)
    =\QQQ^2\frac{(1+c)(1+\rho)}{\rho+c+1}
    <(1+\rho)\QQQ^2.
\]
Since \(X<\mathsf t\), we also have \(X^2-\mathsf t^2<0\).
Then \eqref{new expr for A} implies \(\AAA>0\), which is a contradiction. Therefore
\(\AAA>0\). The positivity of \(\mathcal K\) now follows from
Lemma~\ref{Lem_mathcal K positive}. 
Furthermore, since $0<\mathsf s<X<\mathsf t$, \eqref{zstar compact appendix} gives
$$ z_*=\frac{X+\mathsf t}{\mathsf t-X}>1.$$ 

We now prove that \(f\) is injective on $\{z\in\C:|z|\geq1\}$. Suppose that \(f(z_1)=f(z_2)\) for two distinct points \(z_1,z_2\) with $|z_{1}|\geq 1$, $|z_{2}|\geq 1$. 
Using \eqref{def of f pre trunc main} and dividing $f(z_1)-f(z_2)=0$ by $z_1-z_2$, we obtain
\[
    (z_1-\QQQ)(z_2-\QQQ)
    =-\frac{\mathcal K}{\RRR}.
\]
Thus,
\begin{align}\label{expr for z2 in terms of z1}
z_2
    =
    \QQQ-\frac{\mathcal K}{\RRR(z_1-\QQQ)}
    =
    \frac{\QQQ z_1-\alpha}{z_1-\QQQ},
    \qquad
    \alpha:=\QQQ^2+\frac{\mathcal K}{\RRR}.
\end{align}
By \eqref{Kappa RAQ},
\[
    \alpha
    =
    1-\frac{\AAA\QQQ}{\RRR}(1-\QQQ^2).
\]
Since $\RRR,\AAA,\mathcal K>0$ and $0<\QQQ<1$, the two expressions for $\alpha$ give $ 0<\alpha<1.$ 

Let \(|z_1|=r\geq1\). Then
\[
\begin{split}
    |z_1-\QQQ|^2-|\QQQ z_1-\alpha|^2
    &=
    (1-\QQQ^2)r^2
    -2\QQQ(1-\alpha)\re z_1
    +\QQQ^2-\alpha^2
    \\
    &\geq
    (1-\QQQ^2)r^2
    -2\QQQ(1-\alpha)r
    +\QQQ^2-\alpha^2.
\end{split}
\]
Since \(\mathcal K>0\), we have by \eqref{Kappa RAQ} that \(\AAA\QQQ<\RRR\), and therefore $ \QQQ(1-\alpha)<1-\QQQ^2.$ 
Hence the last quadratic polynomial $(1-\QQQ^2)r^2-2\QQQ(1-\alpha)r+\QQQ^2-\alpha^2$ is strictly increasing for $r\geq1$, since its derivative satisfies
\[
    2(1-\QQQ^2)r-2\QQQ(1-\alpha)
    \geq 2\bigl(1-\QQQ^2-\QQQ(1-\alpha)\bigr)>0.
\]
Its value at \(r=1\) is $ (1-\alpha)(1+\alpha-2\QQQ)>0,$ since $\alpha<1$ and
\[
    1+\alpha-2\QQQ
    =(1-\QQQ)^2+\frac{\mathcal K}{\RRR}>0.
\]
Consequently, $|z_1-\QQQ|^2-|\QQQ z_1-\alpha|^2
    \geq (1-\alpha)(1+\alpha-2\QQQ)>0$, from which it follows (using \eqref{expr for z2 in terms of z1}) that $|z_2|<1$.
Thus two distinct points of \(\{|z|\geq1\}\) cannot have the same
image under \(f\).

By \eqref{def of f pre trunc main}, the critical points of $f$ are $z=\QQQ\pm i(\frac{\mathcal K}{\RRR})^{1/2}$. 
Since $ \QQQ^2+\frac{\mathcal K}{\RRR}=\alpha<1,$ it follows that \(f'\) does not vanish on \(\{|z|\geq1\}\). Since $f$ is also injective on this set, it is conformal on \(\{|z|>1\}\).

It remains to verify that $f(\T)\subset\D(0,\sqrt{1+\rho})$. 
Since $\overline{f(w)}=f(1/w)$ for $|w|=1$, \eqref{factorisation denominator F} and \eqref{Delta positive appendix} give
\[
\begin{split}
    1+\rho-|f(w)|^2
    &=
    \frac{\Delta}{\QQQ^2}
    \frac{(w-z_*)(w-z_*^{-1})}
    {(w-\QQQ)(w-\QQQ^{-1})}
    =
    \frac{\Delta}{\QQQ^2}
    \frac{
        2\cos\theta-(z_*+z_*^{-1})
    }{
        2\cos\theta-(\QQQ+\QQQ^{-1})
    }>0,
    \qquad
    w=e^{i\theta}.
\end{split}
\]
Indeed, \(z_*>1\) and \(0<\QQQ<1\), so both the numerator and the
denominator in the last quotient are strictly negative. Thus $ f(\T)\subset\D(0,\sqrt{1+\rho}).$ 
Since $\D(0,\sqrt{1+\rho})$ is a simply connected domain, the bounded domain enclosed by the Jordan curve \(f(\T)\) is also contained in this disc.

Moreover, \(f(z_*)=a\) and \(z_*>1\) (by \eqref{evaluations of zstar}), so the point charge \(a\) belongs to the exterior of \(S_W\). Finally, since $\PP(Y)=0$, Lemma~\ref{Lem_alg eq for X} gives \eqref{mass one Res eqn}. Lemma~\ref{Lem_Mass one} then shows that the measure in \eqref{def of mu W 2D main} has total mass one.

\section{Formal asymptotic calculations in the pre-critical weak limit}
\label{App_formal precritical weak limit}

This appendix contains the formal calculations used in Subsection~\ref{Subsection_1D msr precritical}. We
first derive the small-$\rho$ expansion of the parameters in
\eqref{small rho expansion of QQQ}. We then compute the coefficients $x_1$ and $x_2$ in \eqref{def of x1 x2} and obtain the candidate density
\eqref{def of eq msr 1D pre} using \eqref{def of Upsilon}.
The proof of Theorem~\ref{Thm_1D eq msr}~(b) in Section~\ref{Section_EL weak} verifies the Euler--Lagrange conditions directly and does not use these
formal calculations.

\subsection{Formal expansion of the pre-critical parameters}\label{subsec:exp param weak pre crit}

We first carry out the calculation for $a<1$ and then obtain
the formulas for $a=1$ by formally letting $a\to1$. For $\rho=0$, the polynomial
$\PP$ in \eqref{def of polynomial PP} factorises as
\[
\begin{split}
    \PP_0(y)+\frac{\PP_1(y)}{1-a^2}
    &=(1-a^2)
    \left(y+\frac{1+a}{1-a}\frac{1+\QQQ}{2}\right)
    \left(y+\frac{1-a}{1+a}\frac{1+\QQQ}{2}\right)
    \\
    &\quad\times
    \left(
        y^2+\frac{1+a}{1-a}\QQQ y
        -\frac{1-\QQQ^2}{4}
    \right)
    \left(
        y^2+\frac{1-a}{1+a}\QQQ y
        -\frac{1-\QQQ^2}{4}
    \right).
\end{split}
\]
The case $a=1$ is understood by continuity. 
The unique zero in
$(\frac{1-\QQQ}{2},
       \frac{1+\QQQ}{2})|_{\rho=0}$
is the positive root of the last quadratic factor, namely
\begin{equation} \label{Y rho 0}
    Y\big|_{\rho=0}
    =
    \frac{
        -(1-a)\QQQ\big|_{\rho=0}
        +\sqrt{(1+a)^2-4a(\QQQ\big|_{\rho=0})^2}
    }{2(1+a)}.
\end{equation}
Consequently, if $\QQ=\lim_{\rho\to0}\QQQ$, then \eqref{relation X and Y} gives
\begin{equation} \label{leading order X rho to 0 lol}
    \lim_{\rho\to0}X
    =
    \frac{1-a}{2}
    \left(
        \sqrt{(1+a)^2-4a\QQ^2}
        -(1-a)\QQ
    \right).
\end{equation}
Substituting \eqref{leading order X rho to 0 lol} into \eqref{RRR in terms of X} and \eqref{AAA in terms of RRRR}, we obtain
\[
    \lim_{\rho\to0}\RRR=\QQ,
    \qquad
    \lim_{\rho\to0}\AAA=0.
\]

To determine the first-order terms, we formally write
\[
    \RRR=\QQ+\widehat{\RRR}\rho+O(\rho^2),
    \qquad
    \AAA=\widehat{\AAA}\rho+O(\rho^2),
    \qquad
    \QQQ=\QQ+\widehat{\QQQ}\rho+O(\rho^2), \qquad \mbox{as } \rho\to0,
\]
where $\widehat{\RRR}$, $\widehat{\AAA}$, and $\widehat{\QQQ}$
are coefficients to be determined.
Expanding \eqref{f'(0) relation} as $\rho\to0$ gives
\begin{equation} \label{lol11}
    \widehat{\QQQ}
    =-\frac{c}{2(c+1)}\QQ
     +\widehat{\RRR}
     +\frac12\QQ(\QQ^2-1)\widehat{\AAA}.
\end{equation}
On the other hand, expanding \eqref{AAA in terms of RRRR} as $\rho\to0$ yields
\begin{equation} \label{lol12}
    \widehat{\AAA}
    =
    \frac{
        \bigl(\sqrt{(1+a)^2-4a\QQ^2}-1+a\bigr)
        \bigl(\QQ+2\widehat{\QQQ}-2\widehat{\RRR}\bigr)
    }{
        2(1-a)\QQ(1-\QQ^2)
    }.
\end{equation}
Substituting \eqref{lol11} into \eqref{lol12} gives
\begin{equation} \label{Ahat lol}
    \widehat{\AAA}
    =
    \frac{
        1-a-\sqrt{(1+a)^2-4a\QQ^2}
    }{
        (c+1)(\QQ^2-1)
        \bigl(1-a+\sqrt{(1+a)^2-4a\QQ^2}\bigr)
    }=
    \frac{4a}{
        (c+1)
        \bigl(1-a+\sqrt{(1+a)^2-4a\QQ^2}\bigr)^2
    }.
\end{equation}
Finally, expanding the mass relation \eqref{mass one Res eqn}
as $\rho\to0$ and using \eqref{lol11} and
\eqref{Ahat lol}, we obtain 
\[
    4a(c+1)^2\QQ^2=(1+a)^2(2c+1).
\]
The unique positive solution is given by \eqref{def of QQ}. Substituting this value into \eqref{Ahat lol} gives
\[
    \widehat{\AAA}
    =\frac{4a(c+1)}{(2c+1-a)^2}.
\]
Equation \eqref{lol11} then simplifies to
\[
    \widehat{\QQQ}-\widehat{\RRR}
    =\frac{1-2c-a}{2(1+2c-a)}\QQ,
\]
which gives \eqref{small rho expansion of QQQ}. 
Substituting the leading terms of
\eqref{small rho expansion of QQQ} into \eqref{Kappa RAQ}
and \eqref{def of f pre trunc}, we obtain the limiting map
\eqref{f pre rho0}.

\subsection{Formal derivation of the limiting density}\label{subsec:radial density weak pre crit}

Here, we derive the formulas \eqref{x2 explicit}--\eqref{x1 explicit} for $x_1(\theta)$ and $x_2(\theta)$
stated in Subsection~\ref{Subsection_1D msr precritical}
from \eqref{radial boundary condition}. 
By Remark~\ref{Rem_Rhat independence}, $x_1$ and $x_2$ are independent of $\widehat{\RRR}$, so we set $\widehat{\RRR}=0$ in the following formal calculation.

For fixed $z\in\C$ and $\theta$ satisfying
$\theta_*<|\theta|\leq\pi$, substituting
$w=(1+\rho z)e^{i\theta}$ into
\eqref{gpm first order expansion} and Taylor-expanding
at $w=e^{i\theta}$ gives 
\begin{align}
\begin{split} 
g_\pm ( (1+\rho z) e^{i \theta} ) &= g_{\pm,0}( (1+\rho z) e^{i \theta} ) + \rho\, g_{\pm,1}( (1+\rho z) e^{i \theta} ) +O(\rho^2) 
\\
&= g_{\pm,0}(  e^{i \theta} ) + \rho\,\Big( g'_{\pm,0}( e^{i\theta} ) z e^{i\theta}+ g_{\pm,1}(  e^{i \theta} ) \Big) +O(\rho^2), \label{lol13} 
\end{split}
\end{align}
as $\rho \to 0$. 
Set $\mathcal G_1=g_{+,0}(e^{i\theta})\in \T$, $\mathcal G_2=g_{-,0}(e^{i\theta}) \in \T$.
Choose $z_1,z_2 \in \mathbb{C}$ such that
\begin{align}\label{gpm z1 z2 expansion rho small}
g_{+}( (1+\rho  z_1) e^{i \theta} ) - \mathcal{G}_1 =O(\rho^2), \qquad 
g_{-}( (1+\rho z_2) e^{i \theta} ) - \mathcal{G}_2 =O(\rho^2), \qquad \mbox{as } \rho \to 0.
\end{align}

Then $x_1=\re z_1$ and $x_2=\re z_2$. Indeed, by \eqref{lol13}, the first condition in
\eqref{gpm z1 z2 expansion rho small} gives
\begin{align}\label{z1 first order relation}
e^{i\theta}g'_{+,0}(e^{i\theta})z_1
    +g_{+,1}(e^{i\theta})=0.
\end{align}
Differentiating the first identity in \eqref{def of varphi}
with respect to $\theta$, we obtain
\begin{equation}\label{gplus angular derivative}
    \frac{e^{i\theta}g'_{+,0}(e^{i\theta})}
         {g_{+,0}(e^{i\theta})}
    =\varphi'(\theta)\in\R.
\end{equation}
Moreover, $\varphi'(\theta)\ne0$, since $g_{+,0}$ is
conformal near $e^{i\theta}$. 
For real $x$, equations \eqref{lol13}, \eqref{z1 first order relation}, and
\eqref{gplus angular derivative} therefore give, as $\rho \to 0$,
\[
\begin{aligned}
    \left|g_+\bigl((1+\rho x)e^{i\theta}\bigr)\right|
    &=\left|1+\rho\,\varphi'(\theta)(x-z_1)
        +O(\rho^2)\right|=1+\rho\,\varphi'(\theta)(x-\re z_1)+O(\rho^2).
\end{aligned}
\]
Thus \eqref{radial boundary condition} holds precisely when
$x=\re z_1$, proving that $x_1=\re z_1$. The same argument gives $x_2=\re z_2$.

By \eqref{def of gpm}, we have 
\begin{align}\label{sum and product for gpm}
g_+(w)+ g_-(w) = \frac{ \QQQ w+ \RRR + \AAA\QQQ^3-\AAA\QQQ   }{ \QQQ\RRR }, \qquad g_+(w)g_-(w) = \frac{ \QQQ }{ \RRR } w,  
\end{align}
and thus, using also \eqref{lol13}, we obtain
\begin{align}
\begin{split} 
&\quad g_{+}( (1+\rho  z_1) e^{i \theta} ) + g_{-}( (1+\rho z_2) e^{i \theta} ) 
\\
&= g_{+}( (1+\rho  z_1) e^{i \theta} )- g_{+}( (1+\rho  z_2) e^{i \theta} ) + \frac{ \QQQ (1+\rho z_2) e^{i \theta} + \RRR + \AAA\QQQ^3-\AAA\QQQ   }{ \QQQ\RRR } 
\\
&=  \rho\, g'_{+,0}( e^{i\theta} ) (z_1-z_2) e^{i\theta}     + \frac{ \QQQ (1+\rho z_2) e^{i \theta} + \RRR + \AAA\QQQ^3-\AAA\QQQ   }{ \QQQ\RRR }+O(\rho^2), \qquad \mbox{as } \rho \to 0, \label{sum gp sum gm expansion}
\end{split}
\end{align}
and  
\begin{align}
\begin{split} 
&\quad g_{+}( (1+\rho  z_1) e^{i \theta} )  g_{-}( (1+\rho z_2) e^{i \theta} ) = \frac{ g_{+}( (1+\rho  z_1) e^{i \theta} )   }{g_{+}( (1+\rho  z_2) e^{i \theta} )    } \frac{\QQQ}{ \RRR } (1+\rho  z_2) e^{i \theta} 
\\
&= \bigg[ 1+ \frac{1}{ g_{+,0}( e^{i\theta} ) } \rho\, g'_{+,0}( e^{i\theta} ) (z_1-z_2) e^{i\theta}  +O(\rho^2) \bigg] \frac{\QQQ}{ \RRR } (1+\rho  z_2) e^{i \theta}, \qquad \mbox{as } \rho \to 0. \label{sum gp prod gm expansion}
\end{split}
\end{align}
Using \eqref{small rho expansion of QQQ} and
\eqref{def of QQ} (with $\widehat{\RRR}=0$), we further expand
\eqref{sum gp sum gm expansion} and
\eqref{sum gp prod gm expansion} as follows:
\begin{align*}
&\quad \rho\, g'_{+,0}( e^{i\theta} ) (z_1-z_2) e^{i\theta}     + \frac{ \QQQ (1+\rho z_2) e^{i \theta} + \RRR + \AAA\QQQ^3-\AAA\QQQ   }{ \QQQ\RRR } 
\\
&=  \frac{1+e^{i\theta}}{\mathcal{Q}} + \rho \bigg[ g'_{+,0}( e^{i\theta} ) (z_1-z_2) e^{i\theta}+ \sqrt{ \frac{a}{2c+1} } \frac{ c(2c+1)+1-a-3ac }{ (1+a)(2c+1-a) } + \sqrt{ \frac{a}{2c+1} } \frac{2(c+1)}{a+1} e^{ i \theta} z_2 \bigg] +O(\rho^2) 
\end{align*} 
and 
\begin{align*}
&\quad \bigg[ 1+ \frac{1}{ g_{+,0}( e^{i\theta} ) } \rho\, g'_{+,0}( e^{i\theta} ) (z_1-z_2) e^{i\theta}   +O(\rho^2) \bigg] \frac{\QQQ}{ \RRR } (1+\rho  z_2) e^{i \theta}
\\
& =e^{i\theta} + \bigg[  \frac{g'_{+,0}( e^{i\theta} )e^{i\theta}}{ g_{+,0}( e^{i\theta} ) }   (z_1-z_2)   + \frac{1-2c-a }{ 2(1+2c-a) } + z_2 \bigg]e^{i\theta}  \rho +O(\rho^2).  
\end{align*} 
By \eqref{gpm z1 z2 expansion rho small}, the terms of order $\rho$ in \eqref{sum gp sum gm expansion} and \eqref{sum gp prod gm expansion} vanish. In particular,  
\begin{align}\label{lol14}
\Big(  \sqrt{ \frac{a}{2c+1} } \frac{2(c+1)}{a+1} e^{ i \theta} -g_{+,0}( e^{i\theta} ) \Big) z_2 = g_{+,0}( e^{i\theta} ) \frac{1-2c-a }{ 2(1+2c-a) } -  \sqrt{ \frac{a}{2c+1} } \frac{ c(2c+1)+1-a-3ac }{ (1+a)(2c+1-a) } . 
\end{align}
Interchanging $z_1$ with $z_2$ and $g_+$ with $g_-$ in the derivation of \eqref{lol14} yields 
\begin{align}\label{lol15}
\Big(  \sqrt{ \frac{a}{2c+1} } \frac{2(c+1)}{a+1} e^{ i \theta} -g_{-,0}( e^{i\theta} ) \Big) z_1 = g_{-,0}( e^{i\theta} ) \frac{1-2c-a }{ 2(1+2c-a) } -  \sqrt{ \frac{a}{2c+1} } \frac{ c(2c+1)+1-a-3ac }{ (1+a)(2c+1-a) } . 
\end{align}

Using \eqref{def of varphi} and \eqref{def of ABC weak pre},
we rewrite \eqref{lol14} and \eqref{lol15} as
\begin{align*}
(A e^{i\theta} - e^{i \varphi})z_{2} = e^{i \varphi}B-C, \qquad (A e^{i\theta} - e^{i (\theta-\varphi)})z_{1} = e^{i (\theta-\varphi)}B-C.
\end{align*}
We have
\begin{align}\label{z2 complex expression}
	\begin{split} 
z_{2} &= \frac{ e^{ i \varphi } B-C }{ A e^{ i\theta } -e^{ i \varphi } }  = \frac{ B- e^{ -i \varphi } C }{ A e^{ i( \theta-\varphi) } - 1} 
\\
& =   \frac{ (B- e^{ -i \varphi } C) (  A e^{ -i( \theta-\varphi) } - 1 ) }{ A^2-2A \cos(\theta-\varphi)+1 } =  \frac{  AB e^{ -i( \theta-\varphi) } - AC e^{ -i \theta } + C e^{ -i\varphi }-B      }{ A^2-2A \cos(\theta-\varphi)+1 }.
\end{split}
\end{align}
Similarly,
\begin{align}\label{z1 complex expression}
\begin{split}
    z_1=
    \frac{e^{i(\theta-\varphi)}B-C}
         {Ae^{i\theta}-e^{i(\theta-\varphi)}}
    =
    \frac{
        AB e^{-i\varphi}
        -AC e^{-i\theta}+C e^{-i(\theta-\varphi)}-B
    }{A^2-2A\cos\varphi+1}.
\end{split}
\end{align} 
Taking real parts in \eqref{z2 complex expression} and
\eqref{z1 complex expression}, and using $x_j=\re z_j$,
gives the first expressions in \eqref{x2 explicit} and
\eqref{x1 explicit}, respectively. 
Moreover, \eqref{def of g0pm} and \eqref{def of varphi} give
\begin{equation}\label{cosine sum weak pre}
    \cos\varphi+\cos(\theta-\varphi)
    =A(1+\cos\theta).
\end{equation}
Using \eqref{cosine sum weak pre} and
$C=A(B+c/(c+1))$, we obtain the second expressions in
\eqref{x2 explicit} and \eqref{x1 explicit}.
Substituting \eqref{x2 explicit} and \eqref{x1 explicit}
into \eqref{def of Upsilon}, we obtain
\eqref{Upsilon in terms of varphi}. 

We now derive \eqref{cos varphi in terms of theta} from
\eqref{eiphi in terms of theta}. It suffices to consider
$\theta_*\leq\theta\leq\pi$, since
\[
    g_{+,0}(e^{-i\theta})
    =\overline{g_{+,0}(e^{i\theta})},
\]
and hence $\cos\varphi$ is even in $\theta$. To simplify \eqref{eiphi in terms of theta}, we use the identity
$$
(e^{i\theta}+1)^2-4\QQ^2e^{i\theta}= 2 (1-2\QQ^2+ \cos \theta)e^{i\theta} = 4 ( \cos^2(\theta/2) -\QQ^2 ) e^{i\theta}. 
$$
Taking the square root with nonnegative real part gives, for $\pi \geq \theta \geq 2\arccos \mathcal{Q}$,
\begin{align*}
	&\quad e^{i\theta}+1 - \sqrt{ (e^{i\theta}+1)^2-4\QQ^2e^{i\theta} } 
	\\
	&= \Big( \cos \theta + 1 - 2  \sqrt{ \QQ^2- \cos^2(\theta/2) } \sin (\theta/2) \Big) +i \Big( \sin \theta + 2 \sqrt{ \QQ^2- \cos^2(\theta/2) } \cos (\theta/2) \Big)
	\\
	&= 2\Big(  \cos^2 (\theta/2) -    \sqrt{ \QQ^2- \cos^2(\theta/2) } \sin (\theta/2) \Big) +i \,2 \Big( \sin (\theta/2)\cos(\theta/2) +  \sqrt{ \QQ^2- \cos^2(\theta/2) } \cos (\theta/2) \Big).  
\end{align*}
Taking real parts in \eqref{eiphi in terms of theta} and
using the evenness of $\cos\varphi$ gives
\eqref{cos varphi in terms of theta}.

Using \eqref{cos varphi in terms of theta} and
\eqref{def of QQ} to simplify
\eqref{Upsilon in terms of varphi}, we obtain,
for $\theta_*\leq\theta\leq\pi$, 
\begin{align*}
  a   (c+1)  -   \sqrt{a} (a+1)    \sqrt{2c+1}   \cos \varphi+    a   (c+1)   \cos\theta 
 = 2a (c+1)   \sqrt{ \QQ^2- \cos^2(\theta/2) } \sin (\theta/2) .
\end{align*}
Similarly, using \eqref{cos varphi in terms of theta} and
\eqref{def of QQ}, we rewrite the two denominator factors
containing $\cos\varphi$ in
\eqref{Upsilon in terms of varphi} as
\begin{align*}
&\quad (a-1)(a(2c+1)-1)       +   2\sqrt{a} (a+1)    \sqrt{2c+1} \cos \varphi -    4a   (c+1)   \cos \theta
\\
& 	=  (a+1)(a(2c+1)+1)      -  4a(c+1) \Big(  \cos^2 (\theta/2)+    \sqrt{ \QQ^2- \cos^2(\theta/2) } \sin (\theta/2) \Big)   
\end{align*} 
and 
\begin{align*}
&\quad 1+a(2c+1)    -      2\sqrt{a}     \sqrt{2c+1}   \cos \varphi 
\\
&= \frac{1}{a+1} \bigg[  (a+1)(a(2c+1)+1)-4a(c+1)   \Big(  \cos^2 (\theta/2) -    \sqrt{ \QQ^2- \cos^2(\theta/2) } \sin (\theta/2) \Big)  \bigg]. 
\end{align*} 
Finally, using the identity
\begin{align*}
&\quad (a(2c+1)-1) ^2 \Big(   (a+1)^2 -4a  \cos^2(\theta/2) \Big)  
\\
&= \Big(  (a+1)(a(2c+1)+1)      -  4a(c+1)   \cos^2 (\theta/2) \Big)^2 - \Big(   4a(c+1)      \sqrt{ \QQ^2- \cos^2(\theta/2) } \sin (\theta/2) \Big)^2,
\end{align*} 
we obtain the desired density \eqref{def of eq msr 1D pre}.




\begin{thebibliography}{999}

\bibitem{Ad18} K. Adhikari, \emph{Hole probabilities for $\beta$-ensembles and determinantal point processes in the complex plane}, Electron. J. Probab. \textbf{23} (2018), no. 23, 1--21.

\bibitem{ABK21} G. Akemann, S.-S. Byun and N.-G. Kang, \emph{A non-Hermitian generalisation of the Marchenko–Pastur distribution: from the circular law to multi-criticality}, Ann. Henri Poincaré \textbf{22} (2021), 1035--1068.
 

\bibitem{AFLS25} M. Allard, P. J. Forrester, S. Lahiry and B.-J. Shen, \emph{Partition function of 2D Coulomb gases with radially symmetric potentials and a hard wall}, arXiv:2506.14738.


\bibitem{AB23} Y. Ameur and S.-S. Byun, \emph{Almost-Hermitian random matrices and bandlimited point processes}, Anal. Math. Phys. \textbf{13} (2023), 52, 57pp.
 
\bibitem{ACC24} Y. Ameur, C. Charlier and J. Cronvall, \emph{Random normal matrices: Eigenvalue correlations near a hard wall}, J. Stat. Phys. \textbf{191} (2024), 98.

\bibitem{ACC26} Y. Ameur, C. Charlier and J. Cronvall, \emph{Free energy and fluctuations in the random normal matrix model with spectral gaps}, Constr. Approx. \textbf{63} (2026), 279--335.


\bibitem{ACCL22} Y. Ameur, C. Charlier, J. Cronvall and J. Lenells, \emph{Disk counting statistics near hard edges of random normal matrices: the multi-component regime}, Adv. Math. \textbf{441} (2024), 109549.

\bibitem{ACCL23} Y. Ameur, C. Charlier, J. Cronvall and J. Lenells, \emph{Exponential moments for disk counting statistics at the hard edge of random normal matrices}, J. Spectr. Theory \textbf{13} (2023), 841--902. 


\bibitem{BBLM15} F. Balogh, M. Bertola, S.-Y. Lee, and K. D. T.-R. McLaughlin, \emph{Strong asymptotics of the orthogonal polynomials with respect to a measure supported on the plane}, Comm. Pure Appl. Math. \textbf{68} (2015), 112--172.


\bibitem{BGM17} F. Balogh, T. Grava and D. Merzi, \emph{Orthogonal polynomials for a class of measures with discrete rotational symmetries in the complex plane}, Constr. Approx. \textbf{46} (2017), 109--169.

\bibitem{BM15} F. Balogh and M. Merzi, \emph{Equilibrium measures for a class of potentials with discrete rotational symmetries}, Constr. Approx. \textbf{42} (2015), 399--424.
 

\bibitem{BKP23} S. Berezin, A. B. J. Kuijlaars and I. Parra, \emph{Planar orthogonal polynomials as type I multiple orthogonal polynomials}, SIGMA Symmetry Integrability Geom. Methods Appl. \textbf{19} (2023), Paper No. 020, 18 pp.

\bibitem{BEG18} M. Bertola, J. G. Elias Rebelo and T. Grava, \emph{Painlevé IV critical asymptotics for orthogonal polynomials in the complex plane}, SIGMA Symmetry Integrability Geom. Methods Appl. \textbf{14} (2018), Paper No. 091, 34pp.
 
\bibitem{BK12} P. M. Bleher and A. B. J. Kuijlaars, \emph{Orthogonal polynomials in the normal matrix model with a cubic potential}, Adv. Math. \textbf{230} (2012), 1272--1321.


\bibitem{BS20} P. M. Bleher and G. L. F. Silva, \emph{The mother body phase transition in the normal matrix model}, Mem. Amer. Math. Soc. \textbf{265} (2020), no. 1289, v+144 pp.

\bibitem{BDHK25} P. Bourgade, G. Dubach, L. Hartung and A. Keles, \emph{Fisher-Hartwig asymptotics for non-Hermitian random matrices}, arXiv:2512.09123.

\bibitem{By24} S.-S. Byun, \emph{Planar equilibrium measure problem in the quadratic fields with a point charge},  Comput. Methods Funct. Theory \textbf{24} (2024), 303--332.

\bibitem{By26} S.-S. Byun, \emph{Anomalous free energy expansions of planar Coulomb gases: multi-component and conformal singularity}, Bull. Lond. Math. Soc. (to appear), arXiv:2508.00316.

\bibitem{BC24} S.-S. Byun and C. Charlier, \textit{On the characteristic polynomial of the eigenvalue moduli of random normal matrices}, Constr. Approx. \textbf{62} (2025), 471--521.

\bibitem{BCMS25} S.-S. Byun, C. Charlier, P. Moreillon and N. Simm, \emph{Precise large deviations in geometric last passage percolation}, Comm. Math. Phys. (to appear), arXiv:2510.17470. 

\bibitem{BF25} S.-S.~Byun and P. J.~Forrester, \emph{Progress on the study of the Ginibre ensembles}, KIAS Springer Ser. Math. \textbf{3} Springer, 2025, 221pp.

\bibitem{BF25a} S.-S.~Byun and P. J.~Forrester, \emph{Electrostatic computations for statistical mechanics and random matrix applications}, 2025 MATRIX Annals Part II, MATRIX Book Series, pp. 211--254.

\bibitem{BFKL26} S.-S. Byun, P. J. Forrester, A. B. J. Kuijlaars and S. Lahiry, \emph{Orthogonal polynomials in the spherical ensemble with two insertions}, SIAM J. Math. Anal. \textbf{58} (2026), no. 4, 3472--3509.


\bibitem{BFL25} S.-S. Byun, P. J. Forrester and S. Lahiry, \emph{Properties of the one-component Coulomb gas on a sphere with two macroscopic external charges}, Pure Appl. Funct. Anal. (to appear), arXiv:2501.05061. 

\bibitem{BKS23} S.-S. Byun, N.-G. Kang and S.-M. Seo, \emph{Partition functions of determinantal and Pfaffian Coulomb gases with radially symmetric potentials}, Comm. Math. Phys. \textbf{401} (2023), 1627--1663.

\bibitem{BKSY25} S.-S. Byun, N.-G. Kang, S.-M. Seo and M. Yang, \emph{Free energy of spherical Coulomb gases with point charges}, J. Lond. Math. Soc. (2) \textbf{112} (2025), e70294.


\bibitem{BP26} S.-S. Byun and S. Park, \emph{Large gap probabilities of complex and symplectic spherical ensembles with point charges}, J. Funct. Anal. \textbf{290} (2026), 111260.

\bibitem{BSY24} S.-S. Byun, S.-M. Seo and M. Yang, \emph{Free energy expansions of a conditional GinUE and large deviations of the smallest eigenvalue of the LUE}, Comm. Pure Appl. Math. \textbf{78} (2025), 2247--2304.


\bibitem{BYY26} S.-S. Byun, M. Yang and E. Yoo, \emph{Free energy expansion of determinantal Coulomb gases in the quadratic fields with a point charge}, arXiv:2605.29594. 

\bibitem{BY26} S.-S. Byun and E. Yoo, \emph{Three topological phases of the elliptic Ginibre ensembles with a point charge}, Constr. Approx. (Online), arXiv:2502.02948.
 


\bibitem{CFTW15} T. Can, P. J. Forrester, G. Téllez and P. Wiegmann, \emph{Exact and asymptotic features of the edge density profile for the one component plasma in two dimensions}, J. Stat. Phys. \textbf{158} (2015), 1147--1180.


 

\bibitem{Ch22} C. Charlier, \emph{Asymptotics of determinants with a rotation-invariant weight and discontinuities along circles}, Adv. Math. \textbf{408} (2022), 108600.

\bibitem{Ch23} C. Charlier, \emph{Large gap asymptotics on annuli in the random normal matrix model}, Math. Ann. \textbf{388} (2024), 3529--3587.

\bibitem{Ch23a} C. Charlier, \emph{Hole probabilities and balayage of measures for planar Coulomb gases}, Dissertationes Math. (to appear), arXiv:2311.15285.  

\bibitem{CL24} C. Charlier and J. Lenells, \emph{Balayage of measures: behavior near a corner}, Rev. Mat. Iberoam. \textbf{42} (2026), 831--854.  

\bibitem{CL24a} C. Charlier and J. Lenells, \emph{Balayage of measures: behavior near a cusp}, Potential Anal. \textbf{63} (2025), 1407--1439. 


\bibitem{CL26} G. Cipolloni and B. Landon, \emph{Gaussian multiplicative chaos for i.i.d.\ matrices}, arXiv:2605.29962.

\bibitem{CK19} J. G. Criado del Rey and A. B. J. Kuijlaars, \emph{An equilibrium problem on the sphere with two equal charges}, arXiv:1907.04801. 

\bibitem{CK22} J. G. Criado del Rey and A. B. J. Kuijlaars, \emph{A vector equilibrium problem for symmetrically located point charges on a sphere}, Constr. Approx. \textbf{55} (2022), 775--827.
 


\bibitem{DMMS25} A. Deaño, K. D. T.-R. McLaughlin, L. Molag and N. Simm, \emph{Asymptotics for a class of planar orthogonal polynomials and truncated unitary matrices}, arXiv:2505.12633.

\bibitem{DS22} A. Deaño and N. Simm, \emph{Characteristic polynomials of complex random matrices and Painlevé transcendents}, Int. Math. Res. Not. \textbf{2022} (2022), 210--264.
 
 

\bibitem{Fo10} P. J. Forrester, \emph{Log-gases and random matrices}, Princeton University Press, Princeton, NJ, 2010.
 

 

 

\bibitem{JMP94}  B. Jancovici, G. Manificat and C. Pisani, \emph{Coulomb systems seen as critical systems: finite-size effects in two dimensions}, J. Stat. Phys. \textbf{76} (1994), 307--329.


\bibitem{Joh00} K. Johansson, \emph{Shape fluctuations and random matrices}, Commun. Math. Phys. \textbf{209} (2000), 437--476.



\bibitem{KKL24} M. Kieburg, A. B. J. Kuijlaars and S. Lahiry, \emph{Orthogonal polynomials in the normal matrix model with two insertions}, Nonlinearity \textbf{38} (2025), no. 6, Paper No. 065013, 66pp. 
 

\bibitem{KLY25} T. Krüger, S.-Y. Lee and M. Yang, \emph{Local statistics in normal matrix models with merging singularity}, Comm. Math. Phys. \textbf{406} (2025), 130.
 
\bibitem{KL15} A. B. J. Kuijlaars and A. López-García, \emph{The normal matrix model with a monomial potential, a vector equilibrium problem, and multiple orthogonal polynomials on a star}, Nonlinearity \textbf{28} (2015), 347--406.
 

\bibitem{KT15} A. B. J. Kuijlaars and A. Tovbis, \emph{The supercritical regime in the normal matrix model with cubic potential}, Adv. Math. \textbf{283} (2015), 530--587.

\bibitem{LM16} S.-Y. Lee and N. Makarov, \emph{Topology of quadrature domains}, J. Amer. Math. Soc. \textbf{29} (2016), 333--369.

\bibitem{LY17} S.-Y. Lee and M. Yang, \emph{Discontinuity in the asymptotic behavior of planar orthogonal polynomials under a perturbation of the Gaussian weight}, Comm. Math. Phys. \textbf{355} (2017), 303--338.

\bibitem{LY19} S.-Y. Lee and M. Yang, \emph{Planar orthogonal polynomials as Type II multiple orthogonal polynomials}, J. Phys. A \textbf{52} (2019), 275202.

\bibitem{LY23} S.-Y. Lee and M. Yang, \emph{Strong asymptotics of planar orthogonal polynomials: Gaussian weight perturbed by finite number of point charges}, Comm. Pure Appl. Math. \textbf{76} (2023), 2888--2956.

\bibitem{LD21} A. Legg and P. Dragnev, \emph{Logarithmic equilibrium on the sphere in the presence of multiple point charges}, Constr. Approx. \textbf{54} (2021), 237--257.



\bibitem{Le22} M. Lewin, \emph{Coulomb and Riesz gases: The known and the unknown}, J. Math. Phys. \textbf{63} (2022), 061101.


\bibitem{MS19} A. Martínez-Finkelshtein and G. L. F. Silva, \emph{Critical measures for vector energy: asymptotics of nondiagonal multiple orthogonal polynomials for a cubic weight}, Adv. Math. \textbf{349} (2019), 246--315.
 

\bibitem{Noda25} K. Noda, \emph{Partition functions of two-dimensional Coulomb gases with circular root- and jump-type singularities}, SIGMA Symmetry Integrability Geom. Methods Appl. (to appear), arXiv:2510.00843. 
 
\bibitem{RM25} Rashmita and S. Mukherjee, \emph{On topology and singularities of quadrature domains}, arXiv:2509.21468. 

 

\bibitem{Rou25} N. Rougerie, \emph{Free-energy variations for determinantal 2D plasmas with holes}, arXiv:2510.01745.

\bibitem{ST97} E. B. Saff and V. Totik, \emph{Logarithmic Potentials with External Fields}, Grundlehren der Mathematischen Wissenschaften, Springer-Verlag, Berlin, 1997.

\bibitem{Sa92} M. Sakai, \emph{Regularity of a boundary having a Schwarz function}, Acta Math. \textbf{166} (1991), 263--297.


\bibitem{Se24} S. Serfaty, \emph{Lectures on Coulomb and Riesz Gases}, Amer. Math. Soc. Colloq. Publ. \textbf{70}, American Mathematical Society, Providence, RI, 2026.
 

\bibitem{TF99} G. Téllez and P. J. Forrester, \emph{Exact finite-size study of the 2D OCP at $\Gamma=4$ and $\Gamma= 6$}, J. Stat. Phys. \textbf{97} (1999), 489--521.


\bibitem{WW19} C. Webb and M. D. Wong, \emph{On the moments of the characteristic polynomial of a Ginibre random matrix}, Proc. Lond. Math. Soc. \textbf{118} (2019), 1017--1056.



\bibitem{ZW06} A. Zabrodin and P. Wiegmann, \emph{Large-$N$ expansion for the 2D Dyson gas}, J. Phys. A \textbf{39} (2006), 8933--8964.


\end{thebibliography}
\end{document}